\documentclass[twocolumn]
{aastex631}
\usepackage{graphicx}
\usepackage{epstopdf}

\usepackage{orcidlink}

\begin{document}

\title[NGC\,5972 feedback]{Jet-mode feedback in NGC\,5972: insights from resolved MUSE, GMRT and VLA observations}

\author{Arshi Ali
\orcidlink{0000-0002-7174-4221}}
\thanks{E-mail: arshiali1701@gmail.com}
\affiliation{Department of Physics, Savitribai Phule Pune University, Pune 411007, India}

\author{Biny Sebastian}
\affiliation{Space Telescope Science Institute, 3700 San Martin Drive, Baltimore, MD 21218, USA}
\affiliation{Department of Physics and Astronomy, University of Manitoba, Winnipeg, MB R3T 2N2, Canada}

\author{Darshan Kakkad}
\affiliation{Centre for Astrophysics Research, University of Hertfordshire, Hatfield, AL10 9AB, UK}
\affiliation{Space Telescope Science Institute, 3700 San Martin Drive, Baltimore, MD 21218, USA}

\author{Sasikumar, Silpa \orcidlink{0000-0003-0667-7074}}
\affiliation{Astronomy Department, Universidad de Concepción, Casilla 160-C, Concepción, Chile}

\author{Preeti  Kharb}
\affiliation{National Centre for Radio Astrophysics (NCRA) - Tata Institute of Fundamental Research (TIFR),\\ S. P. Pune University Campus, Post Bag 3, Ganeshkhind, Pune 411007, India}

\author{Christopher P. O'Dea}
\affiliation{Department of Physics and Astronomy, University of Manitoba, Winnipeg, MB R3T 2N2, Canada}
\affiliation{Center for Space Plasma \& Aeronomic Research, 
University of Alabama in Huntsville,
Huntsville, AL 35899  USA}

\author{Mainak Singha}
\affiliation{Astrophysics Science Division, NASA, Goddard Space Flight Center, Greenbelt, MD 20771, USA}
\affiliation{Department of Physics, The Catholic University of America, Washington, DC 20064, USA}
\affiliation{Center for Research and Exploration in Space Science and Technology, NASA Goddard Space Flight Center, Greenbelt, MD 20771, USA}

\author{K, Rubinur \orcidlink{0000-0001-5574-5104}}
\affiliation{Institute of Theoretical Astrophysics, University of Oslo, P.O. Box 1029, Blindern, 0315 Oslo, Norway}

\author{Stefi A. Baum}
\affiliation{Department of Physics and Astronomy, University of Manitoba, Winnipeg, MB R3T 2N2, Canada}
\affiliation{Center for Space Plasma \& Aeronomic Research, 
University of Alabama in Huntsville,
Huntsville, AL 35899  USA}

\author{Omkar Bait}
\affiliation{SKA Observatory, Jodrell Bank, Lower Withington, Macclesfield, SK11 9FT, UK
}

\author{Sravani Vaddi}
\affiliation{Department of Physics, University of Central Florida, 4111 Libra Drive, Physical Sciences Bldg. 430, Orlando, FL 32816-2385}

\author{Sushma Kurapati}
\affiliation{Department of Astronomy, University of Cape Town, Private Bag X3, Rondebosch 7701, South Africa}
\affiliation{Netherlands Institute for Radio Astronomy (ASTRON), Oude Hoogeveensedijk 4, NL-7991 PD Dwingeloo, the Netherlands}




\begin{abstract}

NGC\,5972, a Voorwerp galaxy, features a helical-shaped extended emission-line region (EELR) with a radius $> 10$ kpc and a S-shaped radio structure spanning about 470 kpc. We use VLT MUSE, GMRT, and VLA to study the stellar and ionized gas kinematics and how the radio jet influences the gas in the galaxy. Our sensitive radio observations detect the southern jet for the first time, roughly coinciding with the southern EELR. 
The VLA images show a continuous inner jet connected to the outer E-W lobe, confirming the jet origin of the radio emission. Our kinematic analysis shows spatial correlations between the radio jet and the outflowing gas, supporting the jet-driven feedback mechanism. More interestingly, we observe enhanced velocity dispersion in the perpendicular direction along with a shell-like structure. Our BPT analysis shows that the [O\,{\small III}] emission overlapping with the radio jet is consistent with the
shock+precursor model, whereas in the perpendicular region, a pure shock model fits well with the observations, indicating jet-induced shocks. Radio observations indicate episodic AGN activity characterized by surface brightness and spectral index discontinuities. Overall, based on our findings, we propose a jet-driven feedback mechanism as one of the key factors in the formation of the EELR in NGC\,5972. Future high-resolution radio observations will be crucial to further investigate the origin of the EELR and quantify the extent to which the jet influences its formation and evolution.

\end{abstract}




\section{Introduction} 
\label{sec:intro}

Active galactic nuclei (AGN) feedback has become an important ingredient in cosmological models for galaxy evolution, as it helps to reproduce the crucial observational properties of the 
galaxies \citep{Bower2006, McCarthy2010, Silk2012, Schaye2015, Choi2018}. Substantial observational evidence also supports AGN feedback independent of these models. For instance, the M-$\sigma_\star$ relation \citep{Ferrarese2000, Gutekin2009} suggests a strong correlation
between black hole growth and the properties of the host galaxy. Additionally, the slower gas cooling rate in massive galaxies points to AGN-driven heating as a major mechanism preventing excessive cooling and regulating star formation \citep{1995MNRAS.276..663B, 2003ApJ...590..207P}. These findings highlight the significance of studying AGN feedback mechanisms.

This paper primarily focuses on studying the jet-driven feedback mechanism, as it provides direct evidence of how energy is transferred from relativistic jets to the surrounding gas, affecting its physical and kinematics properties across various spatial scales \citep{Fabian2012, King2015, Morganti2016, 2018NatAs...2..198H, Hardcastle2020, Krause2023}. Recent studies \citep{2021MNRAS.503.1780J, 2021A&A...648A..17V, Venturi2023, 2022MNRAS.513.4208S, 2022hypa.confE...2G} find that a large fraction of radio-quiet quasars with kpc-scale ionized outflows show small-scale radio jets, and the gas exhibits enhanced velocity dispersion in the direction perpendicular to the jet. This suggests that the jet -interstellar medium (ISM) interaction is a major player in AGN feedback. Conversely, it was also found that most of the powerful radio galaxies exhibit EELR that can extend up to hundreds of kpcs \citep[e.g.,][]{Baum1988,McCarthy1995, Balmaverde2022}. These EELR provide a unique laboratory to study the origin of the AGN activity as well as the various feedback mechanisms \citep{2010ApJ...724L..30S,2012AJ....144...66K,2014ApJ...786....3S,2015AJ....149..155K,2016MNRAS.457.3629S,2020CoSka..50..309K}.

Voorwerp galaxies constitute a distinct category of emission line galaxies, which came to light through the involvement of citizen scientists in the Galaxy Zoo project \citep{2009MNRAS.399..129L,2009A&A...500L..33J,2011AAS...21714208K}. These galaxies are recognized for the prevalence of robust doubly ionized oxygen ([O\,{\small III}]$\lambda$5007, [O\,{\small III}] hereafter) gas. The emission-line ratios observed in Voorwerp galaxies bear a striking resemblance to those found in the EELR \citep{2011AAS...21714207C,2012MNRAS.420..878K}. These sources are suggested to be examples of quasar ionization echoes from previous episodes of AGN activity \citep{2009MNRAS.399..129L}.
The radio imaging of several of these galaxies suggested that the [O\,{\small III}] emission is coincident with radio emission (\textit{viz.,} “Teacup Quasar” \citep{2015ApJ...800...45H, Venturi2023}, Mrk 78 \citep{2004AJ....127..606W}, NGC 4388 \citep{2020MNRAS.499..334S}.  
More recently \citet{2022MNRAS.514.3879S} uncovered the presence of an old relic radio emission from the ``Hanny's Voorwerp" galaxy, IC2497. They argue that the radio jets play a significant role in shaping the ionization structure within their host galaxies. Similarly, a recent study of the Teacup Quasar \citep{Venturi2023} also shows that the jet strongly perturbs the host ISM. According to previous studies, the feedback in the Voorwerp galaxies is primarily caused by AGN photoionization \citep{2009MNRAS.399..129L,2012MNRAS.420..878K,2017ApJ...835..256K}, however, the presence of a radio jet in these galaxies prompts a discussion of whether jet also plays a significant role in the feedback.

Since the majority of the Voorwerp galaxies are Seyfert galaxies, the origin of the radio emission itself is highly debated.
The correlation between [O\,{\small III}] and radio luminosity in Seyfert galaxies has been known for several decades \citep{deBruyn1978,Schmitt2003}. Similar spatial coincidences were also seen in radio galaxies \citep{Baum1989}. 
While such spatial correlations might immediately suggest a jet-related origin, alternate explanations, including shock acceleration due to winds driven by the 
AGN accretion leading to radio emission has been proposed (e.g., \cite{Zakamska2014}) to explain such correlations in radio-quiet systems. 

In this paper, we investigate the impact of radio jets on the ionized gas morphology and kinematics in NGC\,5972, a Voorwerp galaxy, using optical IFS observations from the Multi Unit Spectroscopic Explorer (MUSE) at the Very Large Telescope (VLT), low frequency (610~MHz) observations from the Giant Metrewave Radio Telescope (GMRT), and high frequency (C-band) observations from the Karl G. Jansky Very Large Array (VLA).
This paper is organized as described below. Section~\ref{source} describes the detail about our source, NGC\,5972.
In Section~\ref{obs}, we present the description of MUSE, GMRT, and VLA observations and data analysis. Section~\ref{sec:results} consists of results and in Section~\ref{discussion} we discuss their implications.
The conclusions are summarized in Section~\ref{conclusion}. 

We have assumed $\rm H_0$=73~km~s$^{-1}$Mpc$^{-1}$, $\Omega_{m}=0.27$ and $\Omega_{vac}=0.73$ in this paper. Spectral index $\alpha$ is defined such that flux density at frequency $\nu$ is $\rm S_\nu\propto\nu^\alpha$.


\section{NGC\,5972}
\label{source}
From a sample of 19 Voorwerp galaxies listed in \citet{2012AJ....144...66K}, NGC\,5972 was chosen because of the availability of the science-processed MUSE data cube, and low-redshift ($z$= 0.02964, where $\rm 1^{\prime\prime}=0.593~kpc$) which will enable resolved studies. 
NGC\,5972 is categorized as a Seyfert type-2 galaxy with a distinctive arrangement of ionized gas, featuring a striking double-lobed structure \citep{1995A&A...296..315V}.
It has an EELR with a radius of $\sim$12~kpc, which can be seen in HST images shown in \cite{2015AJ....149..155K}, revealing rich filamentary structures. Furthermore, NGC\,5972 presents compelling findings in the literature that are indicative of past AGN activity \citep{2022ApJ...936...88F,2022arXiv220805915H}. \cite{2022ApJ...936...88F} had
conducted an ellipse fitting analysis of the F621M HST image, which shows multiple asymmetric tidal structures within a few kpc from the center. The formation of these structures is discussed by \cite{2015AJ....149..155K, 2022ApJ...936...88F} as a probable result of past mergers or interactions.
Twisted dust structures can also be observed near the central region in the HST image. \cite{2015AJ....149..155K} conducted an extensive analysis on these intertwined dust lanes, and proposed a deferentially precessing, warped disk model \citep{1992AJ....104.1339S} as the most plausible explanation for these structures.


\section{Observations and Data analysis}
\label{obs}
\begin{table}
\label{sec:obs_table}
  \begin{center}
    \caption{Details for GMRT and VLA data}
    \begin{tabular}{cc}
     
        \hline
        \hline
        
        GMRT\\
        \hline
        Observation Date & 2022-05-23 \\
        $\nu$ (MHz) & 610 \\
        Beam, PA (arcsec$^2$, $\degr$) & 6.5$\times$5.0, 87.97 \\
        Image peak flux density (mJy) & 6.06 \\
        Image r.m.s (mJy~beam$^{-1}$) & 0.075\\
        On source time (min) & 150 \\
        
        \hline
        VLA\\
        \hline

        Array Configuration &
        D \\
        Observation Date & 2015-12-13 \\
        $\nu$ (GHz) & 6.0 \\
        Beam, PA (arcsec$^2$, $\degr$) & 11.3$\times$9.6, 54.11 \\
        Image peak flux density (mJy) & 4.22 \\
        Image r.m.s (mJy~beam$^{-1}$) & 0.034\\
        On source time (min) & 35 \\

        \hline
        VLA\\
        \hline

        Array Configuration &
        B \\
        Observation Date & 2023-01-14 \\
        $\nu$ (GHz) & 5.5 \\
        Beam, PA (arcsec$^2$, $\degr$) & 2.43$\times$1.14, 65.03 \\
        Image peak flux density (mJy) & 4.8 \\
        Image r.m.s (mJy~beam$^{-1}$) & 0.008\\
        On source time (min) & 60\\ \hline

    \end{tabular}
    \label{tab:HI_properties}
  \end{center}
\end{table}


\subsection{MUSE }

\label{muse}

\begin{figure*}
\centering
\includegraphics[height=7.cm,trim = 0 0 0 0 ]{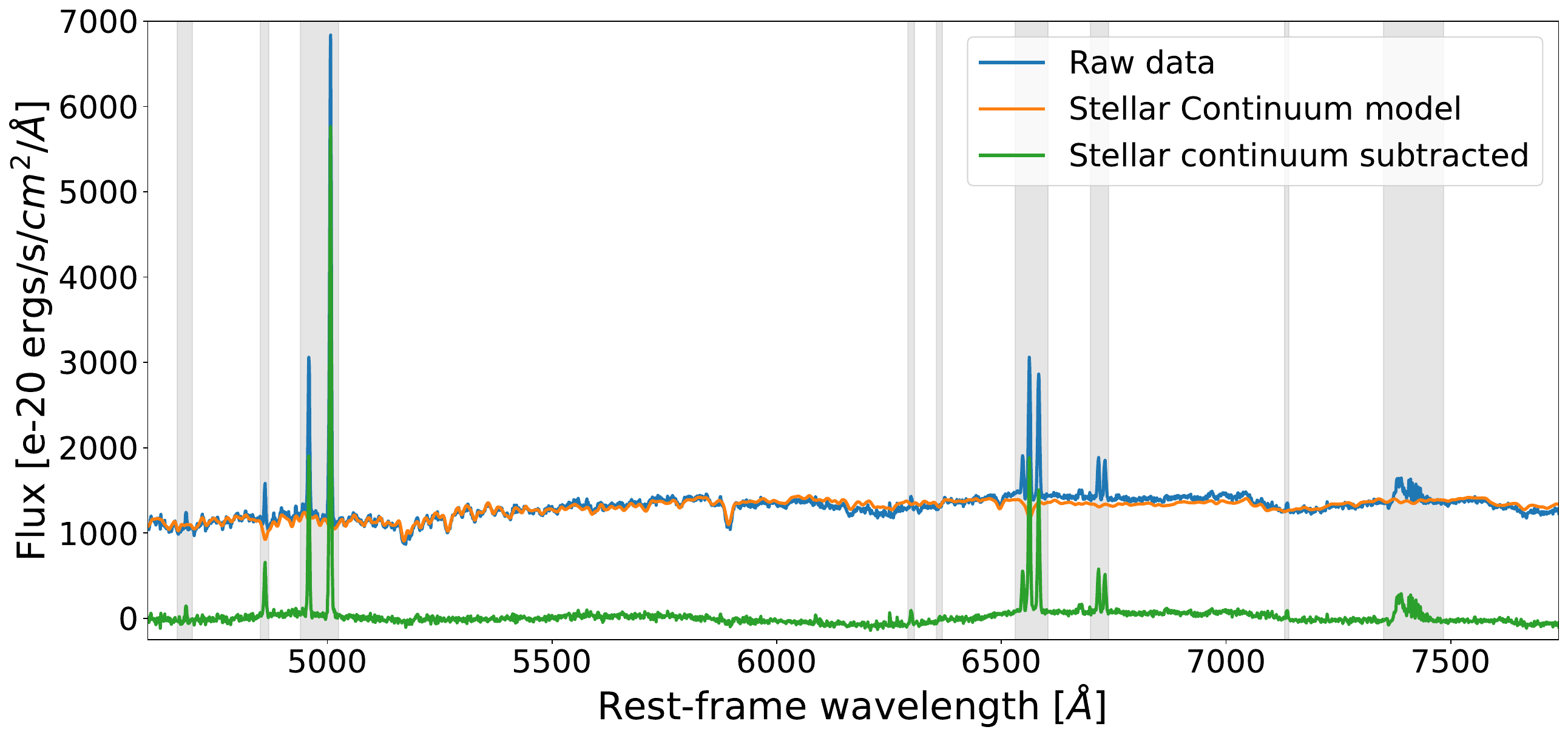}
\includegraphics[height=7cm,trim = 0 0 0 0 ]{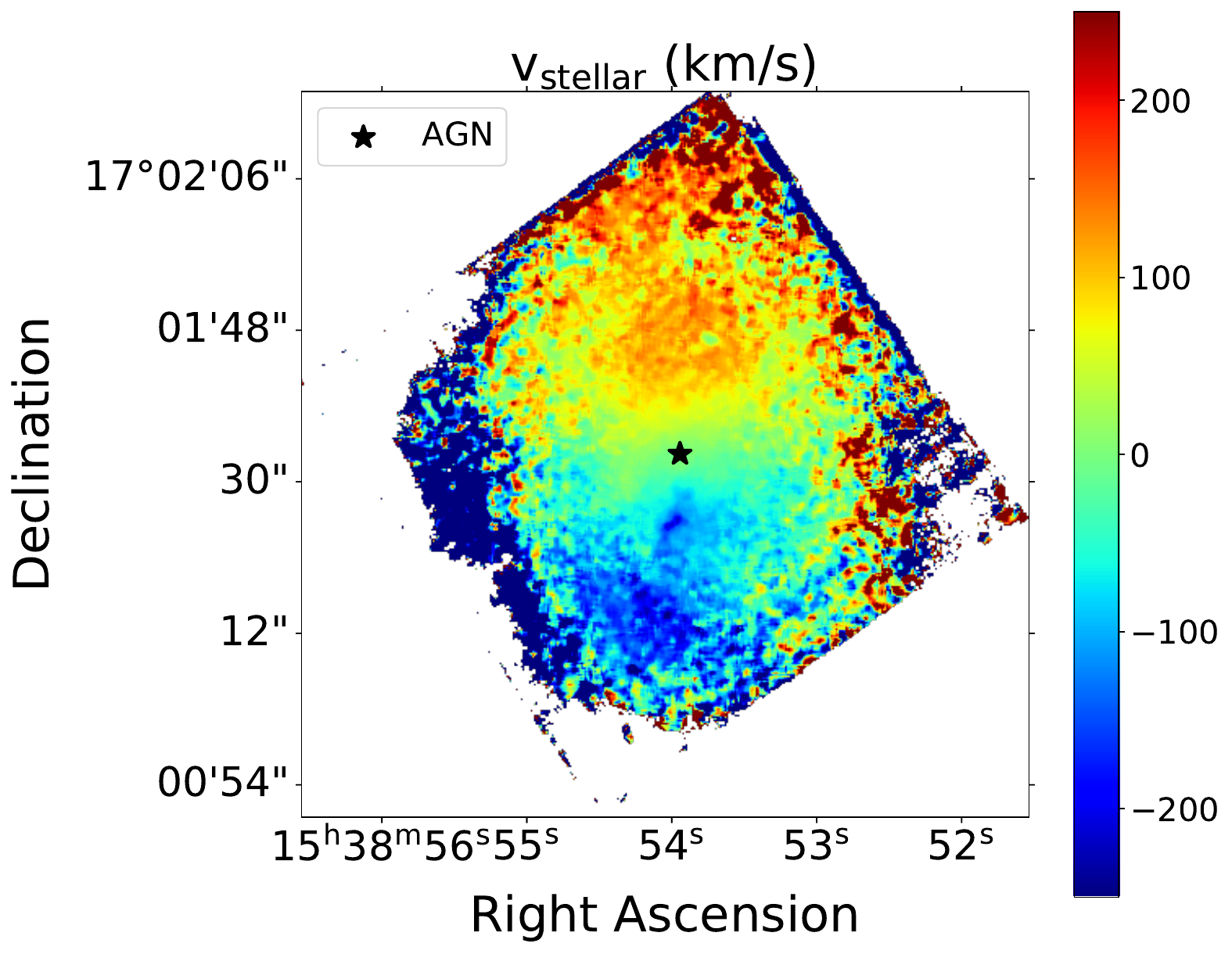}
\includegraphics[height=7cm,trim = 0 0 0 0 ]{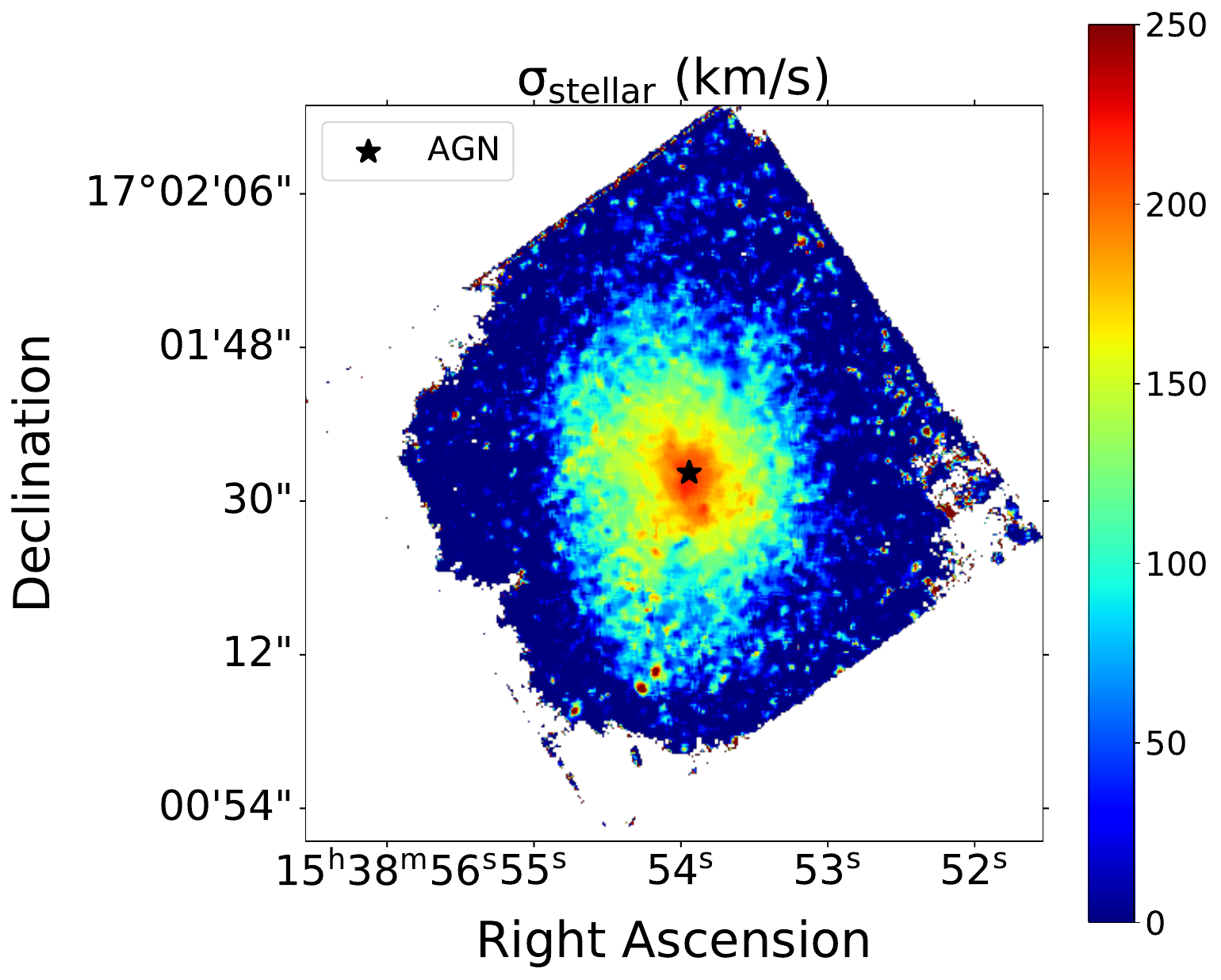}

\caption
{Top panel: Stellar continuum fit for one of the spaxels of the MUSE data. Excluded sky emission lines are highlighted with shaded regions. Bottom panels: line-of-sight stellar velocity and stellar dispersion maps, obtained after the pPXF fitting. 
}
\label{ppxf_fit}
\end{figure*}



\begin{figure*}
\centering

\includegraphics[height=5.5cm,,trim = 0 0 0 0 ]{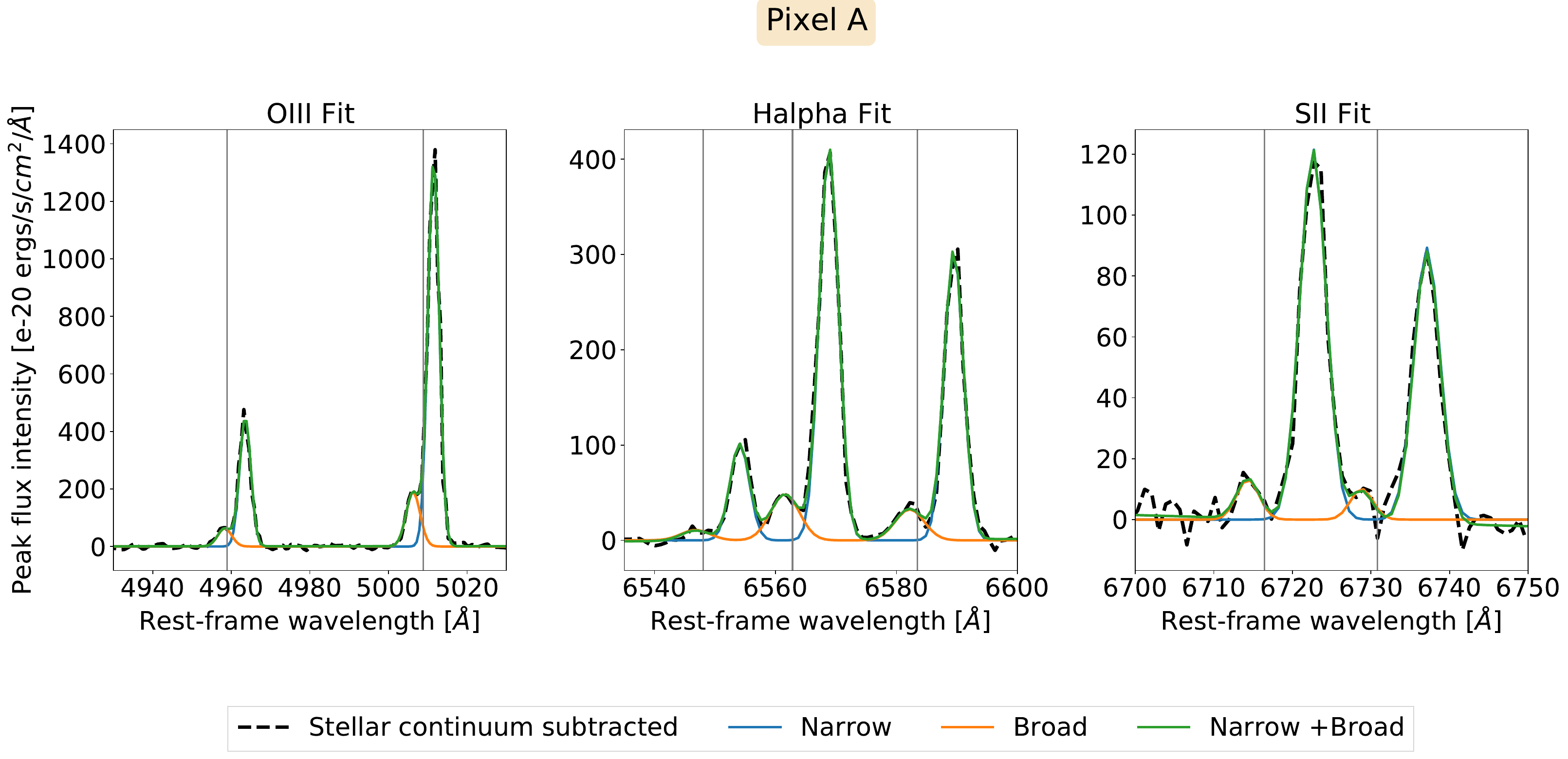}
\includegraphics[height=5.5cm,,trim = 0 0 0 0 ]{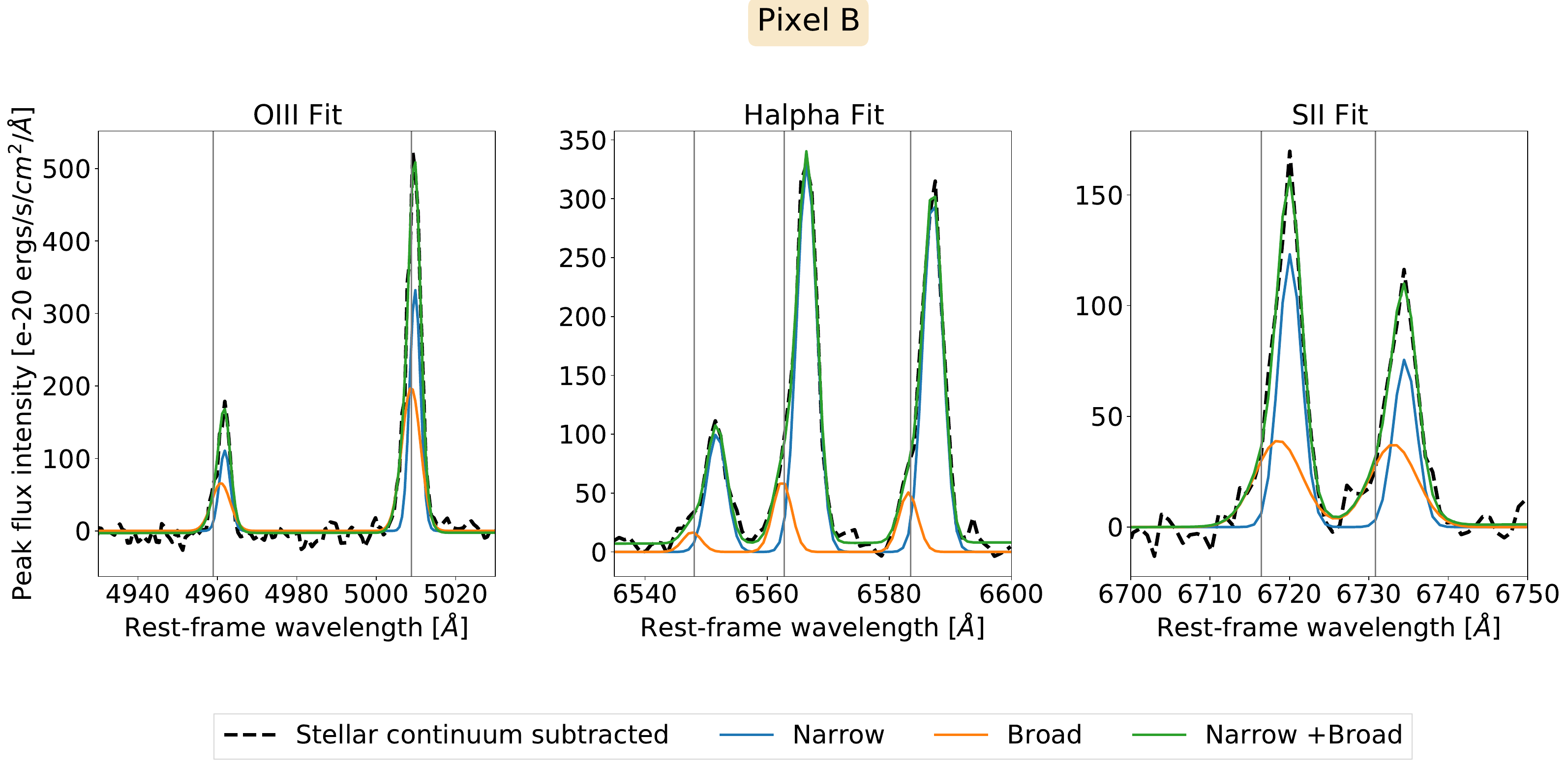}
\includegraphics[height=5.5cm,,trim = 0 0 0 0 ]{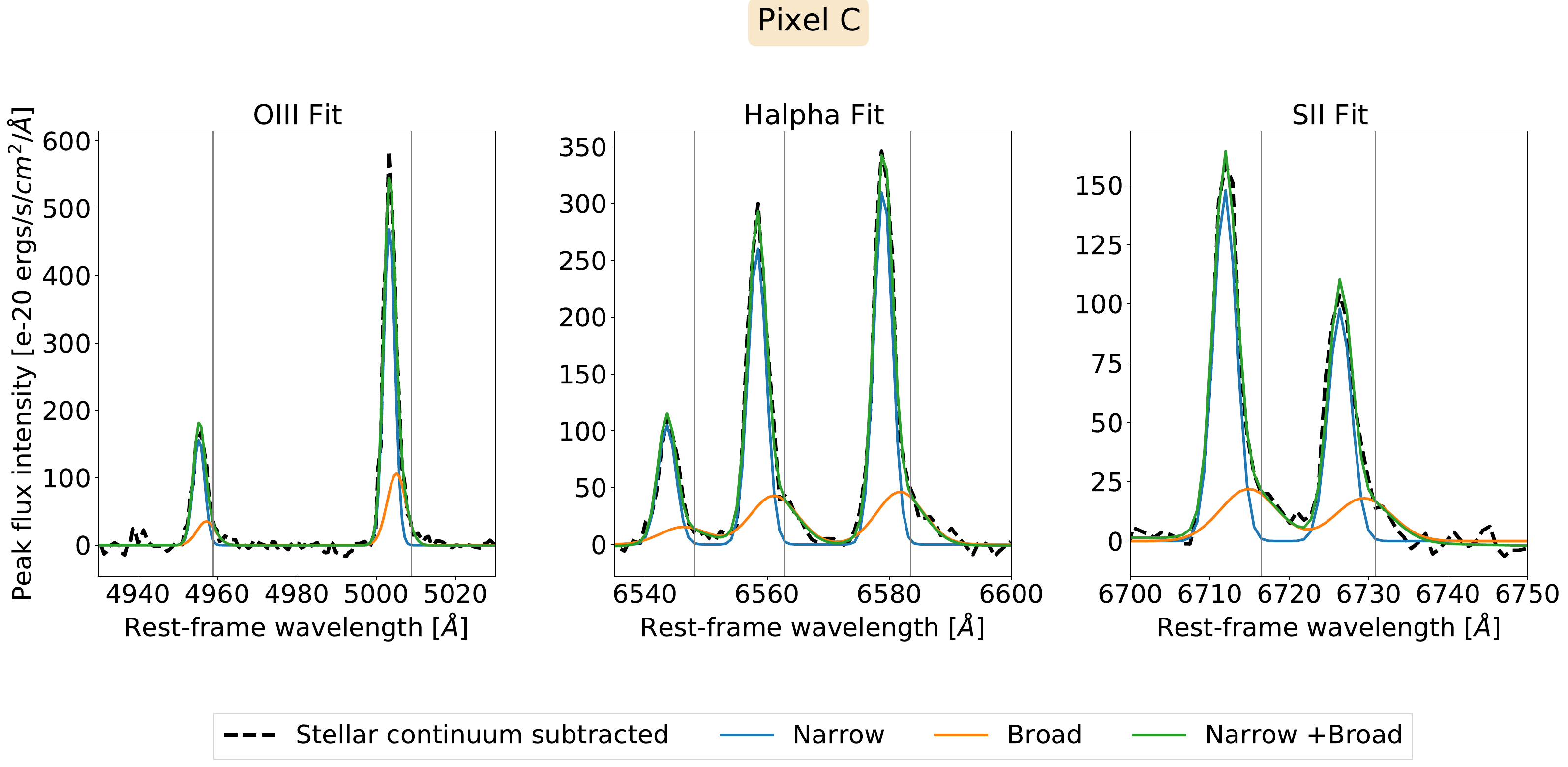}
\includegraphics[height=5.5cm,,trim = 0 0 0 0 ]{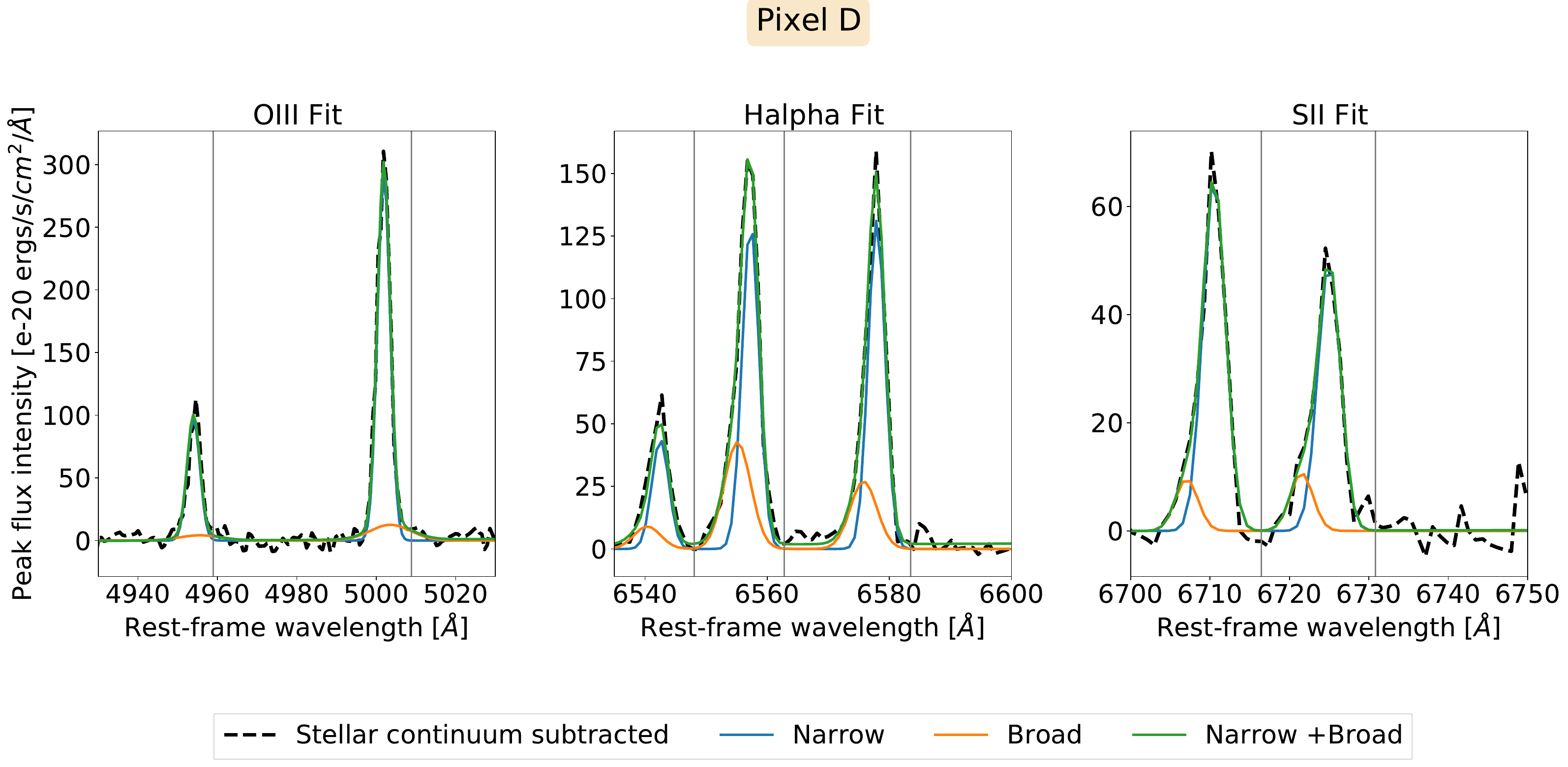}

\caption
{Example of double Gaussian fitting for four distinct locations extracted from the [O\,{\small{III}}] residual fit for Gaussian 1 component, as shown in Figure~\ref{spaxel}. Left panels: [O\,{\small{III}}] $\lambda4958, 5007$ line profile. Middle panels: H$\alpha$ $\lambda6562$, N\,{\small{II}} $\lambda6548,6583$ line profile. Right panels: S\,{\small{II}} $\lambda6716,6730$ line profile. In each plot, the stellar continuum subtracted data is represented by the dashed black curve, the narrow component (Gaussian 1) by the blue curve, the broad component (Gaussian 2) by the orange curve, and the narrow+broad (total double Gaussian) by the green curve. Solid grey lines indicate the rest-frame wavelength of the respective emission lines.
}
\label{gauss_fit}
\end{figure*}

The MUSE \citep{Bacon2010} is an integral field spectrograph located on VLT. NGC\,5972 was observed by the MUSE as part of program 0102.B-0107 (PI: SARTORI) on March 10, 2019. We downloaded the fully reduced and calibrated science data cube from the ESO data archive.\footnote{\url{http://archive.eso.org/}} Observations of NGC\,5972 were conducted in wide-field mode (WFM) with a FoV of 1$\arcmin$x1$\arcmin$, and a pixel scale of 0.2$\arcsec$. More details regarding the observation and data reduction is discussed in \cite{2022ApJ...936...88F}. MUSE data covers a wavelength range of 4600-9300 $\rm \AA$. However, for our purposes, we have only utilized the range between 4600-8800 $\rm \AA$.

To perform the stellar continuum subtraction, we employed the penalized PiXel-Fitting procedure, pPXF \citep{2004PASP..116..138C, 2017MNRAS.466..798C} to analyze the entire MUSE FoV (2.03\arcmin) which encompasses approximately 180,000 spaxels.
In order to ensure accurate results, we masked out regions in the spectra that contained strong skylines and emission lines. Specifically, the following lines were masked: He\,{\small{II}} $\lambda4685$, H$\beta$ $\lambda4861$, O\,{\small{III}} doublets $\lambda4958, 5007$, H$\alpha$ $\lambda6562$, N\,{\small{II}} doublets $\lambda6548,6583$, O\,{\small{I}} $\lambda6300$, S\,{\small{II}} doublets $\lambda6716,6730$, and Ar\,{\small{III}} $\lambda7135$. Figure~\ref{ppxf_fit}: top, shows an example of a stellar continuum fit for one of the pixels, along with the stellar velocity (bottom left) and stellar dispersion velocity (bottom right) respectively.

We subtracted the modeled stellar continuum emission from the raw data, and the resulting continuum-subtracted cube was utilized to perform the single or double Gaussian fitting based on the complexity of the line profiles in different regions. We have used the \texttt{scipy.optimize.curve\_fit} Python package \citep{2020NatMe..17..261V} to perform the fitting procedure. The primary objective of this step is to extract the morphological and kinematic information to study the distribution and motion of ionized gas within the host galaxy. We observed that the region within 6~kpc from the nucleus cannot be fitted using only the single Gaussian component, thus we added another Gaussian component
to account for the additional (outflowing) component.
After running the fitting procedure once, we intentionally adjusted the widths of Gaussian 1 and Gaussian 2 by comparing them. Upon this comparison, we swapped the widths to ensure that Gaussian 1 is identified as the narrow component, while Gaussian 2 is referred to as the broad component. 
The peak intensity of the second component is initialized as half the peak intensity of the first component. 
To guide the fitting process, we introduce certain constraints to the parameters governing the centroids and FWHM of the individual Gaussian components used to model the emission lines. 
The centroids of the following lines: H$\beta$ $\lambda4861$, [O\,{\small{III}}] doublets $\lambda4958, 5007$, H$\alpha$ $\lambda6562$,[N\,{\small{II}}] doublets $\lambda6548,6583$, and [S\,{\small{II}}] doublets $\lambda6716,6730$ were tied together based on their anticipated positions within the rest-frame spectra.

The line fluxes were left unconstrained for all the lines, except for the line ratios  [O{\small{III}}]$\lambda$5007/$\lambda$4958 and [N\,{\small{II}}]$\lambda$6548/$\lambda$6583, which were set at 3 as per established theoretical values \citep{2000MNRAS.312..813S, 2007MNRAS.374.1181D}. Gaussian fitting for four different locations in the EELR (Figure~\ref{spaxel}) are shown in Figure~\ref{gauss_fit}.

Our line ratio maps made using the spaxel-by-spaxel analysis were limited to the high SNR regions (refer to Section~\ref{bpt_sec}). 
 Hence, we opted to transition to Voronoi binning to improve our sensitivity and understand the nature of the weaker emission regions. We have used the Galaxy IFU Spectroscopy Tool\footnote{\url{https://abittner.gitlab.io/thegistpipeline/}}, $\tt{GIST}$ \citep{2019ascl.soft07025B} for this step.
$\tt{GIST}$ uses a python-implemented version of $\tt{pPXF}$ and $\tt{GANDALF}$ \citep{2006MNRAS.366.1151S, 2006MNRAS.369..529F, 2019ascl.soft07025B} to provide stellar kinematics and emission-line properties, respectively.
The pipeline first creates Voronoi bins of the data cube, such that we get a constant SNR across all bins. Bins with continuum SNR$\leq$5 were discarded to reduce the noisy spectra, and a minimum SNR cut of 20 is applied on the emission lines. This process resulted in the division of the galaxy into 2103 Voronoi bins over the galaxy-scale frame.
Using these bins, the pipeline then performs a stellar kinematics routine using $\tt{pPXF}$. We have used the MILES library \citep{2015MNRAS.449.1177V} as a template for stellar population synthesis. 
The stellar spectrum is subtracted from the observed spectra, and the emission line profiles are fitted using $\tt{pyGANDALF}$. The algorithm returns the following list of parameters for each line: flux, amplitude, line-of-sight velocity, and velocity dispersion. For the binning procedure, we have set the wavelength range from 4000\,\text{\AA} - 6800\,\text{\AA} which covers the required range of emission lines used in the ``Baldwin, Phillips \& Terlevich'' (BPT) analysis (refer to Section~\ref{bpt_sec}).

\subsection{VLA}

\subsubsection{Archival data at 6~GHz and new data at 5.5~GHz}
\label{VLA_archive}
We used archival VLA \citep{Napier1983, Perley2011} data
of NGC\,5972 at 6~GHz (C-band), which was observed on 13 Dec 2015 (PI: Schawinski, Project Id: 15B-145). The details of the observations are shown in Table~\ref{tab:HI_properties}. 3C\,286 was used for flux density and bandpass calibration, whereas J1608+1029 was used as the phase calibrator.
Data reduction and calibration of continuum data were performed with the NRAO Common Astronomy Software Applications package (CASA), version 6.2.1‑7, using the calibration and imaging pipeline\footnote{\href{https://science.nrao.edu/facilities/vla/data-processing/pipeline/VIPL}{https://science.nrao.edu/facilities/vla/data-processing/pipeline/VIPL}}. We carried out three rounds of phase-only self-calibration on the data while reducing the `solint' with every iteration.

We observed NGC\,5972 using the VLA at 5.5 GHz in the B-array configuration on 14 Jan 2023 (Project Id: 23A-264, PI: Ali). Details of the observations are shown in Table~\ref{tab:HI_properties}. 3C\,286 was used for flux density and bandpass calibration, whereas J1504+1029 was used as the phase calibrator. The continuum data were calibrated and edited using the NRAO CASA calibration and imaging pipeline. We then carried out the manual execution of the polarization calibration steps. The strongly polarized 3C\,286 was used as the polarization angle calibrator, while the unpolarized calibrator OQ\,208 was used for leakage calibration. 

The polarization calibration steps included: (i) manually setting the polarization model for 3C\,286 using the $\tt{CASA}$ task $\tt{SETJY}$. Parameters such as the reference frequency, the total intensity value at the reference frequency, the spectral index, and the coefficients of the polynomial expansion of fractional polarization and polarization angle as functions of frequency about the reference frequency were provided to define the model; (ii) solving the cross-hand (RL, LR) delays arising from residual delay differences between the right and left circularly polarized signals. 
This step was carried out using 3C\,286 in the {CASA} task $\tt{GAINCAL}$ with $\tt{gaintype=KCROSS}$; (iii) solving instrumental polarization (`D-terms' or antenna leakages) arising from imperfect and non-orthogonal antenna feeds, or cross-talk between the feeds. This step was carried out using OQ\,208 in the $\tt{CASA}$ task $\tt{POLCAL}$ with $\tt{poltype=Df}$\footnote{The parameter $\tt{poltype}$ is set to $\tt{Df+QU}$ if the polarized calibrator is used for leakage calibration. We did not use 3C\,286 for leakage calibration here since we had not acquired multiple scans of 3C\,286 to ensure a good parallactic angle coverage.}. Five antennas were found to have very high leakage. Therefore, they were flagged at the beginning of the polarization calibration steps. The final leakages obtained were typically $<15$\%, and finally, (iv) solving the residual R-L phase difference on the reference antenna. This step was carried out using 3C\,286 in the $\tt{CASA}$ task $\tt{POLCAL}$ with $\tt{poltype=Xf}$.

After applying the calibration solutions to the multisource data set, we extracted the visibility data for NGC\,5972 using the $\tt{CASA}$ task $\tt{SPLIT}$ while also averaging the spectral channels for reduced data volumes without introducing the bandwidth smearing effects. We used the multiterm–multifrequency synthesis \citep[MT-MFS;][]{RauCornwell11} algorithm in the $\tt{TCLEAN}$ task in $\tt{CASA}$ to create the continuum or Stokes I image of NGC\,5972. We carried out three rounds of phase-only self-calibration followed by one round of amplitude and phase self-calibration. The last self-calibrated visibility data was imaged for Stokes Q and U using the same input parameters as for the Stokes I image except for a fewer number of iterations and the Stokes parameter. 

We combined Stokes Q and U images using the $\tt{AIPS}$ task $\tt{COMB}$ with $\tt{opcode=POLC}$ (which corrects for Ricean bias) to create the linear polarized intensity ($P=\sqrt{Q^2+U^2}$; $\tt{PPOL}$) image and with $\tt{opcode=POLA}$ to create the polarization angle ($\chi$ = 0.5 tan$^{-1}$(U/Q); $\tt{PANG}$) image. We blanked the regions with intensity values less than 3 times the rms noise and with angle errors greater than $10^\circ$ while making $\tt{PPOL}$ and $\tt{PANG}$ images, respectively. We created the fractional polarization ($\tt{FPOL=P/I}$) image from the $\tt{PPOL}$ and Stokes I images using the task $\tt{COMB}$ with $\tt{opcode=DIV}$. We blanked the regions with fractional polarization errors $>$10\%.



\subsection{GMRT}
\label{gmrt}

\begin{figure*}
\centering
\includegraphics[height=12cm,trim = 0 0 0 0]{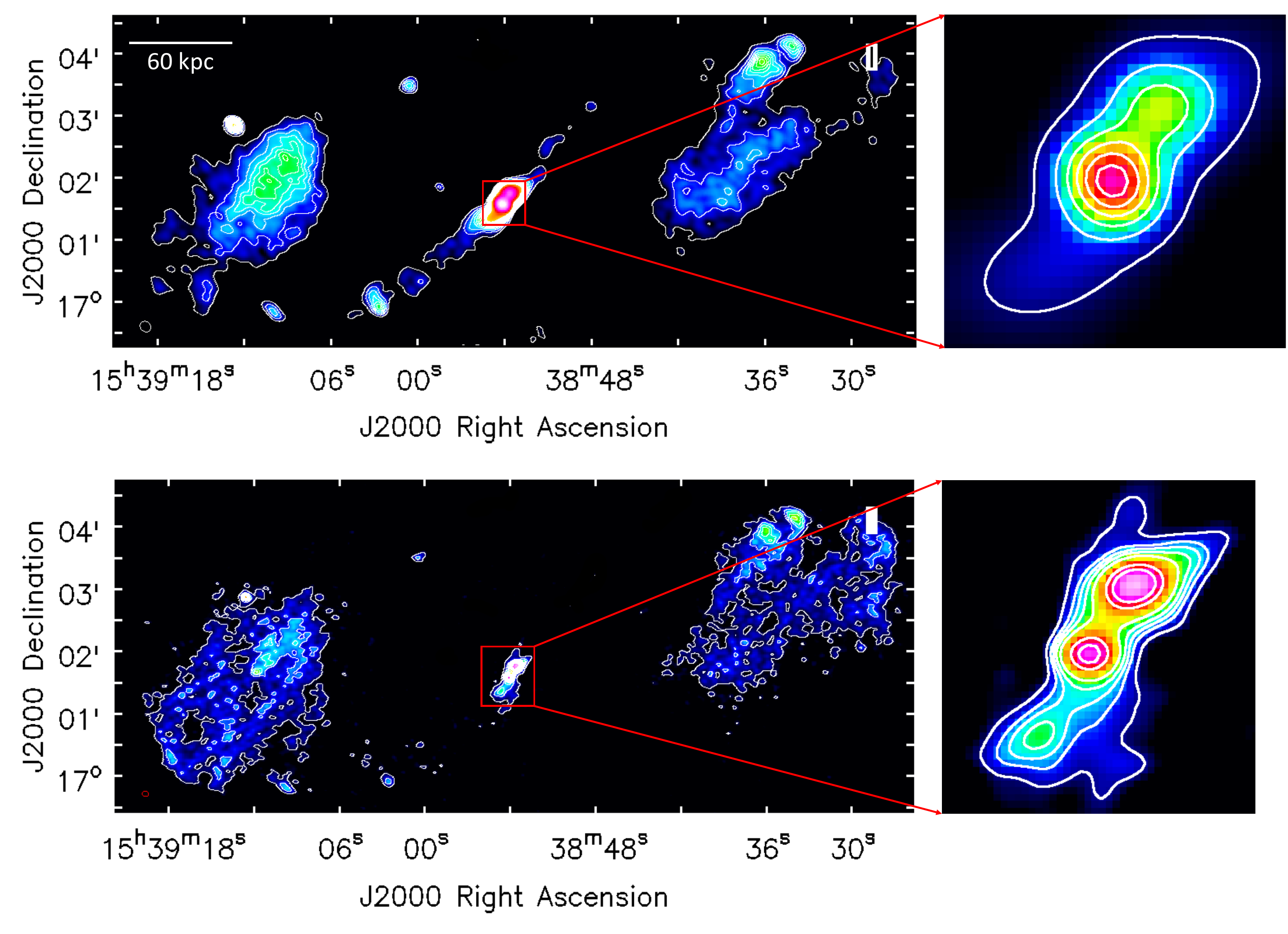}

\caption
{Top: Archival VLA D-array image at 6 GHz. Bottom: GMRT image at 610 MHz. The contour levels used are: 3$\sigma\times$(1, 2, 4, 8, 16, 32, 64), where $\sigma$= 45.5$\mu$Jy beam$^{-1}$ for 6 GHz VLA image, and $\sigma$= 98.2$\mu$Jy beam$^{-1}$ for GMRT 610 MHz image.
}
\label{radio_images}
\end{figure*}

\begin{figure*}
    \centering
    \includegraphics[width=\textwidth]{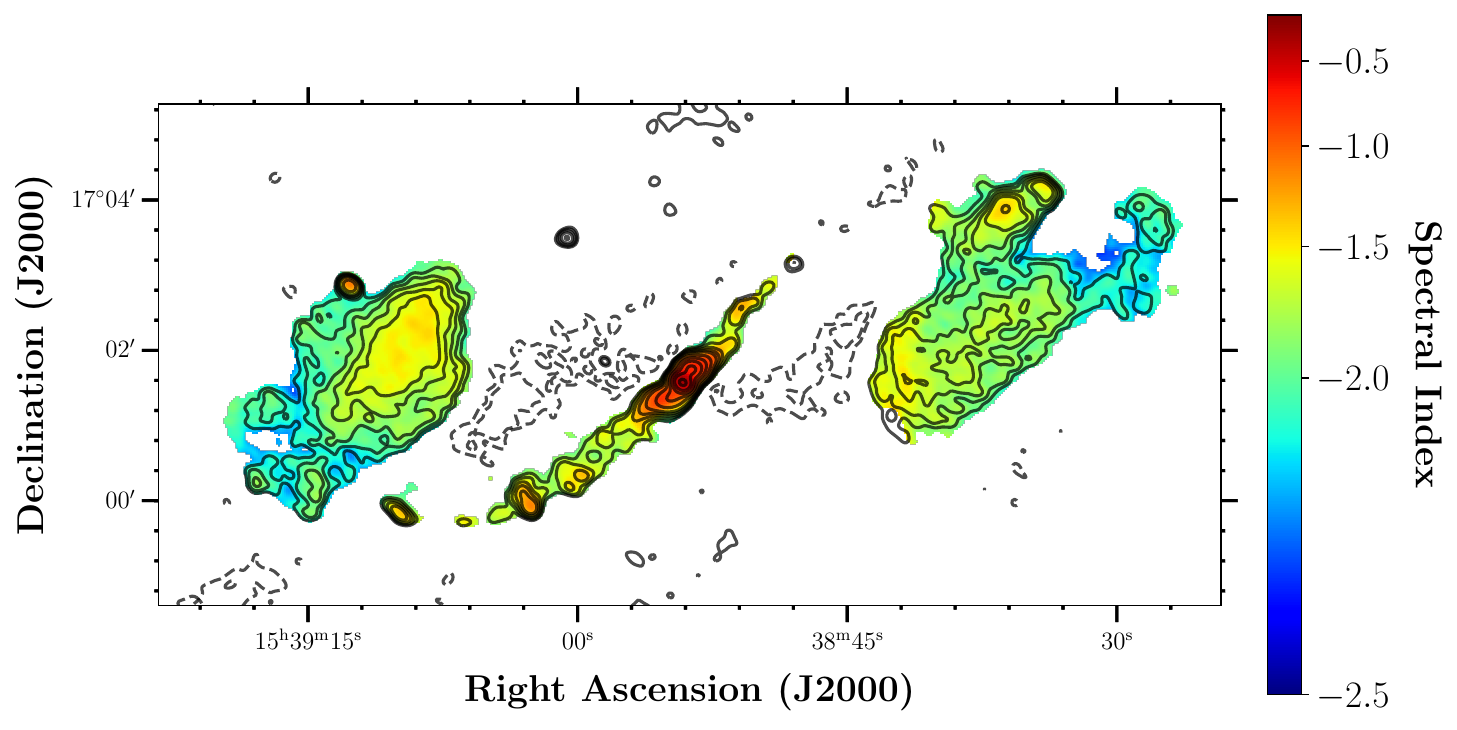}
    \caption{Spectral index image from VLA 6~GHz and GMRT 610~MHz data in color overlaid with 6~GHz total intensity contours. The contour levels are 35$\times(-1.4, -1, 1, 1.4, 2, 2.8, 4, 8, 16, 32, 64, 128, 256, 512, 1024, 2048)$$\rm~\mu Jy~beam^{-1}$.}
    \label{fig:spix}
\end{figure*}
The GMRT observation for NGC\,5972 at 610 MHz (Band-4) was carried out on 23 May 2022 (proposal code: 42\textunderscore015, PI: Ali). For our observations, we have used
3C\,286 (polarized calibrator) as the primary flux calibrator, OQ\,208 (unpolarized calibrator) as the polarization leakage calibrator, and 1347+122 as the phase calibrator. The data analysis was carried out using the GMRT data analysis pipeline {\tt aipsscriptwriter}\footnote{\href{https://github.com/binysebastian/aipsscriptwriter}{https://github.com/binysebastian/aipsscriptwriter}} \citep{Sebastian2024}. It uses both {\tt AIPS }and {\tt CASA} tasks to carry out the initial editing and flagging of bad data.
The pipeline uses standard procedures in {\tt AIPS} to calibrate and image the data. The GMRT image of the galaxy at 610 MHz is shown in Figure~\ref{radio_images} (top).

The spectral index image made using the VLA 6~GHz D-array and GMRT 610~MHz images convolved to the same beam size is presented in Figure~\ref{fig:spix}.
We find that the average spectral index value in the inner lobes is $-1.14\pm0.09$, while the average spectral index in the outer western lobe is $-1.89\pm0.13$ and outer eastern lobe is $-1.89\pm0.17$.


\section{Results }
\label{sec:results}


\subsection{Morphologies}
\label{sec:morphology}

The radio morphology of NGC\,5972 is shown in Figure~\ref{radio_images}. The GMRT 610~MHz image and VLA 6~GHz images display the presence of outer radio lobes, extending up to $\sim$235~kpc in radius, which is remarkable, considering that in typical Seyfert galaxies, the radio lobes can usually be traced out to only a few tens of kpc \citep{Baum1993, Colbert1996, 2006AJ....132..546G, 2020MNRAS.499..334S}, making NGC\,5972 a rare case. 

\begin{figure*}
\centering
\includegraphics[height=6.5cm,trim = 0 0 0 0 ]{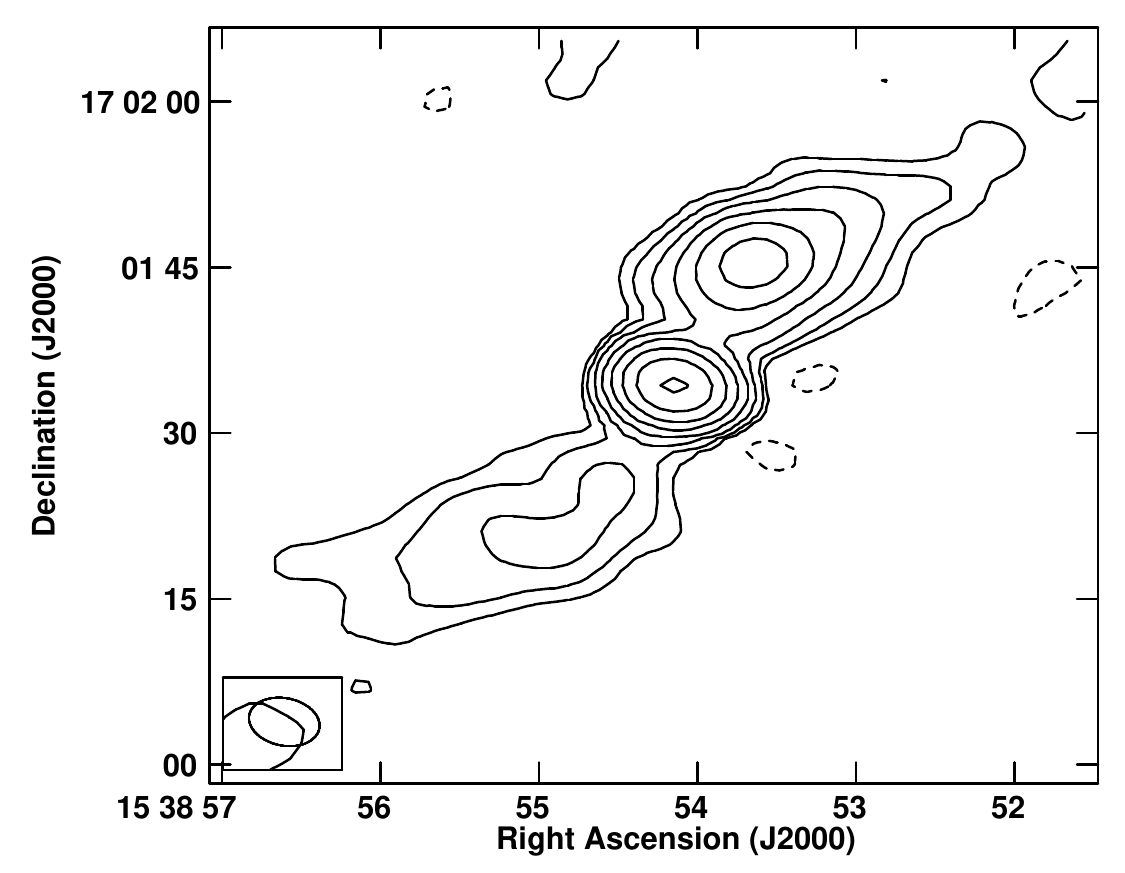}
\includegraphics[height=6.5cm,trim = 0 0 0 0 ]{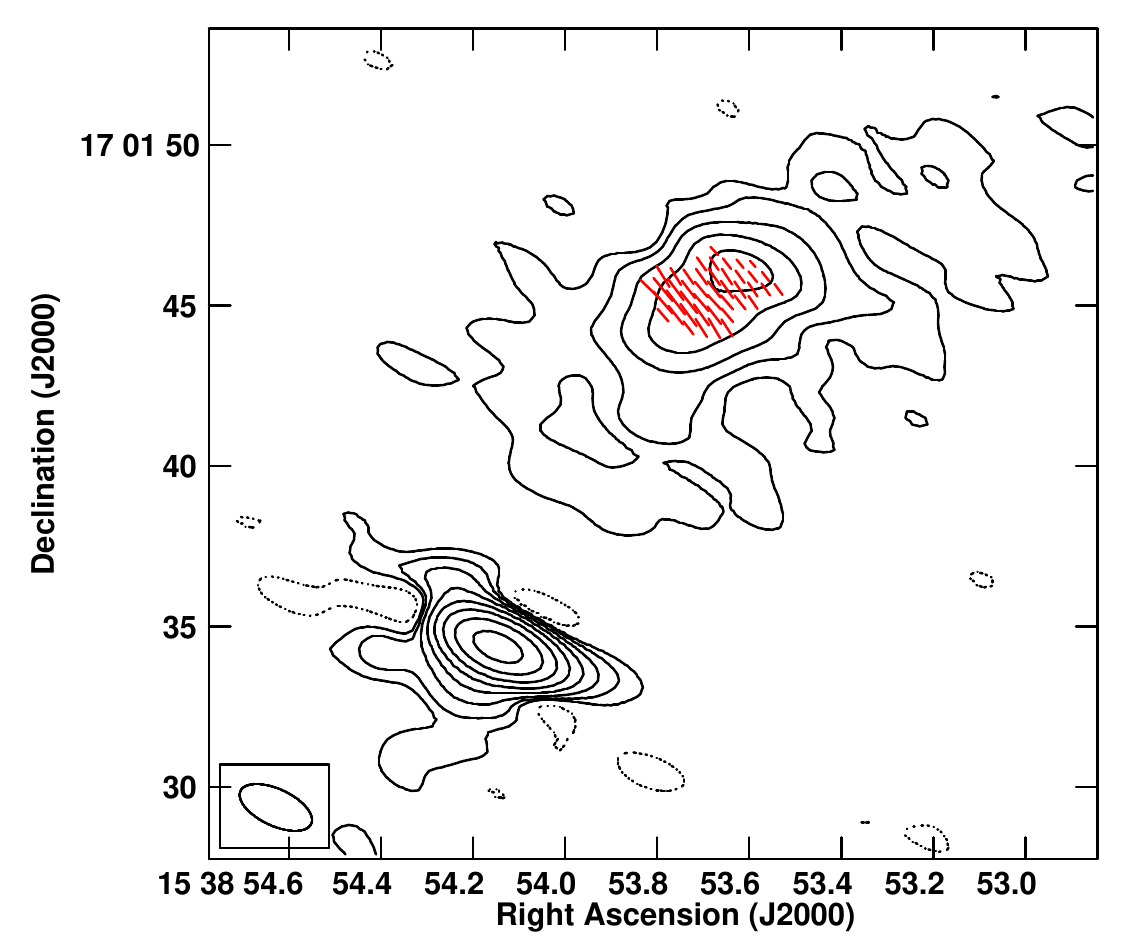}
\caption
{Left: VLA 5.5~GHz B-array {\it uv}-tapered total intensity contour image. Right: VLA 5.5~GHz B-array total intensity contour image superimposed with electric fractional polarization vectors in red. 1$\arcsec$ length of the vector corresponds to 25\% fractional polarization. The peak contour flux density is {\it x} mJy beam$^{-1}$ and the contour levels are {\it y} $\times$ (-1, 1, 2, 4, 8, 16, 32, 64, 128, 256, 512) mJy beam$^{-1}$, where ({\it x},{\it y}) = (5.0;0.0350) for the left panel and (4.9;0.0275) for the right panel.
}
\label{VLA_poln}
\end{figure*}

The left panel of Figure~\ref{VLA_poln} presents the VLA 5.5~GHz B-array {\it uv}-tapered total intensity contour image of NGC\,5972. The {\it uv}-tapering was carried out at 20~k$\lambda$ for the last self-calibrated visibility data (keeping all the antennas) in order to bring out the diffuse emission better. We detect a radio core and a pair of radio lobes extending in the northwest-southeast direction.
We do not detect the outer lobes in the B-array 5.5 GHz image due to the lack of short spacings in the UV-plane, which are crucial for mapping diffuse emission. The total flux density measured from the D-array image at 6~GHz is 27.8~mJy, while the B-array at 5.5~GHz recovers only 12.2~mJy, meaning that more than half of the total flux density is resolved out in the B-array. This missing large-scale emission is further seen as the negative bowls in the contours of the 6 GHz D-array image (see Figure~\ref{fig:spix}).

The right panel of Figure~\ref{VLA_poln} presents the VLA 5.5~GHz B-array total intensity contour image overlaid with electric fractional polarization ($\chi$) vectors in red. We find that the north-western jet/lobe region is highly linearly polarized with a fractional polarization of $15\pm3$\%. According to the synchrotron theory, the magnetic fields are inferred to be perpendicular to the $\chi$ vectors for optically thin regions like jets and lobes, whereas parallel for optically thick regions like the core. In NGC\,5972, the inferred magnetic fields in the jet/lobe region are found to be largely poloidal, i.e. aligned with the jet direction \citep[similar to FRII jets;][]{Bridle94}. We also note that, using different strategies and different calibrators, the core shows around $1.2\pm0.3$\% polarization which needs to be confirmed with additional data.

NGC\,5972 is an excellent source to study the AGN feedback via jets as it hosts a kpc-scale jet that aligns with the EELR (Figure~\ref{kinematics}). NGC\,5972 is classified as a ``radio-loud" galaxy (L$_{1.4}$=$2\times10^{24}$ W Hz$^{-1}$).
According to the radio-loudness parameter (R) as defined in \cite{1989AJ.....98.1195K}, R$\approx$31, suggests that the origin of radio emission can be attributed to jets powered by the central engine.


\subsection{Kinematics} 
\label{kinematics_sec}

\begin{figure*}
\centering
\includegraphics[height=7.00cm,trim = 0 0 0 0 ]{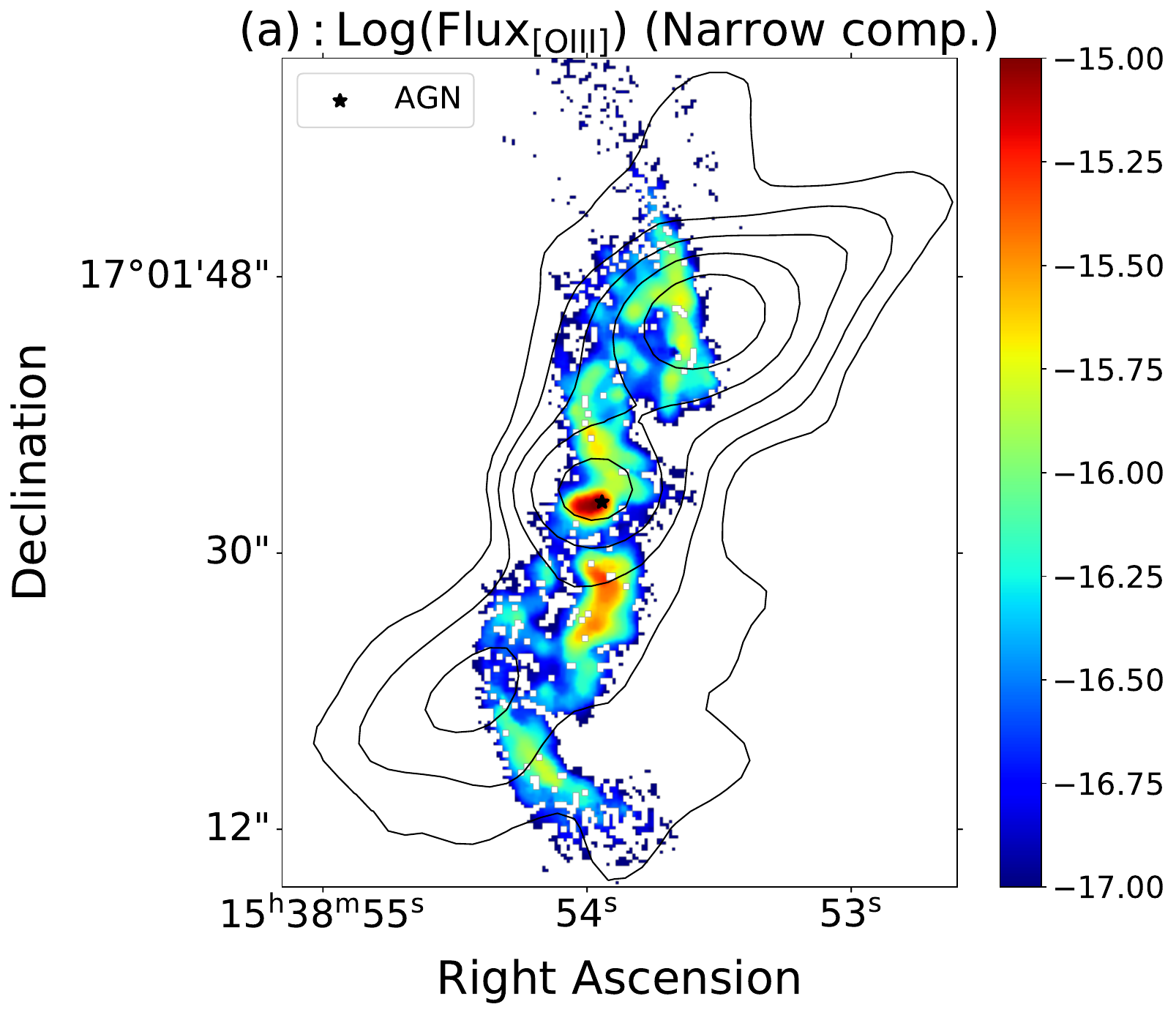}
\includegraphics[height=7.00cm,trim = 0 0 0 0 ]{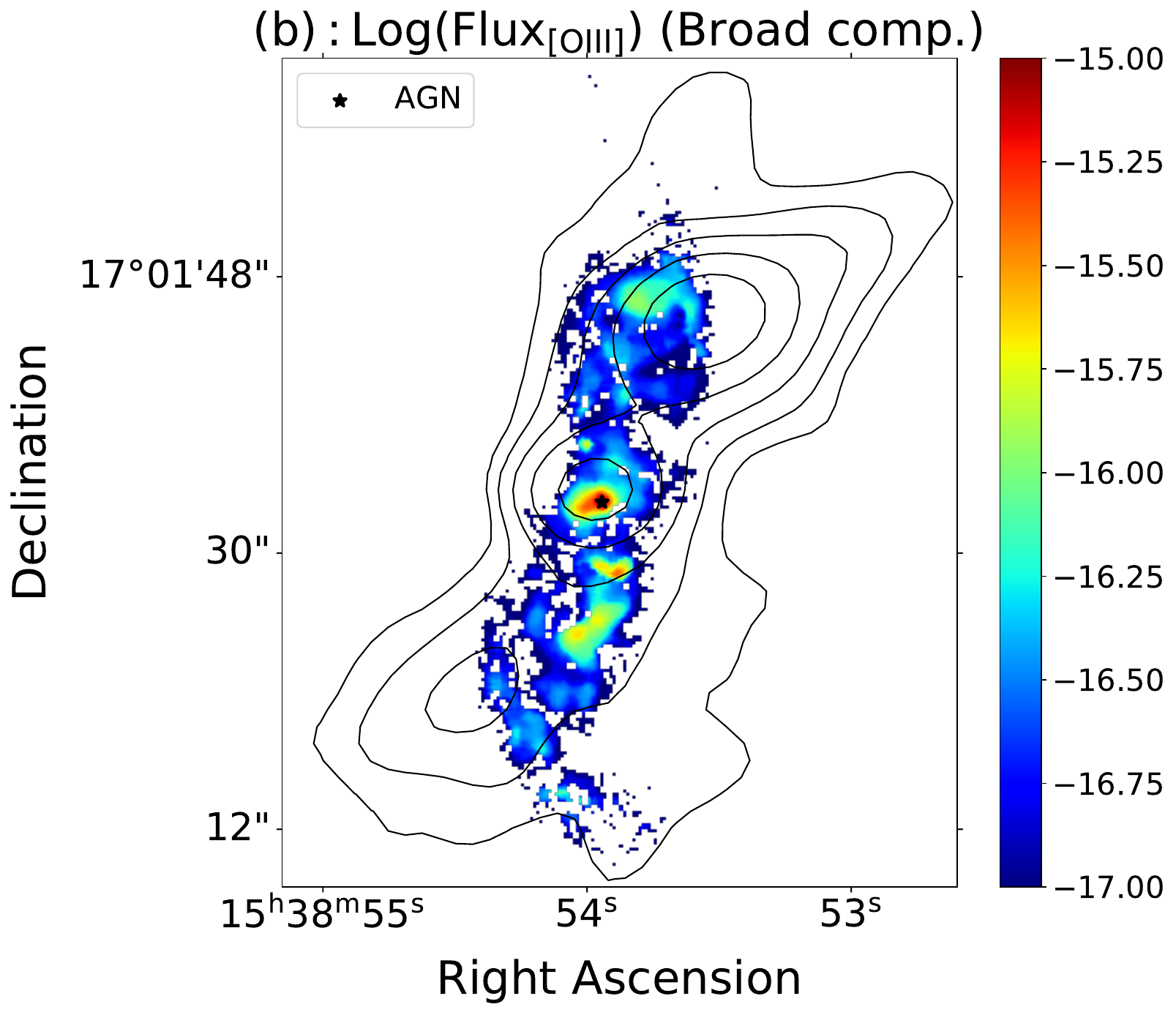}
\includegraphics[height=7.00cm,trim = 0 0 0 0 ]{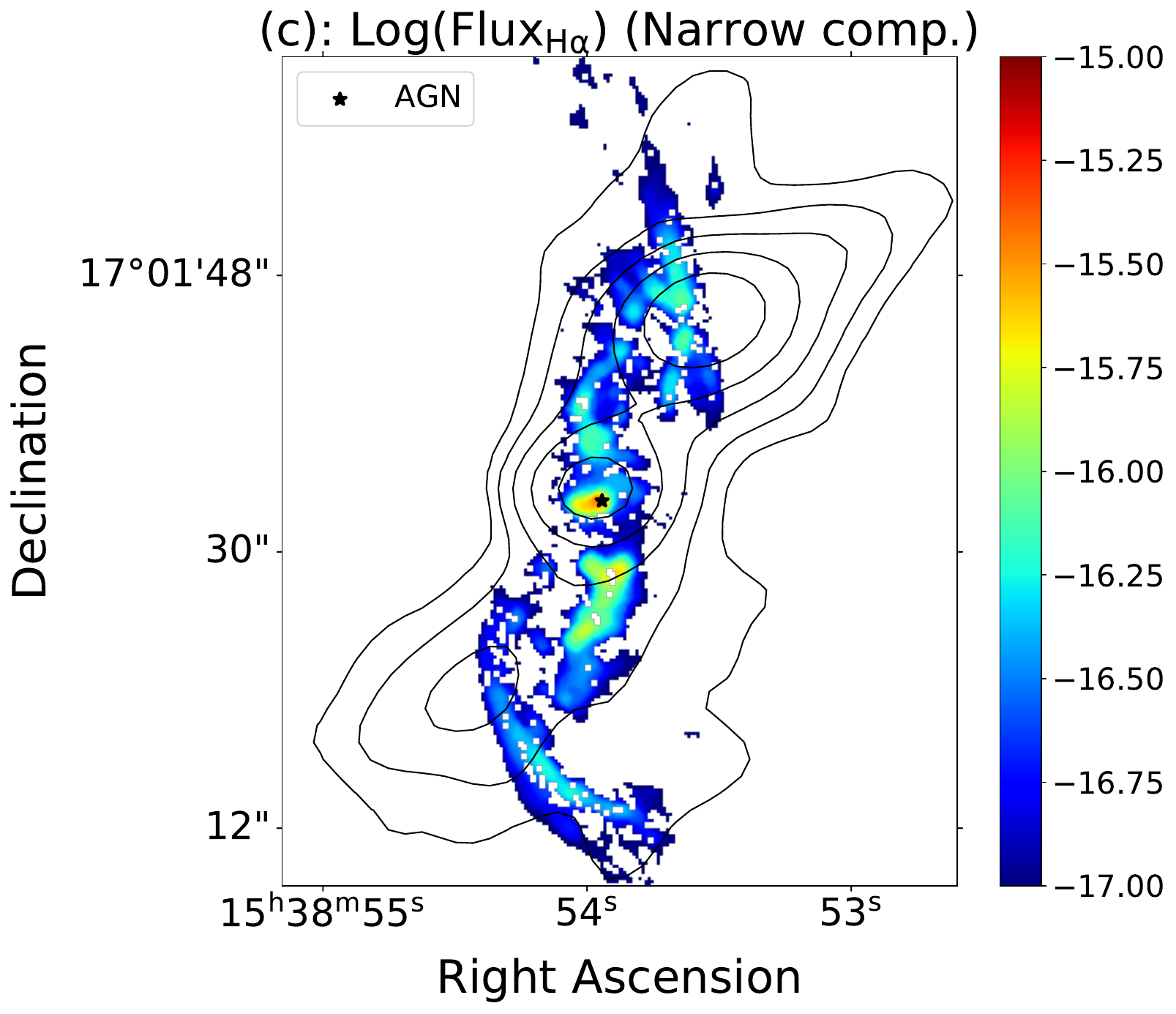}
\includegraphics[height=7.00cm,trim = 0 0 0 0 ]{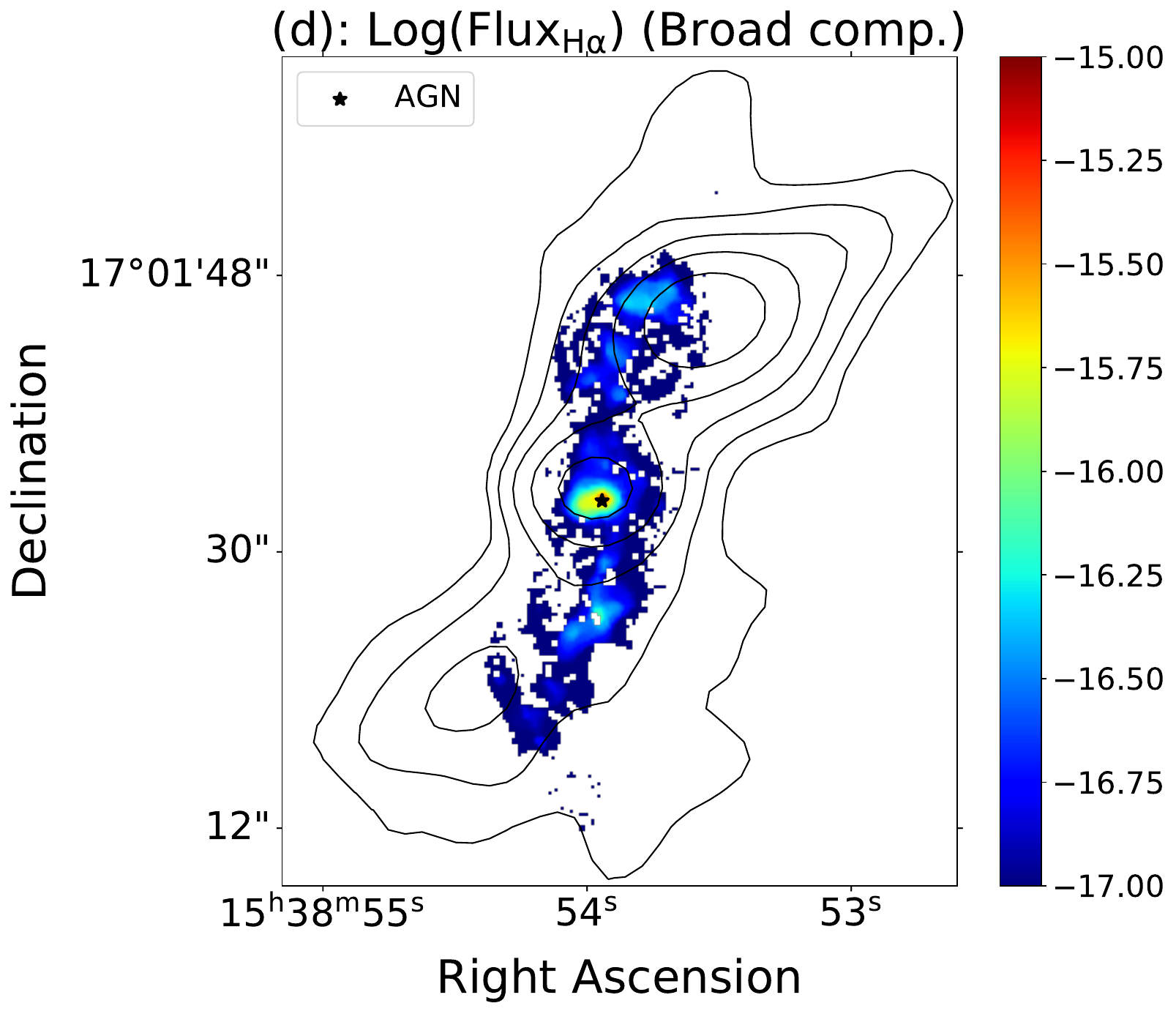}
\includegraphics[height=7.00cm,trim = 0 0 0 0 ]{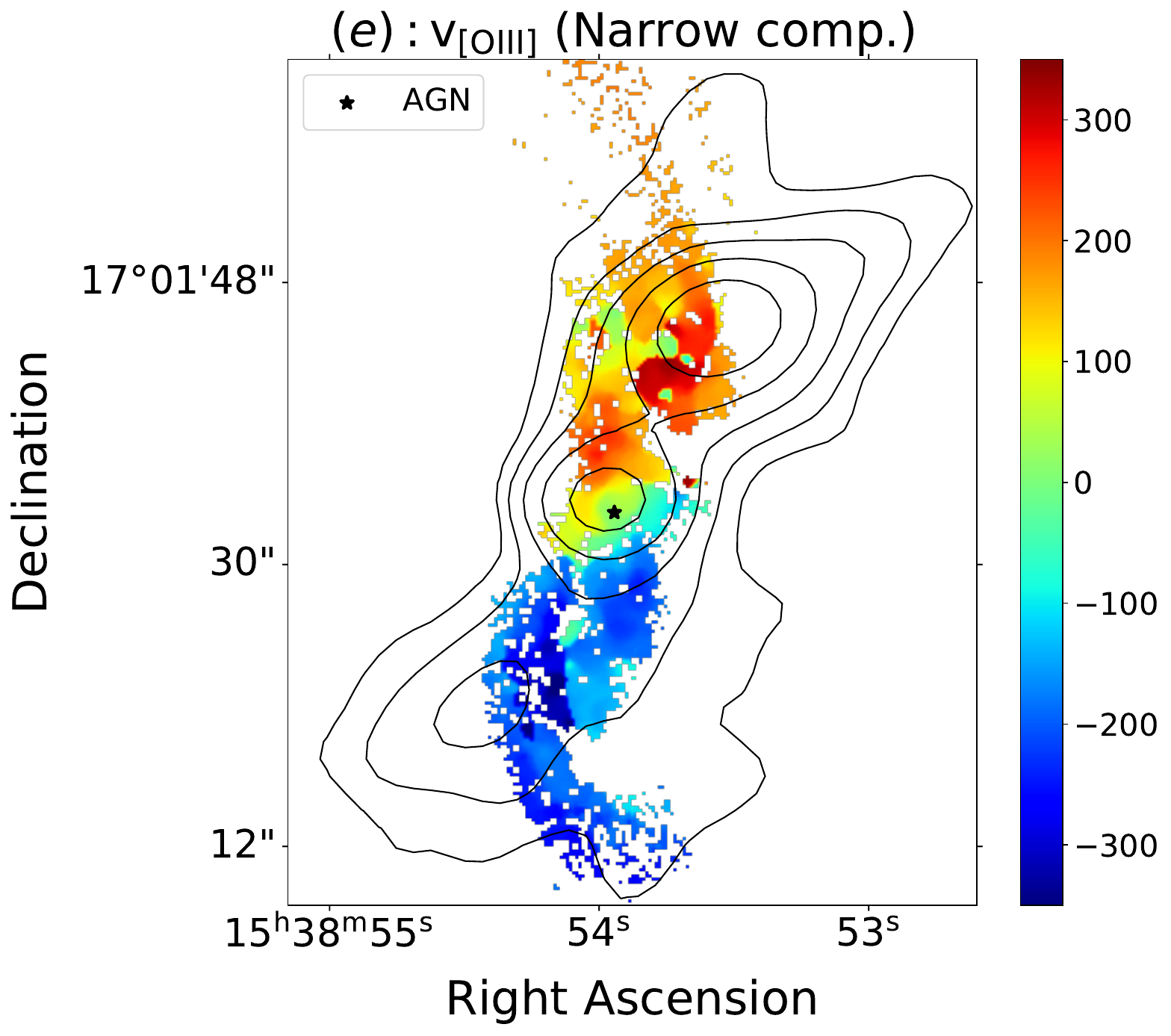}
\includegraphics[height=7.00cm,trim = 0 0 0 0 ]{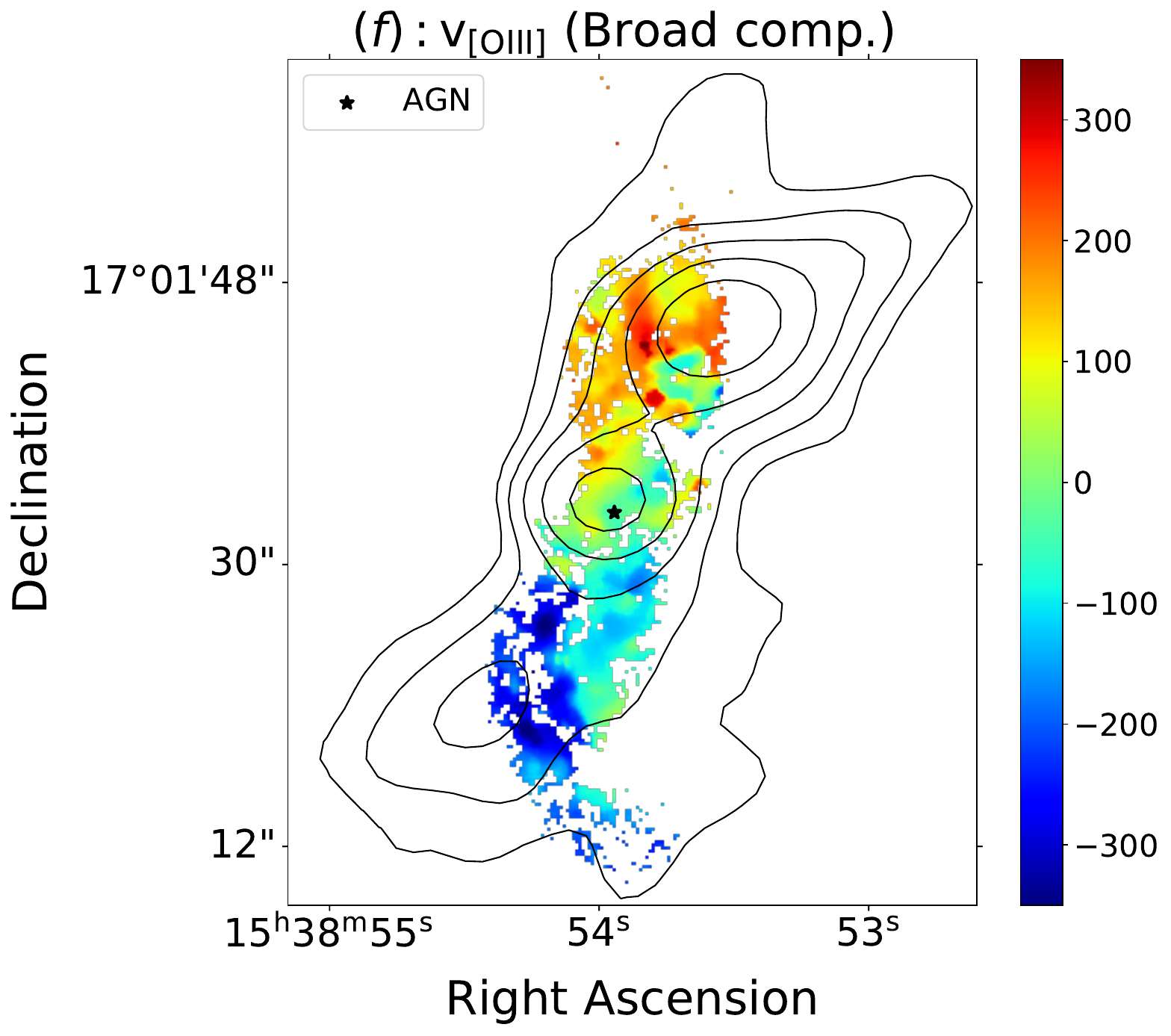}

\caption
{Maps obtained after double Gaussian fitting for [O\,{\scriptsize{III}}] and H\(\alpha\) emission lines. (a) - (d): flux maps. (e) - (h): line-of-sight velocity maps. (i) - (l): residual velocity created using the difference between gas velocities and stellar velocity. The flux maps are in \(\text{erg s}^{-1} \text{cm}^{-2}\), whereas all the velocity maps are in \(\text{km s}^{-1}\).
SNR cut of \(3\sigma\) is applied, where \(\rm \sigma = (15, 10)\times 10^{-20} \text{ erg s}^{-1} \text{cm}^{-2} \AA\) for the [O\,{\scriptsize{III}}] and H\(\alpha\) peak flux density respectively. The black contours trace the radio emission from the jet at \(610 \text{ MHz}\).
}
\label{kinematics}
\end{figure*}

\begin{figure*}
\centering
\addtocounter{figure}{-1}

\includegraphics[height=7.00cm,trim = 0 0 0 0 ]{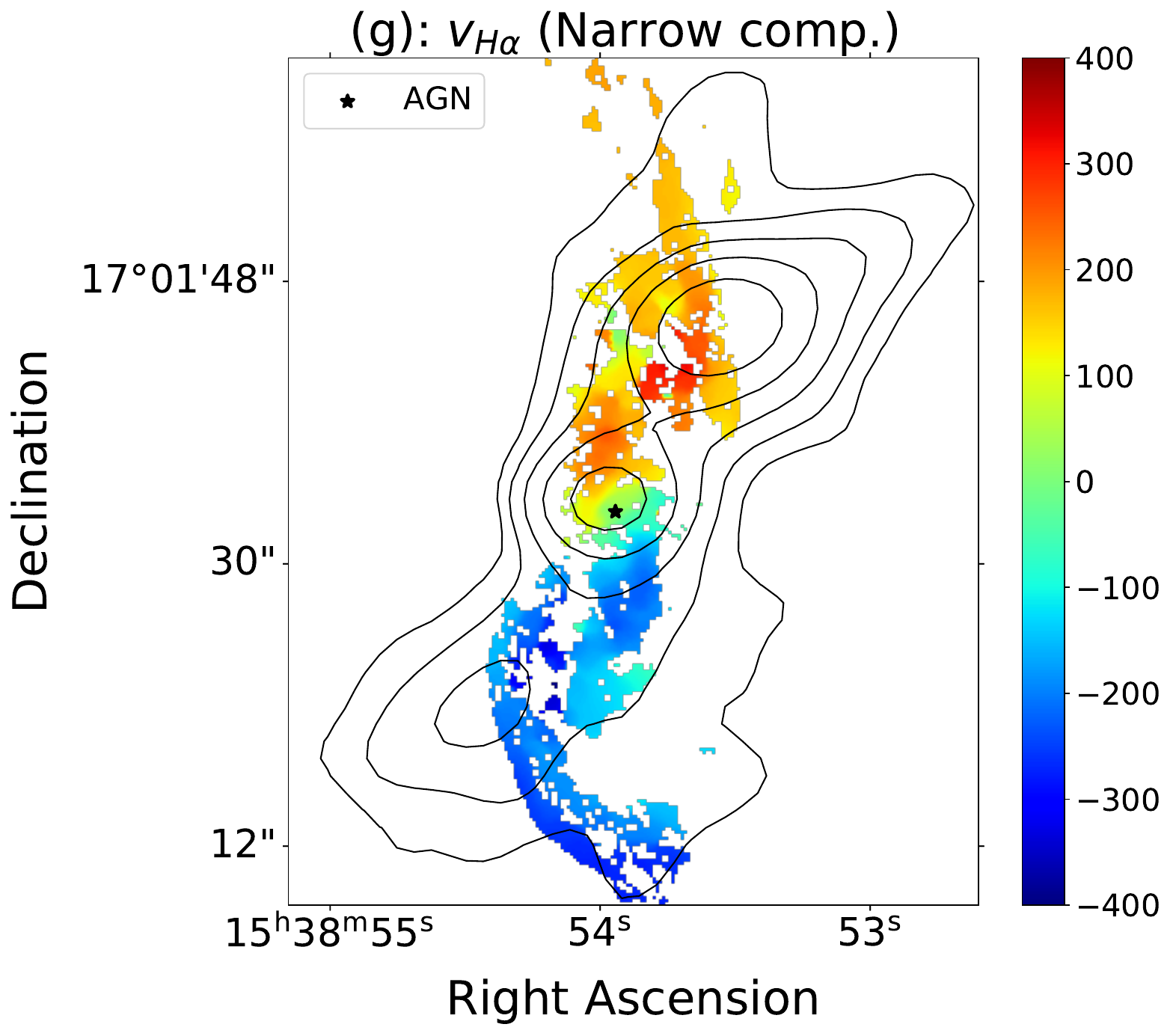}
\includegraphics[height=7.00cm,trim = 0 0 0 0 ]{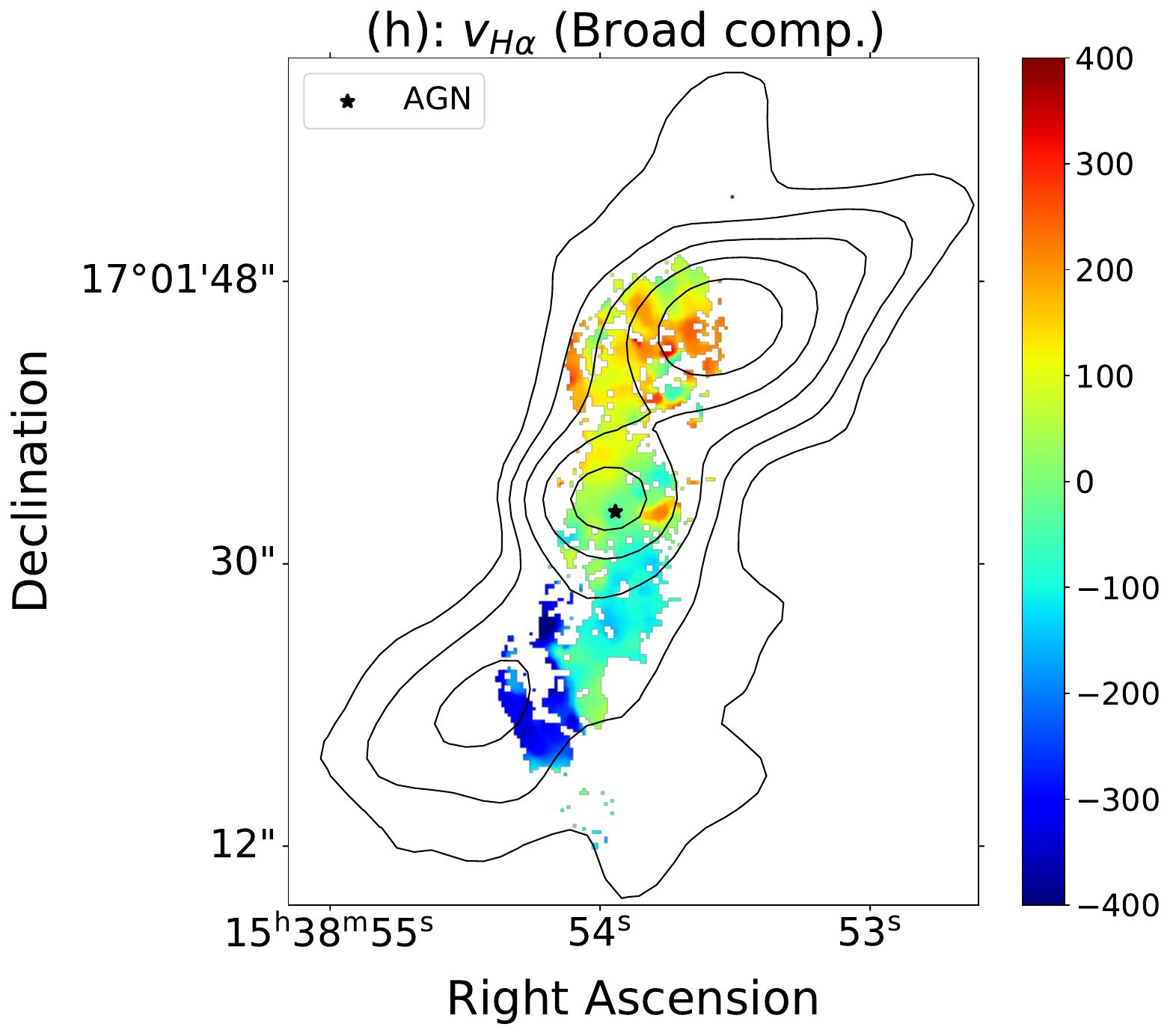}
\includegraphics[height=7.00cm,,trim = 0 0 0 0 ]{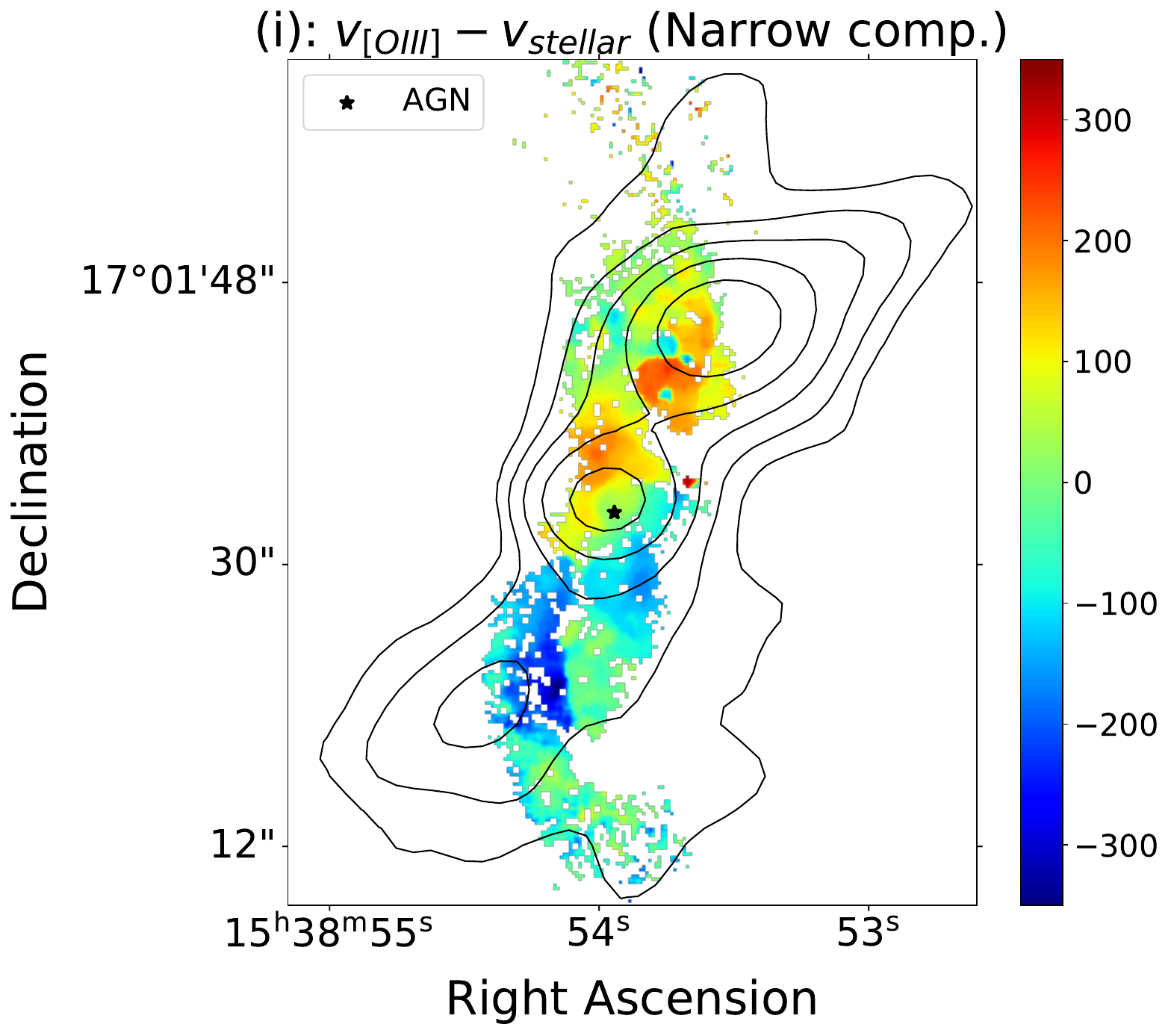}
\includegraphics[height=7.00cm,,trim = 0 0 0 0 ]{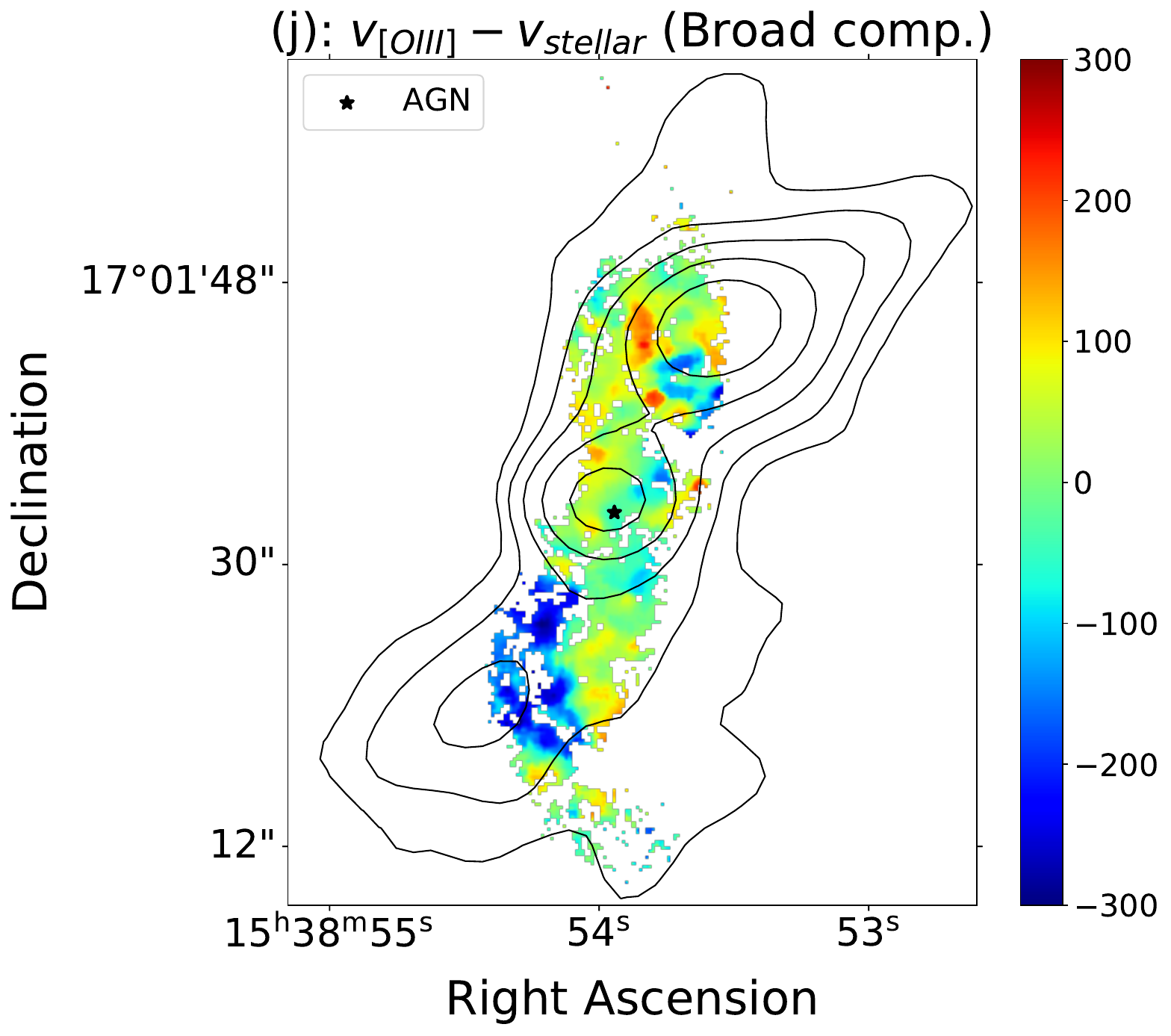}
\includegraphics[height=7.00cm,,trim = 0 0 0 0 ]{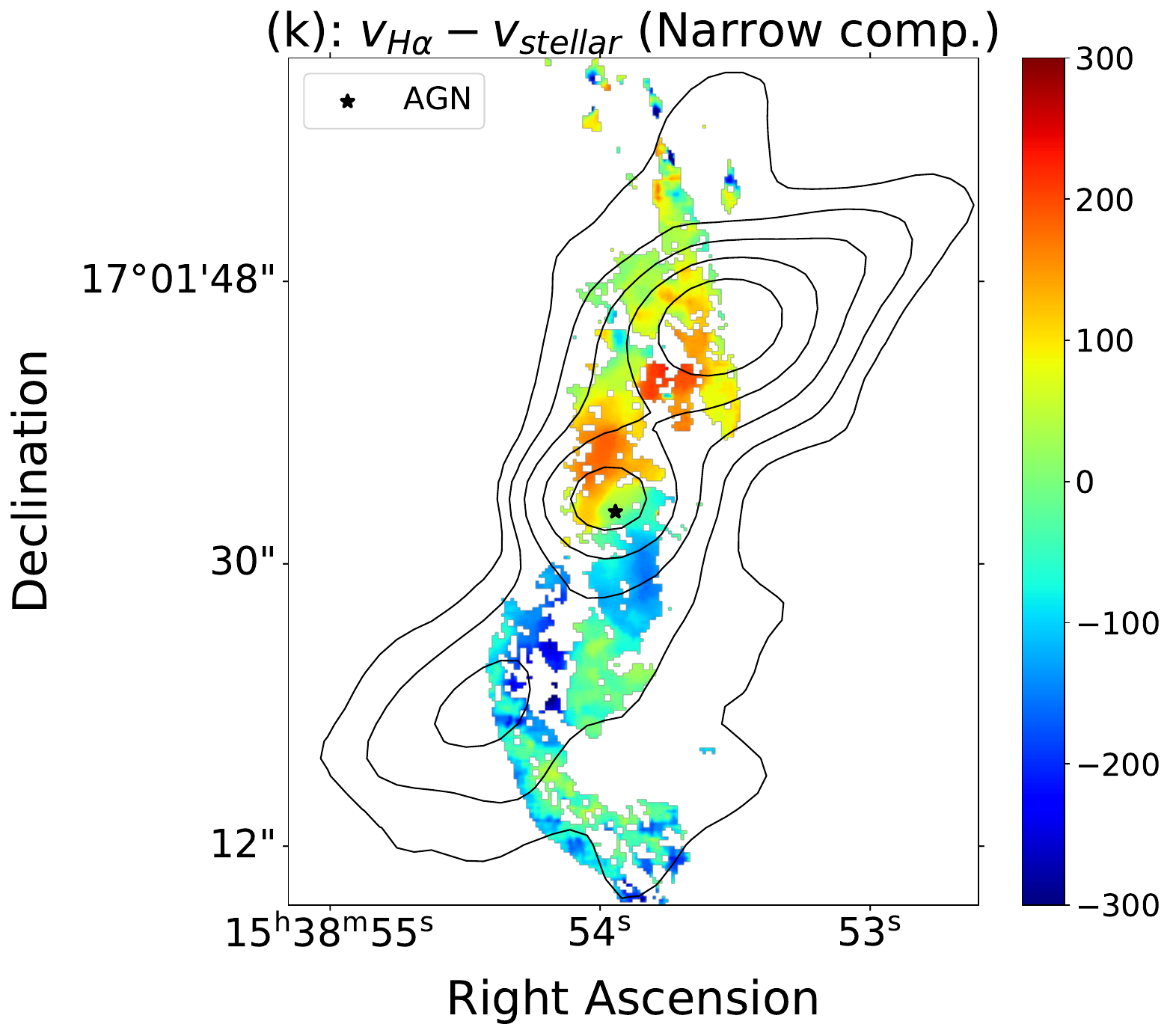}
\includegraphics[height=7.00cm,,trim = 0 0 0 0 ]{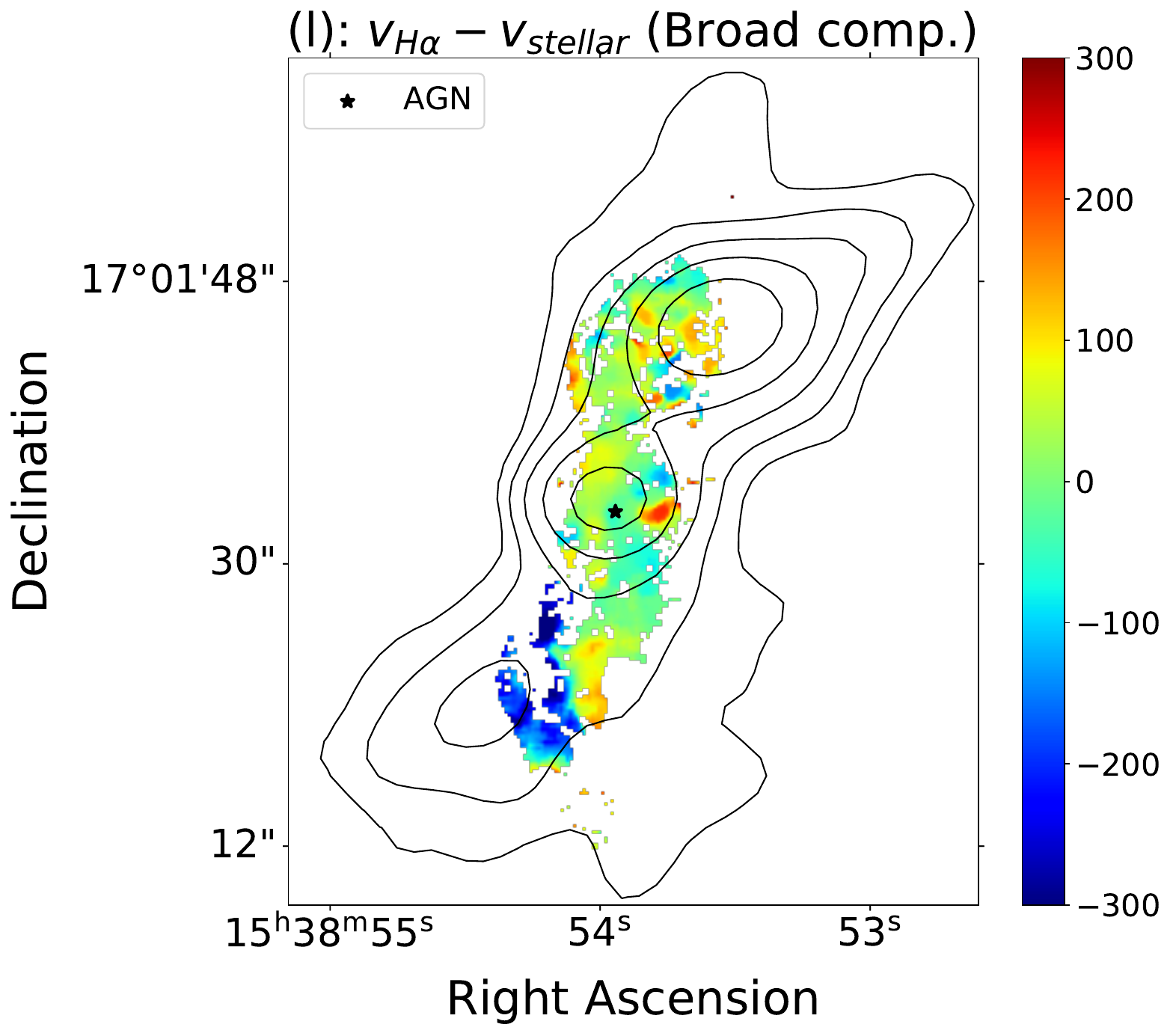}\\

\caption
{
Continued.
}

\end{figure*}

The multi-wavelength IFU data of NGC\,5972 provide valuable insights into various aspects of the galaxy's dynamics, including the outflow patterns, gas distribution, and the correlation between the AGN jet and the surrounding gas.
Results obtained after the Gaussian fitting (refer to Section~\ref{muse}) were used to estimate the gas kinematics. Panels in Figure~\ref{ppxf_fit} show the stellar velocity and stellar dispersion maps and Figure~\ref{kinematics} shows [O\,{\small{III}}] and H$\alpha$ flux maps, gas velocity, and residual velocity for both narrow and broad Gaussian components. The residual maps are created by subtracting the emission line gas velocity and the stellar velocity.

The stellar velocity map (Figure~\ref{ppxf_fit}, bottom left), derived from the pPXF fit, shows a regular rotating system with velocities up to 150$\pm$50 km~s$^{-1}$, consistent with \cite{2022ApJ...936...88F}. Slight blue-shifted features to the south and red-shifted features to the north suggest perturbations in the rotation, possibly due to a recent minor merger or interaction. Using the $\rm \tt{3DBarolo}$ fitting method, \cite{2022ApJ...936...88F} subtracted a best-fit rotation model, which shows a small blue-shifted residual to the south, further supporting the minor merger hypothesis.
The velocity dispersion map (Figure~\ref{ppxf_fit}, bottom right) shows a central peak reaching 220 km~s$^{-1}$, with an irregular, lumpy distribution likely caused by ongoing interactions or merger activity. To evaluate the robustness of these kinematic features, we generated error maps (Figure~\ref{stellar_err_map}). The stellar kinematic maps are created using a 2$\sigma$ cutoff in the amplitude to capture only meaningful structures. Additionally, these features are consistent with those reported by \cite{2022ApJ...936...88F}, further supporting their reliability.

The EELR exhibits a distinct velocity profile aligned with the radio jet in the north-south direction, effectively tracing the path of the jet. From the velocity maps it is evident that the gas demonstrates rotational behavior, as indicated by the observed blue-shift in the lower structure and redshift in the upper structure. For the [O\,{\small{III}}] emission line, the narrow component exhibits average gas velocities of 163$\pm$68 km~s$^{-1}$ in the north and -208$\pm$71 km~s$^{-1}$ in the south. In contrast, the broad component shows average velocities of 132$\pm$51 km~s$^{-1}$ and -182$\pm$28 km~s$^{-1}$ for the north and south regions, respectively. For the H$\alpha$ emission line, the narrow component has average gas velocities of 192$\pm$34 km~s$^{-1}$ in the north and -162$\pm$42 km~s$^{-1}$ in the south, while the broad component suggest average velocities of 115$\pm$47 km~s$^{-1}$ and -129$\pm$23 km~s$^{-1}$ in the north and south regions, respectively. Detailed discussion about the implication of the residual velocity maps are presented in Section~\ref{outflows}.

It is important to note that the gas kinematics in NGC\,5972 demonstrate complex interactions between the ionized gas, the radio jet, and the underlying stellar dynamics. As shown in Figure~\ref{kinematics}, the velocity maps highlight enhanced gas velocities along the radio jet axis, suggesting regions of jet-ISM interaction. However, the spatial alignment between the gas morphology and the radio jet is not exact. Portions of the EELR could be more closely aligned with the stellar kinematic line of nodes or at an intermediate angle between the jet and rotational axis. This misalignment may reflect a matter-bounded configuration, where jet-driven shocks illuminate the gas as the cocoon interacts with it. Alternatively, projection effects (refer to Section~\ref{sec_Clues from  inclination}) could also contribute to the apparent geometry.
We acknowledge this ambiguity and emphasize that the observed gas geometry does not require perfect alignment with either the radio jet or the stellar rotation axis.


\begin{figure}
\centering
\includegraphics[height=7.0cm,trim = 0 0 0 0 ]{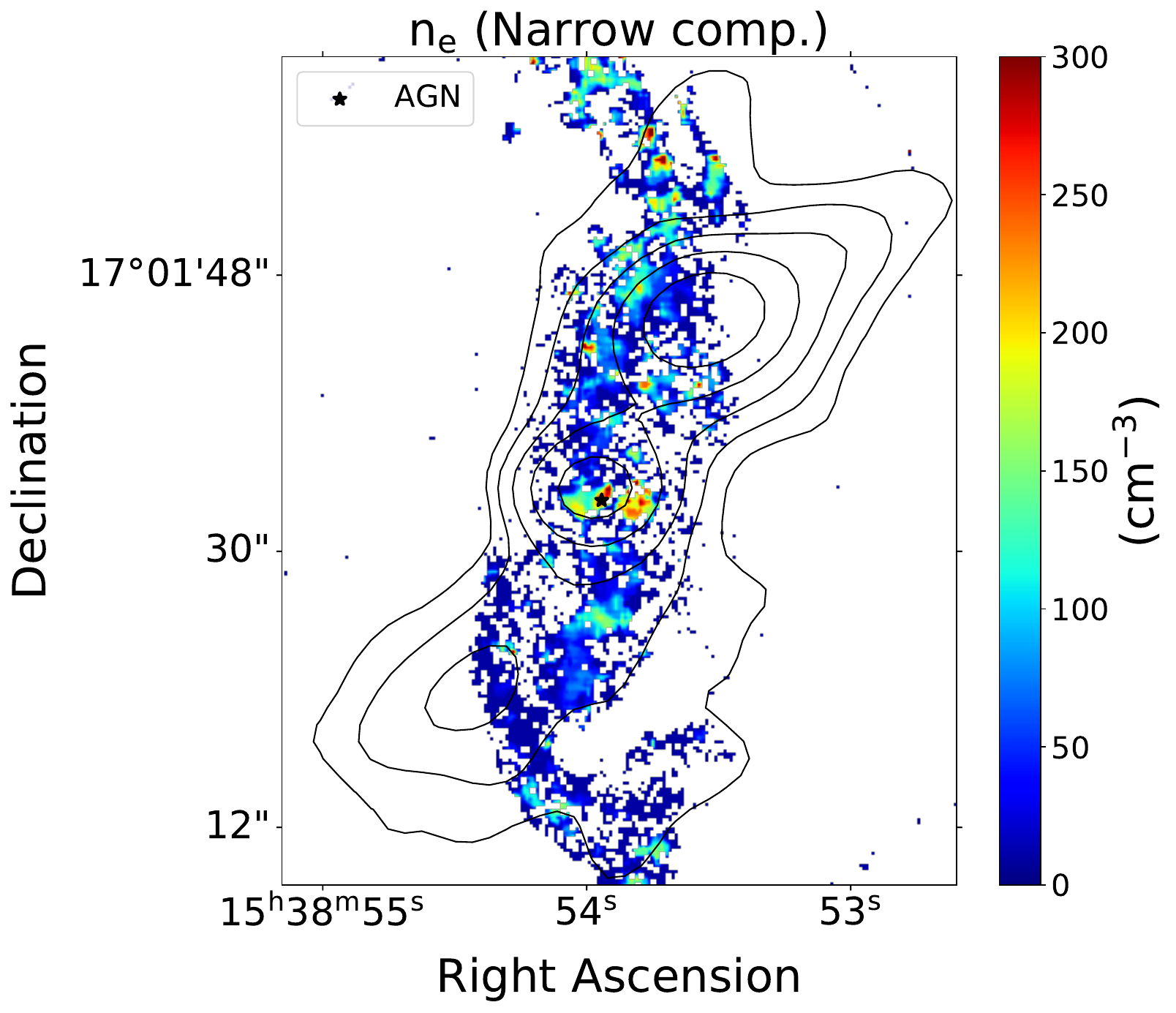}
\hspace{1cm}
\includegraphics[height=7.0cm,trim = 0 0 0 0 ]{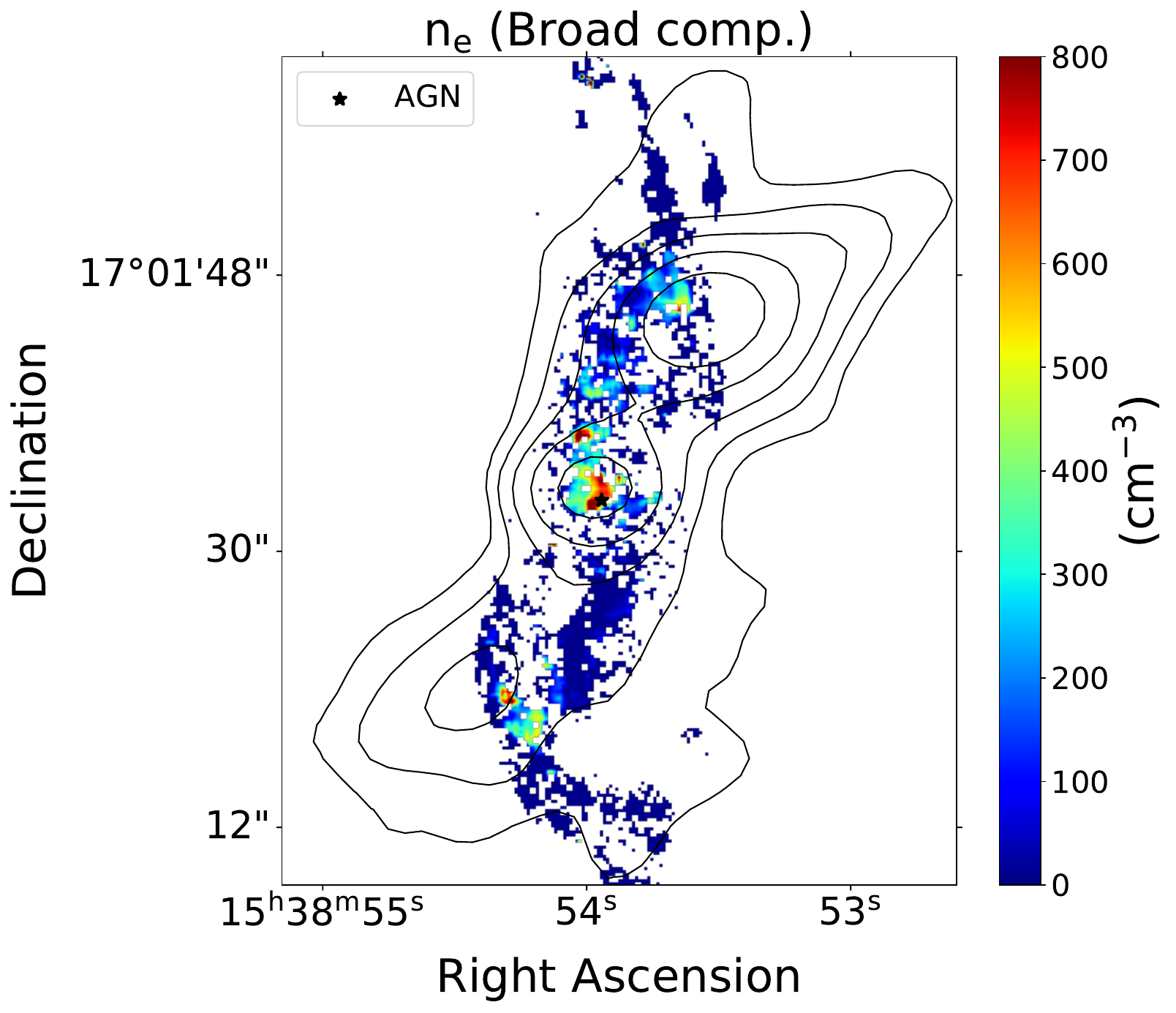}

\caption
{
\small Electron density maps derived from the [S\,{\small{II}}]$\lambda$6716/[S\,{\small{II}}]$\lambda$6731 line ratio. The top panel represents the electron density for Gaussian 1 and the bottom panel represents the electron density for Gaussian 2. The contour corresponds to 610 MHz radio emission.
}
\label{tempden}
\end{figure}
\subsection{Electron density estimates }
\label{ne and te}

Electron density is an important parameter in determining the energy estimates from the AGN. Both mass outflow rates and kinetic luminosity depend significantly on the electron density of the outflow. The optical emission lines such as [S\,{\small{II}}] $\lambda$6716, 6731 (referred to as the [S II] doublet) or [O\,{\small{II}}] $\lambda$3726, 3729), provide direct measurements of this density \citep{2006agna.book.....O, 2016ApJ...816...23S, 2017MNRAS.465.3220K,
2018NatAs...2..198H, 
2018A&A...618A...6K, 2018MNRAS.474..128R}. These emission lines are used because they exhibit relative flux values that solely rely on the electron density occupying specific meta-stable energy levels. For our study, we have used [S\,{\small{II}}] doublets due to the limited spectral resolution of the data.

Figure~\ref{tempden} (left) shows the electron density map derived using the relation given in \cite{2006agna.book.....O}:

\begin{equation}
\label{eq:myequation}
{\rm n_e = \frac{100\sqrt{T_e}(R - 1.49)}{5.61 - 12.8R}}
\end{equation}

where the flux ratio R=f([S\,{\small{II}}]$\lambda$6716/ [S\,{\small{II}}]$\lambda$6731) and $T_e$=10000 K is assumed to be initial temperature condition. In the case of the Gaussian 1, the average electron density is $\leq$300 cm$^{-3}$. While for the Gaussian 2 component, which is assumed to mimic the outflowing motion of the gas (as discussed in Section~\ref{outflow}), we observe that the electron density is notably higher in the region near the galactic core, around 800 cm$^{-3}$, and gradually decreases as it extends away from the center along the trajectory of the jet, reaching approximately 300 cm$^{-3}$. These values closely align with those reported in the study by \cite{2022arXiv220805915H}. Radio contours (610 MHz) are overlaid on top to obtain an understanding of the electron density along the path of the jet. 


\subsection{Resolved BPT diagnostics}
\label{bpt_sec}

Several mechanisms can contribute to the ionization of ISM: photoionization due to star formation or AGN, shock ionization due to stellar winds, jets, or other dynamical processes. To understand the processes at play, optical emission line ratio diagnostics, known as ``BPT diagrams" \citep{1981PASP...93....5B,1987ApJS...63..295V} are used. We have used [O\,{\small{III}}]/H$\beta$ versus [NII]/H$\alpha$ emission line ratios to investigate the ionization mechanism in the galaxy for the individual MUSE bins, which are plotted on the BPT diagram (Figure~\ref{shock and bpt}). The dashed line \citep{2001ApJ...556..121K} represents theoretical starburst models, while the solid line \citep{2003MNRAS.346.1055K} serves as an empirical composite boundary in the [N II]/H$\alpha$ BPT diagram. The region formed in between these boundaries is referred to as, a ``composite region", which represents galaxies that exhibit spectral characteristics indicative of both star formation and AGN activity.
The dash-dotted line represents the separation between Seyfert-2 and LINERs \citep{2007MNRAS.382.1415S}. We have over-plotted the GMRT contours on top of the reconstructed MUSE image which shows that the jet overlaps with the AGN photo-ionized region, while shock ionization is observed in the perpendicular direction to the jet.

The categorization of LINERs itself is challenging due to ongoing debates about their ionization sources. The uncertainty revolves around whether ionization is primarily driven by an AGN \citep{1983ApJ...264..105F}, fast shocks \citep{1981PASP...93....5B,1995ApJ...455..468D}, or due to the ultraviolet radiation emitted by hot old stars \citep{CidF2011, Singh2013}. Therefore, relying solely on the BPT diagram, we cannot determine the specific mechanism responsible for driving feedback in the galaxy.

For NGC\,5972, \cite{2022ApJ...936...88F} demonstrated that the stellar population distribution features older populations at the center and younger populations at larger radii, ruling out the contribution of old stars to LINER emission.
\cite{2022ApJ...936...88F} conducted photoionization modeling of NGC 5972 EELR using CLOUDY, demonstrating that the ionization can not be attributed to the current levels of AGN activity. Their results can only be explained by a significant decline in AGN luminosity over the past \(5 \times 10^4\) years, with the outer EELR reflecting the ionizing influence of a more luminous AGN phase.

While photoionization models effectively explain much of the large-scale ionization, they do not account for kinematic disturbances observed in our study, such as enhanced line widths, and velocity asymmetries along the jet axis (refer to Figure~\ref{kinematics} and \ref{shock and bpt}). Additionally, the elevated velocity dispersions in regions perpendicular to the jet suggest the presence of transverse shocks propagating through the EELR. To address these discrepancies and to complement the photoionization modeling performed by \cite{2022ApJ...936...88F}, we applied shock + precursor models to the NGC\,5972 data (see Section~\ref{sec:model}). This exercise demonstrates that shocks, driven by jet-ISM interactions, contribute to ionization in regions where AGN photoionization alone is insufficient.


\subsection{Shock modeling}
\label{sec:model}

\begin{figure*}
\centering
\includegraphics[height=8.0cm,trim = 0 0 0 0 ]{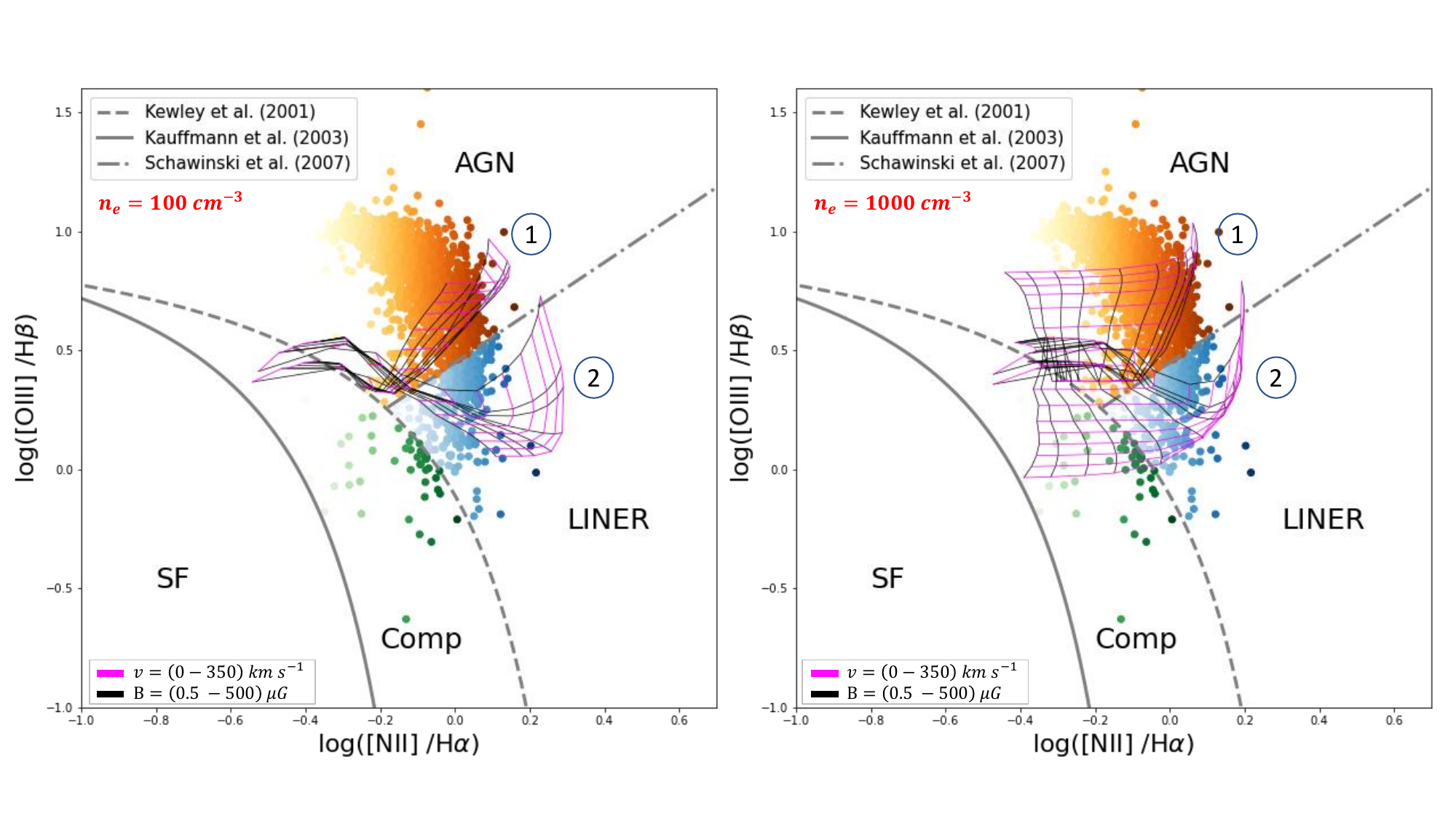}
\includegraphics[height=8.5cm,trim = 0 0 0 0 ]{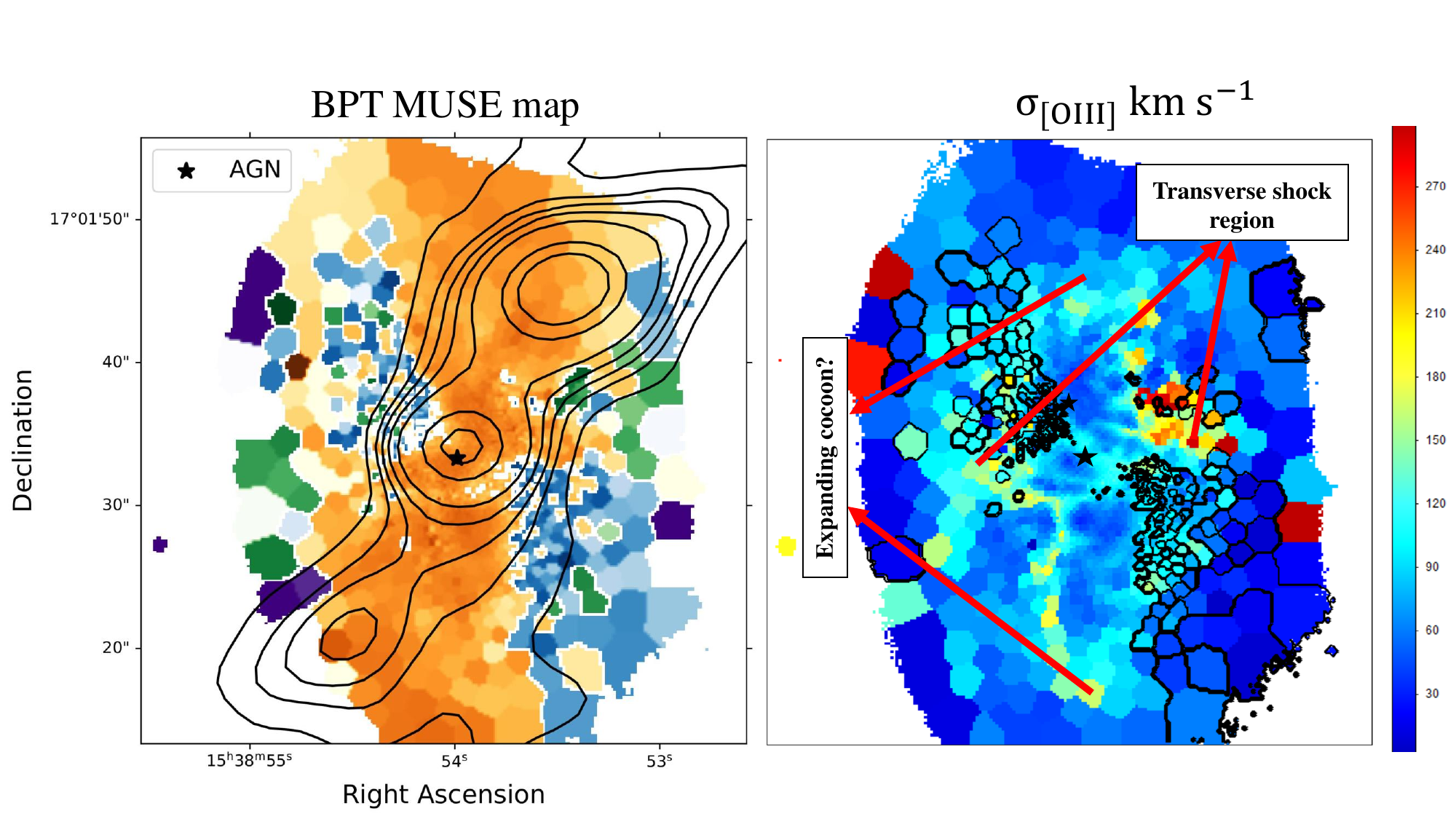}

\caption
{
\small Top: Spatially resolved [O\,{\small{III}}]/H$\beta$ vs [NII]/H$\alpha$ diagnostic diagram. The lines in each panel show the theoretical separation between various line excitation mechanisms (refer to Section~\ref{bpt_sec}). Each point corresponds to a MUSE bin. Shock and shock+precursor models created using MAPPINGS V, are over-plotted on the top. 1:Shock+precursor model, 2:Pure shock model. The magenta grids represent the shock velocities in the range of $\rm v_s $=$(0$-$350)$ km~s$^{-1}$ whereas the black grids represent magnetic field strength within the range $\rm B$=$(0.5-500)$ $\mu$G. The metallicity is assumed to be solar for both models. For the left Figure, the pre-shock electron density is taken as $\rm n_e=$100 cm$^{-3}$, while for the right Figure, it is assumed to be $\rm n_e=$1000 cm$^{-3}$. Bottom left: Muse cube constructed using the corresponding bins from the BPT diagram. 610 MHz radio contours are overlaid on the top. The data points represent the bins created from the $\tt{GIST}$ pipeline from the MUSE cube with an SNR of 20.
Bottom right: Voronoi binned [O\,{\small{III}}] dispersion velocity map. The black regions represent the LINER region from the MUSE BPT map. 
}
\label{shock and bpt}
\end{figure*}

We used the radiative shock model database by \cite{2019RMxAA..55..377A}, calculated with MAPPING V \citep{Sutherl&Dopita2018}, to obtain the shock models. The models include two scenarios: simple shocks and shocks with precursors. In the simple shocks case, shocks arise from the intense interaction between AGN winds/ jets with the surrounding ISM, leading to a collisionally ionized gas. Whereas, in the shock+precursor case, the shock-heated gas produces photons that move upstream, ionizing the gas ahead of the shock front. This model is typically used in situations where both photoionization and shock excitation occur, such as in LINERs (\cite{2018ApJ...864...90M}). Figure~\ref{shock and bpt} shows an overplot of our data on the shock and shock+precursor models for [O\,{\small{III}}]/H$\beta$ vs [NII]/H$\alpha$ BPT diagram. Each data point represents a Voronoi bin of the MUSE cube and is color-coded according to its position on the BPT diagram (Figure~\ref{shock and bpt}).

To obtain the models, we have chosen the input parameters in the following way: the values for the shock velocities ($\rm v_s$) were selected based on the result obtained in Section~\ref{kinematics}. Vacancies were restricted to fall within the range of $\rm v_s= (0-350)$~km~s$^{-1}$. The metallicity was taken as solar as mentioned in \cite{2015AJ....149..155K} for NGC\,5972. The MAPPINGS V models provide a huge range of magnetic field strength, but we have adopted $B=(0.5-500)~\mu$G for all the models since this is a typical range of magnetic fields in Seyfert galaxies \citep{deBruyn1978, 1998ApJ...495..680B, 2016MNRAS.459.1310K}. In general, AGN cover a broad range of radio strengths, and the strength of the resulting magnetic field depends on both the strength and the morphology of the radio emission \citep{2016era..book.....C}. The pre- shock electron density was varied between $10-1000$~cm$^{-3}$ to check how the model performance varies with different density levels. It has been observed that the shock+precursor model with high-density value; $n_e=1000$~cm$^{-3}$ (Figure~\ref{shock and bpt}, right) is partially consistent for the spaxels that fall in the AGN region in the diagnostic diagram. If large-scale shocks indeed ionize the gas in NGC\,5972, then the pure shock models should be able to replicate the LINER fluxes of the emission lines in the BPT diagram. We observe that both low and high-density pure shock models (Figure~\ref{shock and bpt}) are able to reproduce the expected LINER emission.


\section{Discussion}
\label{discussion}

The radio emission observed in Seyfert galaxies are believed to originate primarily due to three main reasons: (i) jet-related activity \citep{Veilleux1991,Spoon2009,Mullaney2013,Morganti2015,Morganti2016,Nesvadba2017, Venturi2023,Singha2023}, (ii) influence of winds leading to the acceleration of thermal electrons to relativistic energies at shocks \citep{Stocke1992,Wang2008,Jiang2010,Ishibashi2011,Faucher-Giguere2012,Zubovas2012,Zakamska2014} and/or (iii) star-forming processes \citep{Rosario2013}.

The main goal of our work is to study the jet-related feedback that leads to the formation of the remarkable EELR in NGC\,5972. We found two main observational signatures of radio-mode feedback: (i) a spatial connection of the radio jet with shocked regions, and (ii) the presence of gas outflows coincident with the radio jets or evidence for acceleration/deceleration of the gas by the jets.

\subsection{Energetics of different components}
\label{energetics}

\subsubsection{Jet kinetic power}
\label{jet power}
There are several studies, such as
\cite{2004ApJ...607..800B, 2008ApJ...686..859B, 2007MNRAS.381..589M, 2010ApJ...720.1066C}
that give an empirical relation between jet power and radio luminosity. These studies mostly utilize lower frequency data ($\nu\leq$ 1.4 GHz) to obtain the flux density of the radio feature. \cite{2007MNRAS.381..589M} have derived the jet power and radio luminosity relation using 5 GHz radio data for a sample of low- luminosity radio galaxies. Since we already acquired the 5.5 GHz VLA data, we have used the equation given by \cite{2007MNRAS.381..589M} to calculate the jet power in the nuclear region:
\begin{equation}
\rm log(P_{jet})=(0.81 \pm 0.11)\,\rm log(L_{5~GHz}) + 11.9^{+4.1}_{-4.4}
\end{equation}
From the VLA 5.5 GHz data, we have estimated the core flux density to be 5.2 mJy (spectral index $\alpha=-0.7$) which corresponds to $\sim10^{38}$ ergs s$^{-1}$ core luminosity. Hence using the above relation, we obtained $\rm log(P_{\rm jet})= 43.23^{+8.36}_{-8.66}$ ergs s$^{-1}$.


\subsubsection{AGN kinetic power}

We assume that a fraction (5\%) of the AGN bolometric luminosity ($\rm L_{\rm bol}$) is converted into AGN kinetic power that drives the jet through the interstellar medium \citep{2005Natur.433..604D, 2017A&A...600A.121N}. For NGC\,5972, L$_{\rm bol}$ $\approx$ 10${^{44}}$ ergs s$^{-1}$ \citep{2017ApJ...835..256K}, therefore $\rm \dot{E}_{AGN}\approx10^{42}$ ergs s$^{-1}$ .

\subsubsection{SFR mechanical power output}
\label{sfr}

We have estimated the star formation rate (SFR) using IRAS 60$\rm \mu$m and 100$\rm \mu$m fluxes. Note that the SFR derived from these fluxes includes contributions from the AGN at FIR, as it has not been corrected for AGN influence. We have taken this into account for our subsequent calculations. The relation used to derive the SFR is given by \cite{1998ARA&A..36..189K}:
\begin{equation}
\rm SFR\,(M_\odot\, yr^{-1})=\,4.5\times10^{-44}L_{FIR}
\end{equation}

where \( \rm L_{\text{FIR}}= 4 \pi D^2 \times \text{FIR} \times L_\odot \), and \(\rm \text{FIR} = 1.26 (2.58 f_{60} + f_{100})\times 10^{-14} \) \citep{Helou1988}. The SFR comes out to be $4.17~\rm M_\odot\, yr^{-1}$. We have also estimated the SFR from the 1.4 GHz NVSS data using the relation given in \cite{1992ARA&A..30..575C} (equation 21), which comes out to be 73 $\rm M_\odot\, yr^{-1}$. The difference in SFR estimated from IR and radio data suggests that the excess radio emission likely stems from sources other than star formation, such as radio jets or winds. Given that the galaxy is ``radio-loud" with radio emission extending to 550 kpcs as observed in Figure~\ref{radio_images}, it is unlikely that this excess radio emission is driven by the winds.

To estimate the net mechanical power output from star formation, we followed the method by \cite{2008MNRAS.383.1210S}, which suggests that if the kinetic energy injected per solar mass of stars formed is $\epsilon_{\text{SN}} \approx 1.8 \times 10^{49}$ ergs $\rm M_\odot^{-1}$, then about 40$\%$ of this energy is carried away by the winds, while the rest is radiated away. Thus, the net mechanical injection rate into the galaxy from star formation $(\rm \dot{E}_{SFR})$ is $0.72 \times 10^{49} \times \text{SFR}$ ergs s$^{-1}$.
For NGC 5972, $\rm \dot{E}_{SFR}$ derived from IR data is $9.5\times10^{41}$ ergs s$^{-1}$.


\subsubsection{Can the jet inflate the EELR?}
\label{workdone}

The velocity dispersion map (Figure~\ref{shock and bpt}, right) shows significant turbulence both along and perpendicular to the jet direction, suggesting that the gas may be inflated or displaced by the jet's energy. To evaluate whether the jet can displace gas at scales of tens of kpc and create such turbulence, we estimated the \( \rm PV \) work required to ionize or displace the gas within the EELR \citep{Rao2023}.

The pressure acting on the gas was calculated as \( P = n k_B T \), where \( k_B \) is the Boltzmann constant, \( T = 10^4 \, \mathrm{K} \) is the assumed gas temperature, and \( n \approx 100 \, \mathrm{cm}^{-3} \) is the average electron density in the EELR, estimated from Section~\ref{ne and te}. To determine the volume, the geometry of the EELR was modeled as a bi-cone with a height of 10.54~kpc and a radius of 4.28~kpc. Using these parameters, the \(\rm PV \) work is estimated to be \( 1.64 \times 10^{57} \, \mathrm{ergs} \). Now assuming $\rm P_{jet}\approx 10^{43}\,ergs~s^{-1}$ (Section~\ref{jet power}) and the spectral age of the inner radio jet to be 20~Myr (refer to Section~\ref{episodic activity}), we have estimated the time-averaged power of the jet, which comes out to be $\rm 1.07\times10^{58}\,ergs$. These results indicate that the jet is capable of inflating the gas, with a transfer efficiency (i.e., the ratio of PV work to the mechanical energy of the jet) of $\approx$15$\%$. The remaining 85$\%$ of the jet’s energy is likely dissipated through other processes such as radiative losses or heating.

\subsection{Jet-medium interaction}

In the following sections, we discuss the evidence of jet-ism interaction and how jet influences the medium.
\subsubsection{Transverse shock and enhanced velocity dispersion}
\label{transverse shock}

When the relativistic jet interacts with the surrounding gas, rapid shocks can induce ionization in the gas medium. This ionized gas emits distinctive spectra and can provide insights into the physical characteristics of the shocks. Typically, line ratios within the narrow-line region suggesting the presence of shocks are positioned within the LINER region of optical BPT diagrams \citep{2019A&A...622A.146M, 2020A&A...643A.139P, Cazzoli2022}. In our study, we have detected LINER excitation in the region perpendicular to the jet (Section~\ref{sec:model}). The emergence of these transverse LINER regions, as illustrated in the resolved BPT diagram (see Figure~\ref{shock and bpt}, bottom left), may signify the existence of a jet-induced shock propagating perpendicularly to the jet axis. Furthermore, the increased velocity dispersion observed perpendicular to the jet axis (Figure~\ref{shock and bpt}, bottom right) aligns with previous studies of low-redshift AGN, where similar features are linked to jet-ISM interactions or lateral gas expansion around the radio jet \citep[e.g.,][]{Couto2013, Riffel2014, Lena2015, 2021A&A...648A..17V, Cazzoli2022}. Using VLT MUSE observations, \cite{2021A&A...648A..17V} found extended emission ($\geq$1 kpc) with elevated velocity dispersion (W70 $\geq$ 800–1000 km s\(^{-1}\)) perpendicular to the jet in nearby Seyfert galaxies. Their BPT analysis also indicated that the gas excitation in these regions is consistent with shock ionization.
While our observations do not show such high-velocity dispersions, we do observe a shocked region with an inflated-shell-like structure, showing a higher value of velocity dispersion compared to the inner region. The average [O III] dispersion is $158\pm55$ km~s$^{-1}$ in the shocked region and $63\pm13$ km~s$^{-1}$ in the inner region.

Besides observational evidence, radio-mode feedback models also support this scenario. As discussed in \citep{2016MNRAS.461..967M, 2018MNRAS.476...80M}, the jet power and its inclination relative to the galaxy disk play a crucial role in shaping the interaction between the jet and the ISM. Jets inclined closer to the disk induce stronger interactions, producing shock-driven, wide-angle outflows along the disk's minor axis and increasing turbulent dispersion within the disk. In contrast, jets oriented perpendicular to the disk have a weaker impact on the surrounding gas.


\subsubsection{Outflows: Enhanced velocities along the jet}
\label{outflows}

The gas dynamics within the EELR can be shaped by multiple contributing factors. Firstly, the gravitational potential of the host galaxy plays a pivotal role, it not only dictates the motion of stars but also exerts its influence on the gas in the EELR \citep{1986ApJ...302...81S, 1996ApJ...465...96N}. Furthermore,
AGN-driven winds, powered by radiation pressure or Magneto-hydrodynamic (MHD) mechanisms \citep{Krolik1984, Emmering1992, 1999MNRAS.308L..39F, 2005ApJ...618..569M, 2015MNRAS.449..147T, Chan2016}, or thermal pressure arising from heated gas, originating from the central AGN/ starburst \citep{1979RvMP...51..715D, 2020ApJ...893L..34D} or even AGN jets
\citep{1989MNRAS.240..225T,  2008MNRAS.387..639H, 2013MNRAS.431.2350I, 2016MNRAS.455.2453M, 2017MNRAS.472.4659V, 2019MNRAS.485.2710J} can trigger the ejection of gas, resulting in outflows. To distinguish between these contributions, we study the residual maps, which is the difference between emission line gas velocity and stellar velocity. Figure~\ref{kinematics}~(i) - (l), shows the residual velocity maps created for the [O\,{\small III}] and H$\alpha$ emission lines for both Gaussian components. 
While several studies typically use the broad component to characterize outflows, e.g., \citep{Harrison2014, Singha2022CARS}, our broad and narrow component residual maps suggest that the broad component closely follows the stellar velocity, whereas the narrow component exhibits a significant offset. This suggests that the narrow component has additional velocity beyond what is expected from the galaxy's gravitational effects. This offset could be due to the outflowing gas, which may be experiencing a systematic velocity shift as it is pushed away by the jet. Hence, we refer to the narrow Gaussian component as the outflowing component of the galaxy.

\cite{2022ApJ...936...88F} also find [O\,{\small III}] velocities in excess to the rotating disk component from their $\rm \tt{3DBarolo}$ fitting and position velocity diagram analysis. They suggested that these high-velocity components could be associated with extra-planar gas caused by tidal debris. In either case, there appears to be at least two components of [O\,{\small III}] gas. The outflowing or the extra-planar component seems to be spatially coinciding with the jet.

Furthermore, \cite{2022arXiv220805915H} also reports the presence of a nuclear outflow (referred to as ``[O\,{\small III}] bubble") in the north-east direction, within a proximity of $2^{\prime\prime}$ (1.2 kpc) from the center. They found the outflow velocities reaching up to 300 km~s$^{-1}$. The average gas velocity calculated from our narrow component [O\,{\small III}] and H$\rm alpha$ residual maps, comes out out to be 212$\pm$33 km~s$^{-1}$ and 130$\pm$18 km~s$^{-1}$ respectively. Since, we are measuring line-of-sight velocities, we suspect that the actual outflow velocities might be substantially higher, depending on the inclination of the EELR.

\subsubsection{Outflow energetics}
\label{outflow}
\begin{figure*}
\centering
\includegraphics[height=7.5cm,trim = 0 0 0 0 ]{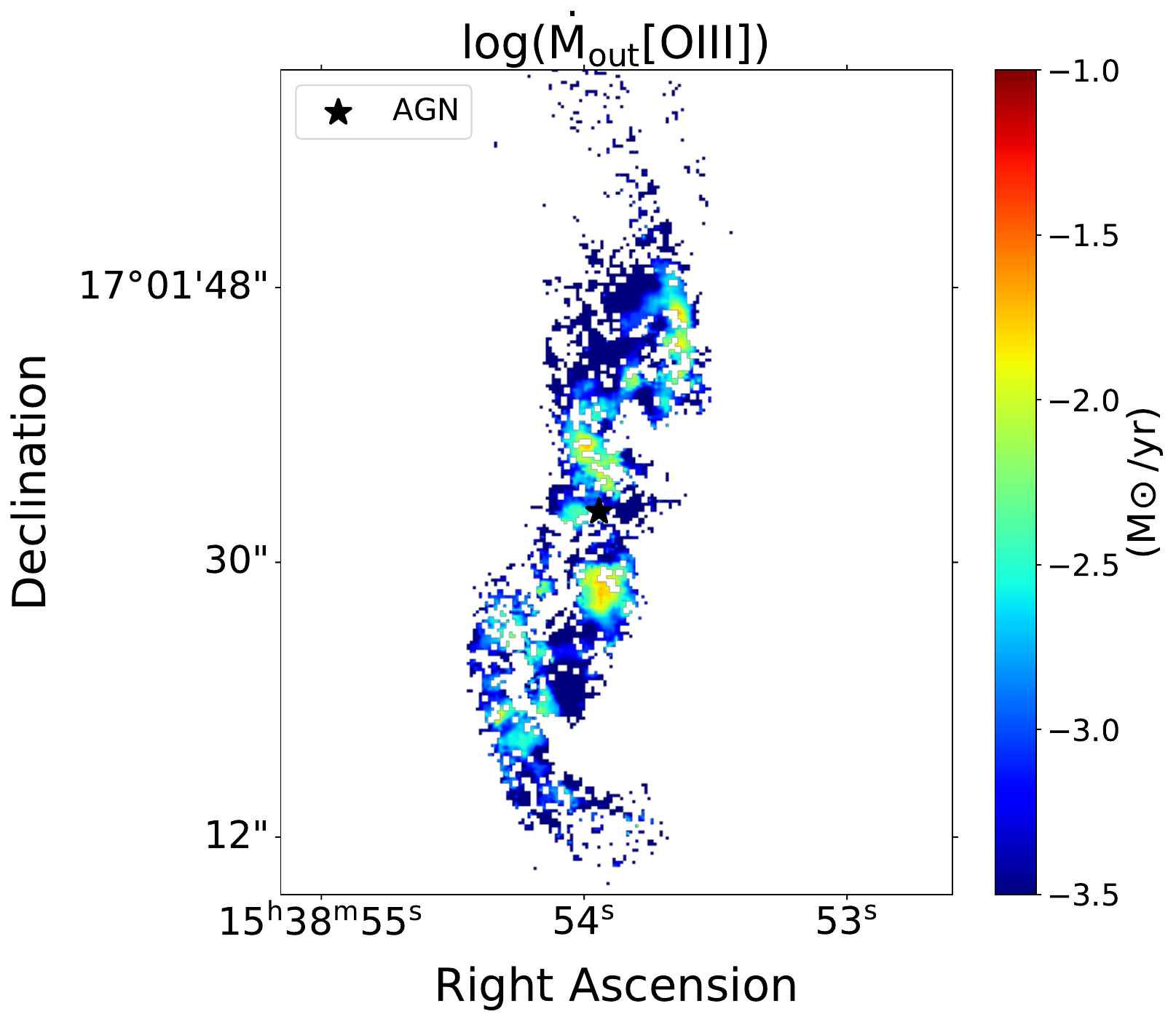}
\includegraphics[height=7.5cm,trim = 0 0 0 0 ]{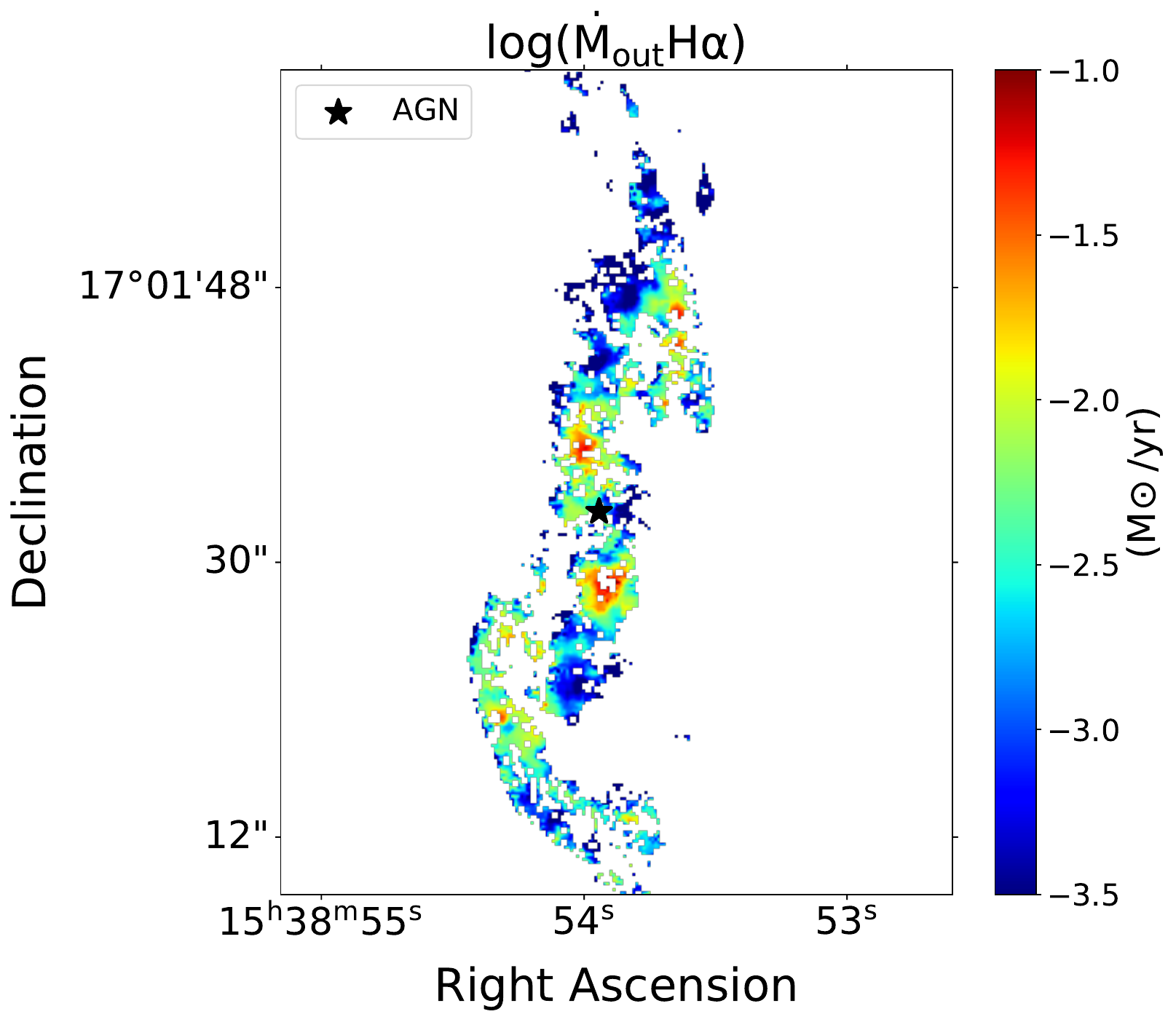}
\caption
{\small Plots showing trends of mass outflow rate derived for [O\,{\small III}] and H$\rm alpha$ gas respectively.
}
\label{outflow plots}
\end{figure*}

\begin{figure*}
\centering
\includegraphics[height=7.5
cm,trim = 0 0 0 0 ]{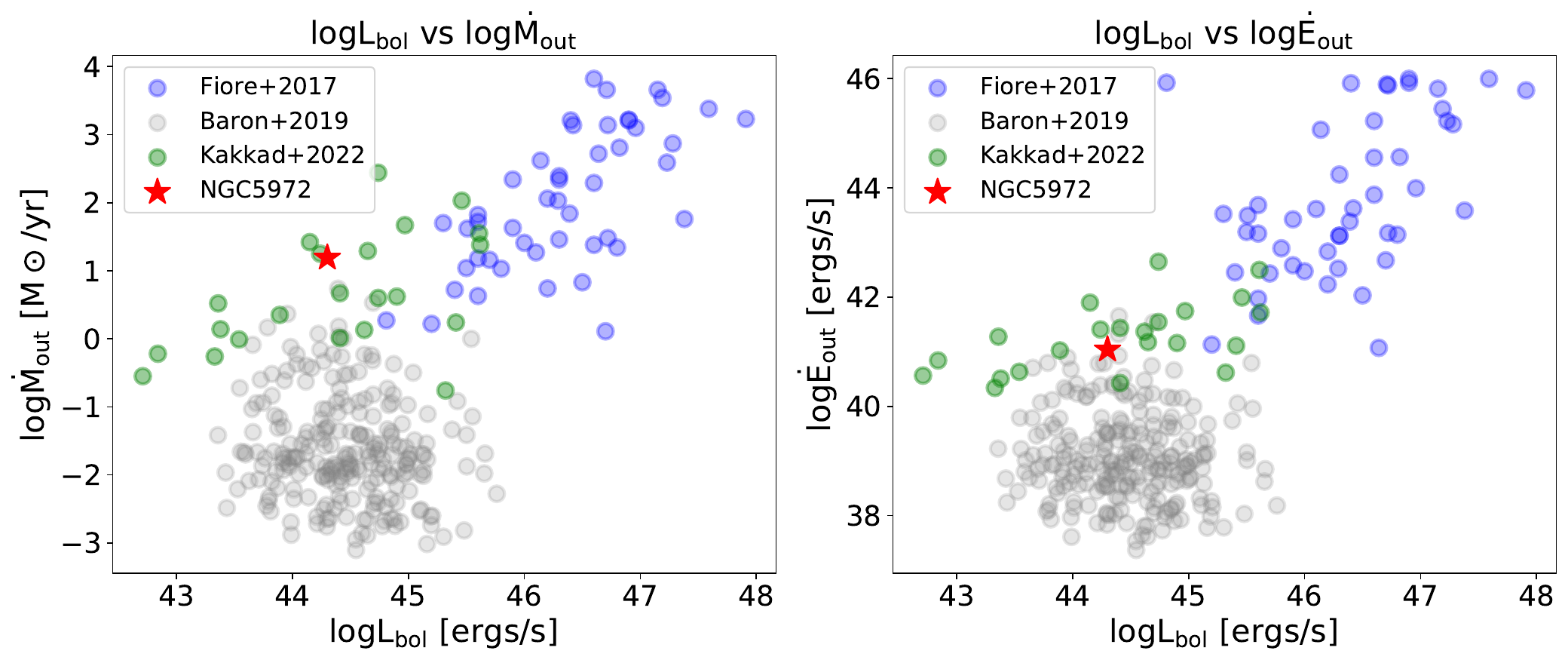}
\caption
{
\small Left: Mass outflow rate as a function of AGN bolometric luminosity. Right: outflow kinetic power as a function of AGN bolometric luminosity. The blue data points represent the ionized outflow measurements reported by \cite{2017A&A...601A.143F}. The green data points represent the ionized outflow measurements reported by \cite{2022MNRAS.511.2105K}. The grey data points represent samples from \cite{2019MNRAS.486.4290B}, while the red star is our target, NGC\,5972.
}
\label{outflow corr_plots}
\end{figure*}

To understand the outflow characteristics, we study the mass outflow rate of the galaxy. We focus on the physical parameters associated with the narrow Gaussian component to determine the outflow properties.
Initially, we calculated the gas mass within the outflow using [O\,{\small{III}}] and H$\alpha$ emission lines, employing the equations outlined in \cite{Venturi2023}, which rely on the relationship previously established by \cite{2015A&A...580A.102C} and \cite{2017A&A...601A.143F}: 
\begin{equation}
\label{outflow_mass}
\rm \frac{M_{\text{out,H}\alpha}}{M_\odot} = 0.6 \times 10^9 C\left(\frac{L_{\text{H}\alpha}}{10^{44} \, \text{erg s}^{-1}}\right)\left(\frac{n_e}{500 \, \text{cm}^{-3}}\right)^{-1}
\end{equation}

\begin{equation}
\label{outflow_mass2}
\rm \frac{M_{\text{out,[O\,{III}]}}}{M_\odot} = 0.8 \times 10^8 \left(\frac{C}{10^x}\right)\left(\frac{L_{\text{[O\,{III}]}}}{10^{44} \, \text{erg s}^{-1}}\right)
\left(\frac{n_e}{500 \, \text{cm}^{-3}}\right)^{-1}
\end{equation}

Wherein, we set $\rm C=\langle {n}{e}{\rangle }^{2}/\langle {n}{e}^{2}\rangle $ to unity. $\rm x={[O/H] - [O/H]_{\odot}}$, where we consider [O/H] to be the solar oxygen abundance. We assume the gas temperature T$\simeq$10$^4$ K, and electron density $\rm n_e$ is calculated using equation~\ref{eq:myequation}.

Assuming that the density \( \rho(r) \) and the outflow velocity \( v(r) \) is constant within a spaxel of thickness \( \Delta R \), the average outflow rate across each spaxel can be calculated. This leads to a simplified expression for the radial average mass outflow rate as:
\begin{equation}
\rm \dot{M}_{\text{out}} = \frac{M_{\text{out}}v_{\text{out}}}{\Delta R}
\end{equation}
Where $\rm M_{\text{out}}$ is the mass of the outflow gas calculated using equations~\ref{outflow_mass} and \ref{outflow_mass2}, $\rm \Delta R$ is the width of the spaxel ($\sim$0.1 kpc), and $\rm v_{\text{out}}$ is the outflow velocity, $\rm v_{\text{out}} \approx v_{res}$, where $\rm v_{res}$=$\rm v_{[O\,{\small{III}}]}-v_{stellar}$ is the residual velocity of the narrow component used as a proxy for the outflow velocity. From Figure~\ref{outflow plots}, it is observed that both outflows are concentrated near the center. However, [O\,{\small{III}}] outflows are more tightly confined within approximately 5 kpc radius, whereas, H$\alpha$ emission extends out up to $\sim10$ kpc radius.

We have also computed the kinetic power of the outflow as:
\begin{equation}
\rm \dot{E}_{\text{out}} = \frac{\dot {M}_{\text{out}}v_{\text{out}}^2}{2}
\end{equation}
We found that the value of $\rm \dot{E}_{\text{out}}$ using H$\alpha$ emission 
line is $\rm 4.9\times 10^{41} ergs$ $\rm s^{-1}$, whereas $\rm \dot{E}_{\text{out}}$ using [O\,{\small{III}}] emission line is $\rm 1.09\times10^{41} ergs$ $\rm s^{-1}$. Comparing the outflow kinetic power with the radio jet power $\rm P_{\rm jet}\approx10^{43}$ ergs s$^{-1}$, suggests that the jet has a substantial amount of energy available to drive or influence the outflows.

We have compared our results with the literature like \citet{2017A&A...601A.143F}, \cite{2019MNRAS.486.4290B} and \cite{2022MNRAS.511.2105K}. All these studies shows a dependence of outflow properties on the AGN bolometric luminosity, with a sample coverage of $\rm L_{bol} \sim 10^{42} - 10^{48}$ ergs $\rm s^{-1}$. For NGC~5972, we adopted the AGN bolometric luminosity value calculated by \citet{2022ApJ...936...88F}, i.e., $\rm 2 \times 10^{44}$ ergs $\rm s^{-1}$ based on the Gemini IFU observations. In Figure~\ref{outflow corr_plots}, the left panel shows the mass outflow rate and the right panel shows the kinetic power as a function of AGN bolometric luminosity. The red star represents NGC\,5972, where the outflow properties were calculated from [O\,{\small{III}}] gas. The blue dots represent samples from \citet{2017A&A...601A.143F}, which include AGN ionized winds traced by high-velocity [O\,{\small{III}}], H$\alpha$, and/or H$\beta$. The green dots represent samples from \cite{2022MNRAS.511.2105K}, which feature sub-kiloparsec scale [O\,{\small{III}}] ionized gas in low-redshift (z $\leq$ 0.1) X-ray AGN. The grey dots represent samples of warm ionized outflows of low-to-moderate luminosity type-II AGN within z $\le 0.15$, studied by \cite{2019MNRAS.486.4290B}. Based on the plots, we observe that NGC\,5972 shows trends similar to the \cite{2022MNRAS.511.2105K} samples with outflow mechanisms at a lower luminosity AGN and outflow strength.


\subsubsection{Clues from  inclination}
\label{sec_Clues from  inclination}

Although there is a remarkable alignment between the inner jet and the EELR in projection, they need not be spatially coincident in the sky. We try to constrain the inclination angle of the jet in comparison with the EELR in the sky.

The disk rotation of NGC\,5972 (Figure~\ref{ppxf_fit}) suggests that the southern part of the galaxy is aligned towards us. Using rotation curve fitting, \cite{2022ApJ...936...88F} estimated an inclination angle in the range 40$^{\circ}$-50$^{\circ}$ for the stellar emission. The [O\,III] emission line gas has a similar orientation as the galaxy (Figure~\ref{kinematics}) although the mean inclination angle determined by \cite{2022ApJ...936...88F} is around 15$^{\circ}$.
They also report that the [O\,{\small III}] emission profile is complex and does not fit with a purely rotational model. 

The radio image of the jet can provide additional insight into the orientation of the jet. Based on the surface brightness contrast between the north-western and south-eastern lobes (Figure~\ref{radio_images}), we expect that the northern jet/lobe is the approaching one if we assume the jet is relativistic. However, it should be noted that such asymmetries can also stem from differences in environments. Indeed, the gas in the northern region appears to be denser than in the southern region. For the broad component, the mean electron density in the north is \( \rm 140 \pm 25 \, cm^{-3} \) compared to \( \rm 71 \pm 16 \, cm^{-3} \) in the south, and for the narrow component, it is \( \rm 252 \pm 20 \, cm^{-3} \) in the north and \( \rm  198 \pm 38 \, cm^{-3} \) in the south.

Another clue comes from the Laing-Garrington effect \citep{1988Natur.331..149L, 1988Natur.331..147G}, which hypothesizes that the lobe showing higher fractional polarization is pointed towards us due to lesser intervening medium causing lower depolarization. While we detect polarization in the northern lobe with high fractional polarization ($\sim15\%$), the upper limit on the fractional polarization in the southern lobe is 30\% based on a 3$\sigma$ limit. Hence this method is also inconclusive in the absence of deeper data. Therefore, at this point, we are not able to confirm or reject the spatial coincidence of the jet and EELR in the sky.

\begin{figure}
    \centering
    \includegraphics[width=0.45\textwidth]{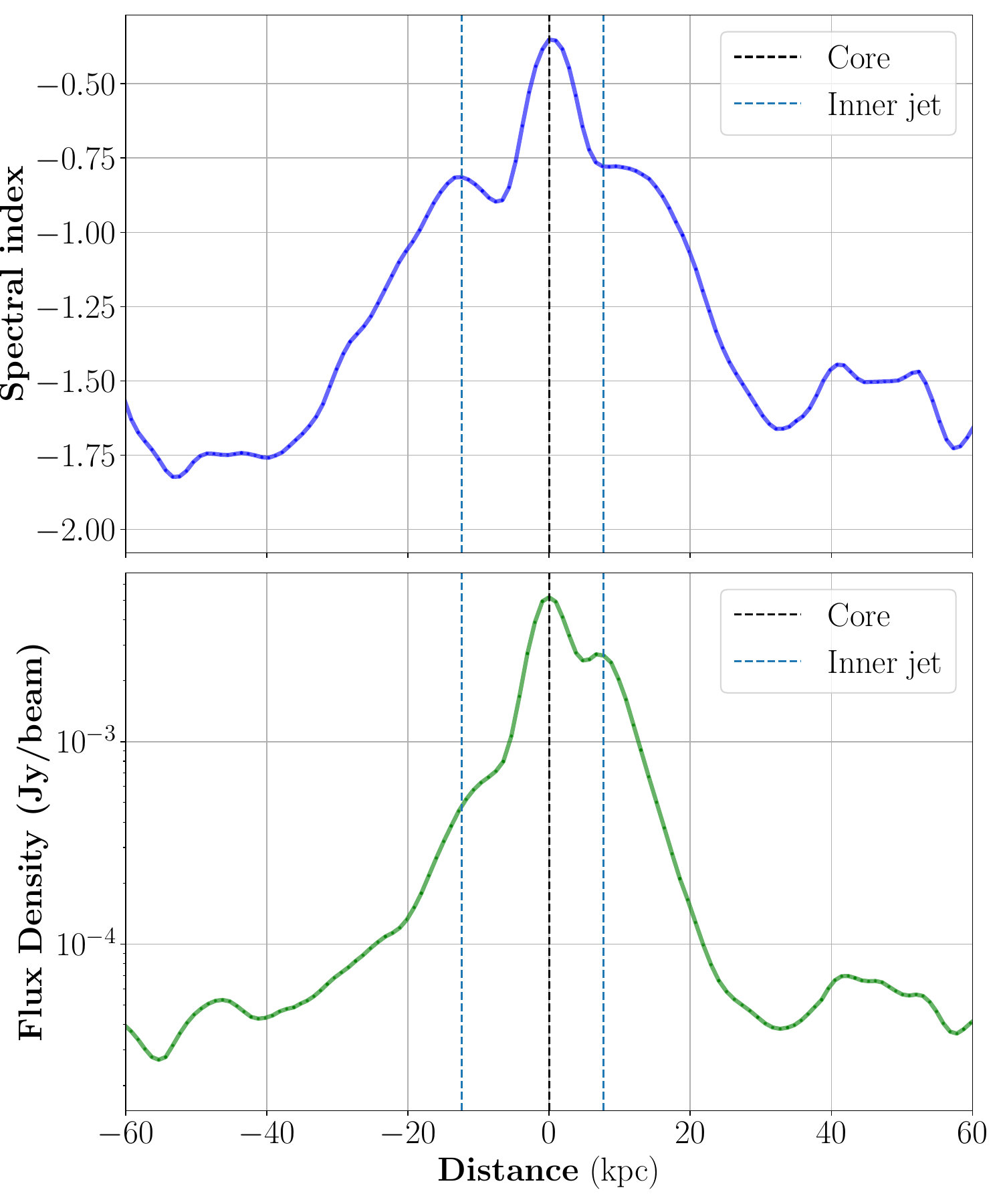}
    \caption{Spatial variation of flux density and spectral index from the core.}
    \label{fig:SB_spix_profile}
\end{figure}


\subsection{Evidence for episodic activity}
\label{episodic activity}
The radio images also show evidence of episodic activity. The surface brightness and the spectral index profile show a clear discontinuity along the jet (see~Figure~\ref{fig:SB_spix_profile}). The flattening of the spectral index at the location of the surface brightness peak on either side of the core is typical of hotspots. The steeper spectral index just beyond the apparent inner hotspot suggests that the inner jet is being straddled by the outer jet which consists of relativistic plasma that underwent acceleration before the inner pair. These features are typical for double-double radio galaxies. It has to be noted, however, that the magnetic field orientation is not typical for hotspot regions. Usually, it is aligned along the edge of the hotspot (due to compression and magnetic field amplification at the edge), whereas in this source the magnetic fields are primarily aligned along the direction of the jet (see Figure~\ref{VLA_poln}). Such an alignment might be because the inner jet does not face much obstruction on its path as the outer jet has cleared most of the material out in the previous episode. The inner jets aligning with the outer jets is indeed commonplace in powerful double-double radio galaxies \citep{Sebastian2018,Marecki2023}, although this is not the case in weaker jet systems \citep{Kharb2006,Sebastian2019,Rao2023}. Another interesting observation is that the core is bright with a relatively flat spectrum  ($\sim-0.35$) pointing to a currently active radio core. 

The leading explanation for the origin of Voorwerp galaxies is that these are Seyfert galaxies showcasing episodic activity. The `on' and `off' timescales in NGC\,5972 were studied in detail by \cite{2022ApJ...936...88F}. They studied the radial dependence of the ionization state and estimated a clear increase in L$_{\rm bol}$ with radius. Hence, they argue that the Quasar faded gradually by a factor of 100 over 10000~years. 

We estimated the dynamical and spectral ages for the inner and the outer lobes to compare with those estimated by \cite{2022ApJ...936...88F}.
We used the equations (1) $-$ (5) from \cite{Odea1987} and \cite{Perez2007} to estimate the equipartition magnetic field parameters and the spectral age, respectively.
The spectral age calculation is fraught with several uncertainties. Parameters, such as the filling factor and the ratio of proton to electron number densities, both of which were assumed to be unity, remain uncertain and also affect the equipartition magnetic field values, which can, in turn, affect the spectral ages. More importantly, the spectral ages will critically depend on the break frequency. We lack the multi-frequency coverage at similar resolutions to constrain the break frequency accurately. We obtain spectral ages of approximately 20 Myr and 40 Myr by assuming break frequencies of 5.5 GHz and 1.5 GHz, respectively, for both the inner and the outer lobes. The inner lobes likely have a much higher break frequency compared to the outer ones, suggesting the outer lobes have undergone more radiative losses over time, consistent with their older age \citep[e.g.,][]{Nandi2019,Marecki2016,Marecki2021}.

In the absence of the relevant data, we opt to estimate the dynamical ages using a more simplistic approach. Typical  FR\,II hotspots have a mildly relativistic advance speed of 0.1~c to 0.5~c \citep{Odea2009}, which remains constant over the lifetime. This advance speed translates to an age range of 1.6-8~Myr and 35-170~Myr for the inner and the outer hotspots, respectively. Note that these ages are lower limits and can be higher depending on the inclination of the lobes in the sky. Furthermore, the current activity of the radio core and the age of the inner hotspots being tens of Myr makes it challenging to align with a scenario where the quasar gradually faded over 10,000 years.

Figure~\ref{radio_images} shows that the inner radio jets are not aligned with the outer lobes, indicating a shift in the ejection axis of the radio jet. This shift could be the result of a merger between two galaxies and their central black holes, as suggested by \cite{2002Sci...297.1310M} for X-shaped or \cite{Rubinur2017} for S-shaped radio sources. A plausible scenario is that the jets were initially oriented east-west, powering the lobes. Following the merger, the jet axis shifted. This scenario aligns with the merger evidence discussed in Section~\ref{sec:morphology}. It also implies that the radio lobes are currently aging without a fresh supply of relativistic particles. Furthermore, based on the clear contrast in radio spectral index between the inner and the outer lobe (Figure~\ref{fig:spix}),
we propose that the inner radio jet originates from a more recent episode of AGN activity compared to the larger structure, which resembles an FRII narrow-line radio galaxy. The inner structure, with its S-shaped jet, resembles the lobes of a Seyfert galaxy, such as NGC\,3516 \citep[e.g.,][]{Baum1993}.

\subsection{UV and X-ray emission associated with the EELR: evidence for shock ionization?} 
\label{uv-xray}

\begin{figure}
\centering
\includegraphics[width=0.48\textwidth,trim = 30 0 0 0 ]{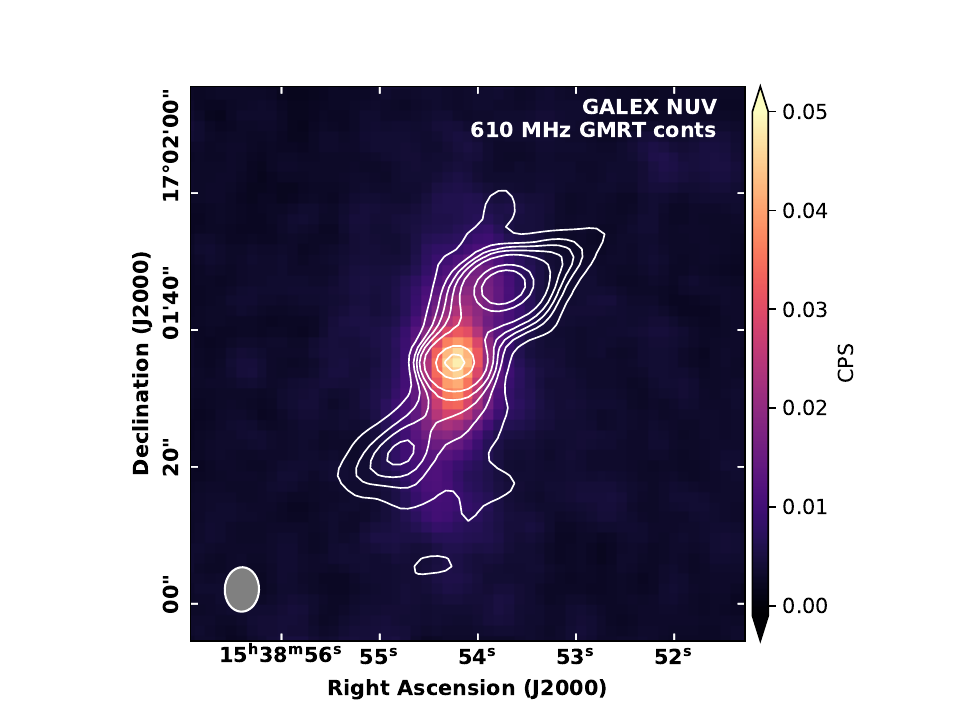}
\includegraphics[width=0.48\textwidth,trim = 30 0 0 0 ]{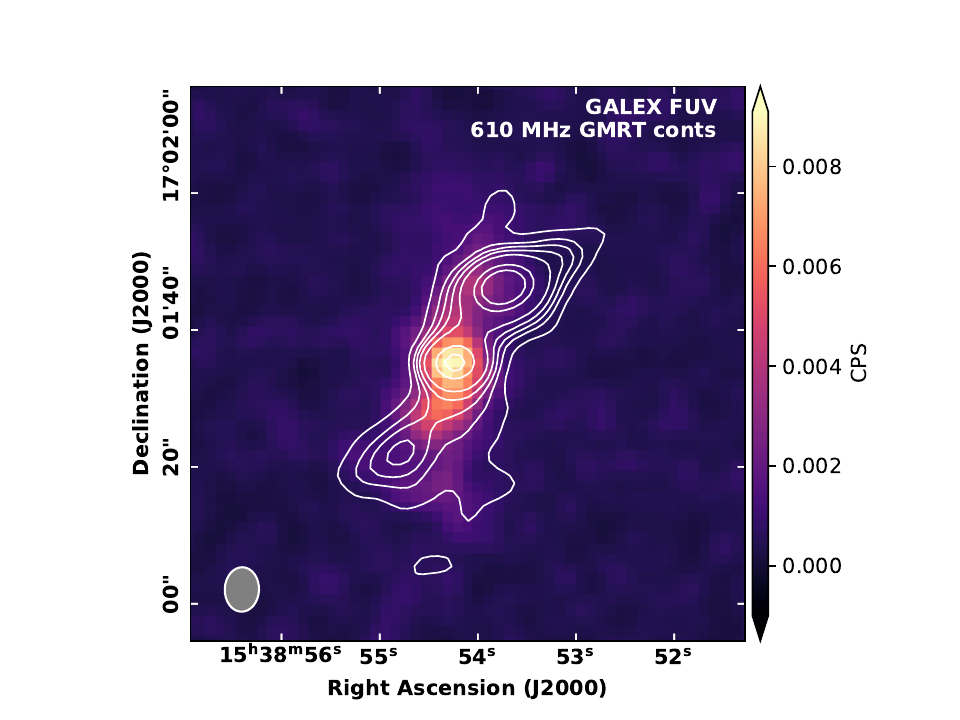}

\caption{\small GALEX FUV and NUV band image of NGC\,5972 with 610 MHz contours overlaid on the top. 
}
\label{uv+hst}
\end{figure}

Archival GALEX data of both FUV and NUV emission is available for NGC\,5972 (Figure~\ref{uv+hst}). The UV images reveal an extended structure aligned with the radio jet. Although photoionization from the central AGN is a common source of UV emission, it does not fully explain the observed features in NGC\,5972. Specifically, the UV emission is neither isotropic nor bipolar and does not decrease monotonically with radius as expected from a purely photoionized model.
An alternative explanation for the UV emission could be star-forming activities induced by the jet \citep{Gaibler2012, Chetna2021, Chetna2023}. However, this scenario does not explain the X-ray emission coincident with the jet observed by \cite{2022arXiv220805915H}.

\textit{Chandra} observations of NGC\,5972 shows an extended soft X-ray emission coincident with the [O\,{\small III}] emission, which may be attributed to either shock from a jet \citep{Sutherland&Bicknell2007, Lanz2015}, a hot wind \citep{Nims2015}, or from AGN photoionized line emission \citep{Sambruna2001}.  \cite{2022arXiv220805915H} noted some alignment of X-ray regions with the radio emission inferred from 3 GHz VLASS data, hinting at a possible connection between the radio jet and X-ray emission. They also observed that the X-ray emission in the southern EELR follows the curved [O\,{\small III}] tail, with the X-ray peak located at a larger distance, suggesting weak shock-induced X-ray emission. Since our 610~MHz radio data provides a better image of the extended radio jet and reveals the southern jet, which shows alignment with UV and optical [O\,{\small III}] emission, it suggests that the observed X-ray emission could also be attributed to jet-induced shocks. Additionally, APEC modeling by \cite{2022arXiv220805915H} indicates EELR temperatures in the range from 0.6 to 3.4 keV ($\rm\approx10^6 - 10^7~K$), which further indicates that such high-temperature gas could be a result of shock heating.

Overall, the shock model provides a simpler and more cohesive explanation for the observed X-ray and UV emission and is consistent with the optical emission line ratios discussed in Section~\ref{bpt_sec}. 
While the fading quasar model cannot be entirely ruled out, we propose a combined shock+precursor model as the most plausible explanation for ionizing the EELR in NGC\,5972. This model accounts for all the multi-wavelength observations and warrants further investigation.


\begin{figure}
\centering
\includegraphics[height=6.4cm,trim = 0 0 0 0 ]{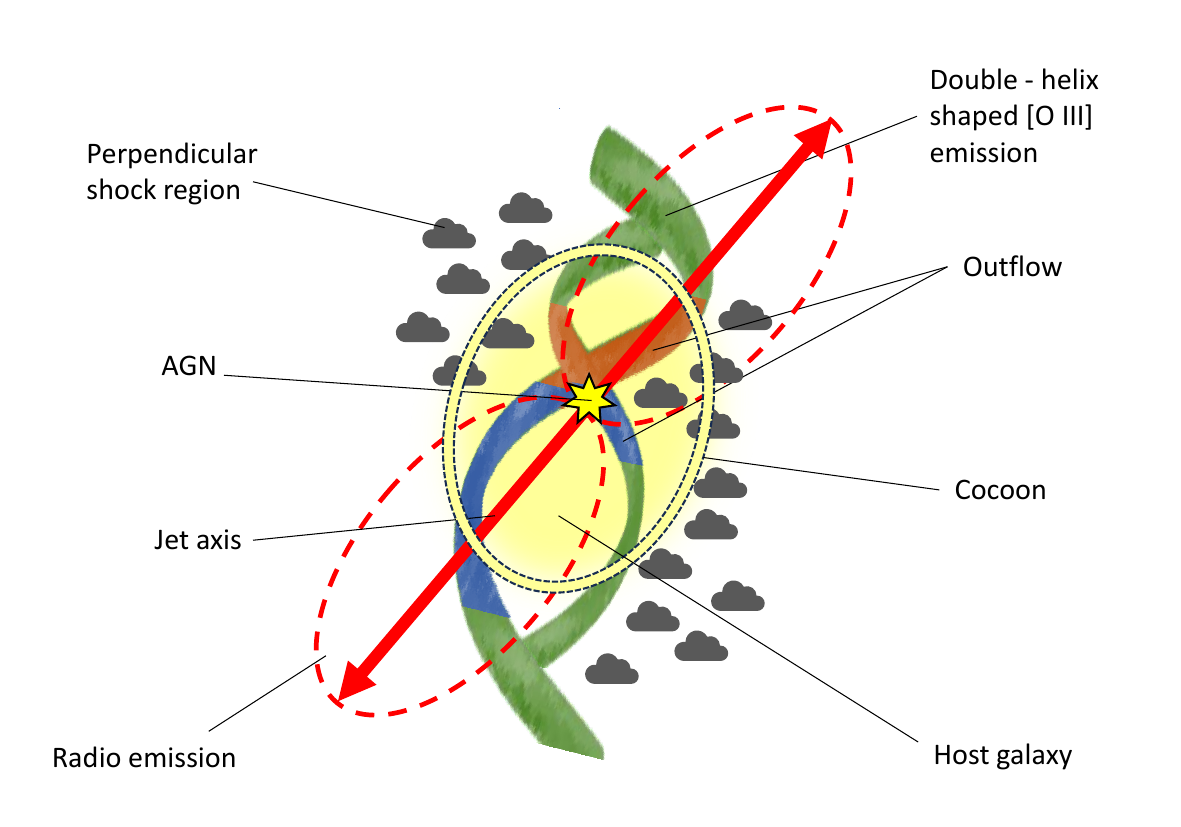}

\caption
{\small Cartoon schematic diagram showing various mechanisms at play in NGC\,5972.
The green helix represents the 
[O\,{\small III}] emission as observed from the HST image. The jet axis is indicated by the red line, and the radio emission is depicted by the dotted red lobes. The black clouds denote the shock regions perpendicular to the jet, as determined by BPT analysis. The red and blue structures represent the outflow region in the galaxy. The cocoon shell is shown in yellow.
}
\label{schematic}
\end{figure}

\section{Conclusions}
\label{conclusion}

We present a detailed study of NGC\,5972, a noteworthy active galaxy having kpc-scale EELR, using VLA L-band and C-band, GMRT 610~MHz radio observations, and IFS VLT/MUSE archival data. Despite previous research suggesting that the EELR is merely a consequence of AGN photoionization, our observations show that the radio jet also plays a significant role. A comprehensive overview of our results is listed below.
\begin{enumerate}

    \item The morphology of NGC\,5972 suggests a dynamic history of the galaxy. A twisted dust lane and a helical-shaped EELR can be explained as a result of a past merger activity or by a precessing disk model as discussed in \cite{2015AJ....149..155K}. Additionally, NGC 5972 showcases an impressive S-shaped radio structure spanning about 235 kpc in radius (Figure~\ref{radio_images}), surpassing the typical extents seen in Seyfert galaxies, where radio lobes generally cover only tens of kpc, suggesting a prolonged AGN activity.

    \item The velocity maps of [O{\small{III}}] and H$\alpha$ emission lines provide intriguing insights into the gas kinematics in the galaxy. The gas velocities for [O{\small{III}}] and H$\alpha$ emission lines show significant differences between narrow and broad components. For [O{\small{III}}], narrow components have velocities of 163$\pm$68 km s$^{-1}$ (north) and -208$\pm$71 km s$^{-1}$ (south), while broad components show 132$\pm$51 km s$^{-1}$ and -182$\pm$28 km s$^{-1}$, respectively. For H$\alpha$, the narrow components have velocities of 192$\pm$34 km s$^{-1}$ (north) and -162$\pm$42 km s$^{-1}$ (south), and broad components have 115$\pm$47 km s$^{-1}$ (north) and -129$\pm$23 km s$^{-1}$ (south). The higher values of velocity for the narrow component compared to the broad component suggest that the narrow component is tracing the fast-moving gas in the galaxy. The residual velocity maps (Figure~\ref{kinematics} (i) to (l)) show that the narrow component exhibits a significant velocity offset from the stellar motion, and likely represents the outflowing gas influenced by the central AGN or jet. The measured average outflow velocities are 212$\pm$33 km s$^{-1}$ for [O\,{\small III}] and 130$\pm$18 km s$^{-1}$ for H$\alpha$.

    \item The outflow characteristics suggest that the [O\,{\small III}]-derived outflows are confined within 5 kpc, while H$\alpha$-derived outflowing gas extends up to 10 kpc. The kinetic power for H$\alpha$-derived outflows is $4.9 \times 10^{41} \, \text{ergs} \,\, \text{s}^{-1}$ and [O\,{\small III}]-derived outflows is $1.09 \times 10^{41} \, \text{ergs} \,\, \text{s}^{-1}$. Comparing the jet and the outflow energetics, we find that the radio jet is capable of driving the outflows. Furthermore, comparing the correlation between mass outflow rates, their energetics, and AGN bolometric luminosity with other similar samples in the literature, we find that NGC 5972 exhibits trends consistent with lower luminosity AGN.
    
    \item The results from BPT analysis show that the gas along the jet axis is AGN ionized, whereas, the gas that is perpendicular to the jet is dominated by LINER-like emission (Figure~\ref{shock and bpt}). This is consistent with results presented in \cite{2022ApJ...936...88F}.
    We have also performed MAPPINGS V shock modeling which shows that the best-fit model onto the AGN ionized region is shock+precursor whereas the region where we observe LINER-like emission shows pure shock ionization (see Section~\ref{sec:model}). Our results complement the work of \cite{2022ApJ...936...88F} by highlighting the role of jet-induced shocks in sustaining or enhancing ionization in the EELR. The dual mechanism provides a more complete view of AGN feedback, capturing the interplay between fading AGN photoionization and ongoing shock excitation driven by the jet.

    \item We observe a region of enhanced [O\,{\small III}] velocity dispersion perpendicular to the jet axis and in a shell-like structure around the jet. The region is partially aligned with the LINER-like emission area ( Figure~\ref{shock and bpt}, bottom right), suggesting it may be a result of jet-induced shocks (see Section~\ref{transverse shock}). The
    finding is consistent with previous studies of low-redshift AGN and radio-mode feedback models.

    \item Our analysis indicates that NGC\,5972 hosts a kpc-scale radio jet with an estimated power of \( \rm P_{\rm jet} \approx 10^{43} \) ergs s\(^{-1}\), which aligns with the [O\,{\small III}] emission-line region (Figure~\ref{kinematics}). The jet power exceeds both the AGN kinetic power (\(\approx 10^{42}\) ergs s\(^{-1}\)) and the mechanical power from star formation (\(\approx 9.5 \times 10^{41}\) ergs s\(^{-1}\)). Additionally, the jet has sufficient energy to drive the gas, with a transfer efficiency of approximately \(\approx 15\%\) (see Section~\ref{workdone}). Thus, this indicates that the jet plays a significant role in the galaxy's energy dynamics.

    \item The radio images of the galaxy indicate episodic activity, marked by surface brightness and spectral index discontinuities typical of double-double radio galaxies. Spectral age estimates, though uncertain, indicate the inner lobes are approximately 20 Myr old, while the outer lobes are around 40 Myr old, pointing to greater radiative losses in the outer lobes. Dynamical age estimates, based on typical FR II hotspot advance speeds, place the inner hotspots at 1.6-8 Myr and the outer hotspots at 35-170 Myr. The misalignment of the inner jets with the outer lobes suggests a merger event that shifted the jet axis. This scenario is supported by contrasting radio spectral indices, implying the inner jets are from a more recent AGN activity episode.

\end{enumerate}

In summary, we propose a jet-driven feedback mechanism as an alternative explanation for the formation of the EELR in the Voorwerp galaxy, NGC\,5972. Figure~\ref{schematic} shows a cartoon representation summarizing the proposed scenarios to explain the structures and alignment of the radio jet with the EELR. 
While we cannot completely dismiss the possibility of the faded Quasar model, the influence of the AGN jet effectively accounts for all the multi-wavelength properties and is a much simpler explanation. Future observations with high-resolution facilities like ALMA and JWST \citep{Audibert2023, Esposito2024, Zhang2024, Esparza-Arredondo2025, Ogle2025} will allow us to study the multiscale, multiphase nature of jet feedback and will be crucial in further understanding the role of jet-driven feedback in Voorwerp galaxies.

\section*{Acknowledgements}
We would like to thank Chetna Duggal for useful suggestions and discussions. PK acknowledges the support of the Department of Atomic Energy, Government of India, under the project 12-R\&D-TFR-5.02-0700. SB, BS, and CO acknowledge support from the Natural Sciences and Engineering Research Council (NSERC) of Canada. SS acknowledges funding from ANID through Fondecyt Postdoctorado (project code 3250762), Millenium Nucleus NCN23{\_}002 (TITANs), and Comité Mixto ESO-Chile. O. Bait (OB) is supported by the {\em AstroSignals} Sinergia Project funded by the Swiss National Science Foundation. 

This research has made use of the services of the ESO Science Archive Facility.
This research is based partly on observations made with the VLA/ National Radio Astronomy Observatory (NRAO). We thank the NRAO for providing access to the archival data and for supporting new observations (Project Id:
23A-264). The NRAO is a facility of the National Science Foundation operated under cooperative agreement by Associated Universities, Inc. 
The GALEX data presented in this paper were obtained from the Mikulski Archive for Space Telescopes (MAST) at the Space Telescope Science Institute. The specific observations analyzed can be accessed via \url{ http://dx.doi.org/10.17909/vpzv-2076}. STScI is operated by the Association of Universities for Research in Astronomy, Inc., under NASA contract NAS5–26555. This research is based partly on observations made with the GMRT. We thank the staff of the GMRT that made these observations possible. GMRT is run by the National Centre for Radio Astrophysics of the Tata Institute of Fundamental Research.

%






\bibliography{sample631}{}

\begin{thebibliography}{}
\expandafter\ifx\csname natexlab\endcsname\relax\def\natexlab#1{#1}\fi
\providecommand{\url}[1]{\href{#1}{#1}}
\providecommand{\dodoi}[1]{doi:~\href{http://doi.org/#1}{\nolinkurl{#1}}}
\providecommand{\doeprint}[1]{\href{http://ascl.net/#1}{\nolinkurl{http://ascl.net/#1}}}
\providecommand{\doarXiv}[1]{\href{https://arxiv.org/abs/#1}{\nolinkurl{https://arxiv.org/abs/#1}}}

\bibitem[{{Alarie} \& {Morisset}(2019)}]{2019RMxAA..55..377A}
{Alarie}, A., \& {Morisset}, C. 2019, \rmxaa, 55, 377, \dodoi{10.22201/ia.01851101p.2019.55.02.21}

\bibitem[{{Audibert} {et~al.}(2023){Audibert}, {Ramos Almeida}, {Garc{\'\i}a-Burillo}, {Combes}, {Bischetti}, {Meenakshi}, {Mukherjee}, {Bicknell}, \& {Wagner}}]{Audibert2023}
{Audibert}, A., {Ramos Almeida}, C., {Garc{\'\i}a-Burillo}, S., {et~al.} 2023, \aap, 671, L12, \dodoi{10.1051/0004-6361/202345964}

\bibitem[{{Bacon} {et~al.}(2010){Bacon}, {Accardo}, {Adjali}, {Anwand}, {Bauer}, {Biswas}, {Blaizot}, {Boudon}, {Brau-Nogue}, {Brinchmann}, {Caillier}, {Capoani}, {Carollo}, {Contini}, {Couderc}, {Daguis{\'e}}, {Deiries}, {Delabre}, {Dreizler}, {Dubois}, {Dupieux}, {Dupuy}, {Emsellem}, {Fechner}, {Fleischmann}, {Fran{\c{c}}ois}, {Gallou}, {Gharsa}, {Glindemann}, {Gojak}, {Guiderdoni}, {Hansali}, {Hahn}, {Jarno}, {Kelz}, {Koehler}, {Kosmalski}, {Laurent}, {Le Floch}, {Lilly}, {Lizon}, {Loupias}, {Manescau}, {Monstein}, {Nicklas}, {Olaya}, {Pares}, {Pasquini}, {P{\'e}contal-Rousset}, {Pell{\'o}}, {Petit}, {Popow}, {Reiss}, {Remillieux}, {Renault}, {Roth}, {Rupprecht}, {Serre}, {Schaye}, {Soucail}, {Steinmetz}, {Streicher}, {Stuik}, {Valentin}, {Vernet}, {Weilbacher}, {Wisotzki}, \& {Yerle}}]{Bacon2010}
{Bacon}, R., {Accardo}, M., {Adjali}, L., {et~al.} 2010, in Society of Photo-Optical Instrumentation Engineers (SPIE) Conference Series, Vol. 7735, Ground-based and Airborne Instrumentation for Astronomy III, ed. I.~S. {McLean}, S.~K. {Ramsay}, \& H.~{Takami}, 773508, \dodoi{10.1117/12.856027}

\bibitem[{{Baldwin} {et~al.}(1981){Baldwin}, {Phillips}, \& {Terlevich}}]{1981PASP...93....5B}
{Baldwin}, J.~A., {Phillips}, M.~M., \& {Terlevich}, R. 1981, \pasp, 93, 5, \dodoi{10.1086/130766}

\bibitem[{{Balmaverde} {et~al.}(2022){Balmaverde}, {Capetti}, {Baldi}, {Baum}, {Chiaberge}, {Gilli}, {Jimenez-Gallardo}, {Marconi}, {Massaro}, {Meyer}, {O'Dea}, {Speranza}, {Torresi}, \& {Venturi}}]{Balmaverde2022}
{Balmaverde}, B., {Capetti}, A., {Baldi}, R.~D., {et~al.} 2022, \aap, 662, A23, \dodoi{10.1051/0004-6361/202142823}

\bibitem[{{Baron} \& {Netzer}(2019)}]{2019MNRAS.486.4290B}
{Baron}, D., \& {Netzer}, H. 2019, \mnras, 486, 4290, \dodoi{10.1093/mnras/stz1070}

\bibitem[{{Baum} \& {Heckman}(1989)}]{Baum1989}
{Baum}, S.~A., \& {Heckman}, T. 1989, \apj, 336, 681, \dodoi{10.1086/167043}

\bibitem[{{Baum} {et~al.}(1988){Baum}, {Heckman}, {Bridle}, {van Breugel}, \& {Miley}}]{Baum1988}
{Baum}, S.~A., {Heckman}, T.~M., {Bridle}, A., {van Breugel}, W. J.~M., \& {Miley}, G.~K. 1988, \apjs, 68, 643, \dodoi{10.1086/191301}

\bibitem[{{Baum} {et~al.}(1993){Baum}, {O'Dea}, {Dallacassa}, {de Bruyn}, \& {Pedlar}}]{Baum1993}
{Baum}, S.~A., {O'Dea}, C.~P., {Dallacassa}, D., {de Bruyn}, A.~G., \& {Pedlar}, A. 1993, \apj, 419, 553, \dodoi{10.1086/173508}

\bibitem[{{Bicknell} {et~al.}(1998){Bicknell}, {Dopita}, {Tsvetanov}, \& {Sutherland}}]{1998ApJ...495..680B}
{Bicknell}, G.~V., {Dopita}, M.~A., {Tsvetanov}, Z.~I., \& {Sutherland}, R.~S. 1998, \apj, 495, 680, \dodoi{10.1086/305336}

\bibitem[{{Binney} \& {Tabor}(1995)}]{1995MNRAS.276..663B}
{Binney}, J., \& {Tabor}, G. 1995, \mnras, 276, 663, \dodoi{10.1093/mnras/276.2.663}

\bibitem[{{B{\^\i}rzan} {et~al.}(2008){B{\^\i}rzan}, {McNamara}, {Nulsen}, {Carilli}, \& {Wise}}]{2008ApJ...686..859B}
{B{\^\i}rzan}, L., {McNamara}, B.~R., {Nulsen}, P.~E.~J., {Carilli}, C.~L., \& {Wise}, M.~W. 2008, \apj, 686, 859, \dodoi{10.1086/591416}

\bibitem[{{B{\^\i}rzan} {et~al.}(2004){B{\^\i}rzan}, {Rafferty}, {McNamara}, {Wise}, \& {Nulsen}}]{2004ApJ...607..800B}
{B{\^\i}rzan}, L., {Rafferty}, D.~A., {McNamara}, B.~R., {Wise}, M.~W., \& {Nulsen}, P.~E.~J. 2004, \apj, 607, 800, \dodoi{10.1086/383519}

\bibitem[{{Bittner} {et~al.}(2019){Bittner}, {Falc{\'o}n-Barroso}, {Nedelchev}, {Dorta}, {Gadotti}, {Sarzi}, {Molaeinezhad}, {Iodice}, {Rosado-Belza}, {de Lorenzo-C{\'a}ceres}, {Fragkoudi}, {Gal{\'a}n-de Anta}, {Husemann}, {M{\'e}ndez-Abreu}, {Neumann}, {Pinna}, {Querejeta}, {S{\'a}nchez-Bl{\'a}zquez}, \& {Seidel}}]{2019ascl.soft07025B}
{Bittner}, A., {Falc{\'o}n-Barroso}, J., {Nedelchev}, B., {et~al.} 2019, {GIST: Galaxy IFU Spectroscopy Tool}, Astrophysics Source Code Library, record ascl:1907.025.
\newblock \doeprint{1907.025}

\bibitem[{{Bower} {et~al.}(2006){Bower}, {Benson}, {Malbon}, {Helly}, {Frenk}, {Baugh}, {Cole}, \& {Lacey}}]{Bower2006}
{Bower}, R.~G., {Benson}, A.~J., {Malbon}, R., {et~al.} 2006, \mnras, 370, 645, \dodoi{10.1111/j.1365-2966.2006.10519.x}

\bibitem[{{Bridle} {et~al.}(1994){Bridle}, {Hough}, {Lonsdale}, {Burns}, \& {Laing}}]{Bridle94}
{Bridle}, A.~H., {Hough}, D.~H., {Lonsdale}, C.~J., {Burns}, J.~O., \& {Laing}, R.~A. 1994, \aj, 108, 766, \dodoi{10.1086/117112}

\bibitem[{{Cappellari}(2017)}]{2017MNRAS.466..798C}
{Cappellari}, M. 2017, \mnras, 466, 798, \dodoi{10.1093/mnras/stw3020}

\bibitem[{{Cappellari} \& {Emsellem}(2004)}]{2004PASP..116..138C}
{Cappellari}, M., \& {Emsellem}, E. 2004, \pasp, 116, 138, \dodoi{10.1086/381875}

\bibitem[{{Carniani} {et~al.}(2015){Carniani}, {Marconi}, {Maiolino}, {Balmaverde}, {Brusa}, {Cano-D{\'\i}az}, {Cicone}, {Comastri}, {Cresci}, {Fiore}, {Feruglio}, {La Franca}, {Mainieri}, {Mannucci}, {Nagao}, {Netzer}, {Piconcelli}, {Risaliti}, {Schneider}, \& {Shemmer}}]{2015A&A...580A.102C}
{Carniani}, S., {Marconi}, A., {Maiolino}, R., {et~al.} 2015, \aap, 580, A102, \dodoi{10.1051/0004-6361/201526557}

\bibitem[{{Cavagnolo} {et~al.}(2010){Cavagnolo}, {McNamara}, {Nulsen}, {Carilli}, {Jones}, \& {B{\^\i}rzan}}]{2010ApJ...720.1066C}
{Cavagnolo}, K.~W., {McNamara}, B.~R., {Nulsen}, P.~E.~J., {et~al.} 2010, \apj, 720, 1066, \dodoi{10.1088/0004-637X/720/2/1066}

\bibitem[{{Cazzoli} {et~al.}(2022){Cazzoli}, {Hermosa Mu{\~n}oz}, {M{\'a}rquez}, {Masegosa}, {Castillo-Morales}, {Gil de Paz}, {Hern{\'a}ndez-Garc{\'\i}a}, {La Franca}, \& {Ramos Almeida}}]{Cazzoli2022}
{Cazzoli}, S., {Hermosa Mu{\~n}oz}, L., {M{\'a}rquez}, I., {et~al.} 2022, \aap, 664, A135, \dodoi{10.1051/0004-6361/202142695}

\bibitem[{{Chan} \& {Krolik}(2016)}]{Chan2016}
{Chan}, C.-H., \& {Krolik}, J.~H. 2016, \apj, 825, 67, \dodoi{10.3847/0004-637X/825/1/67}

\bibitem[{{Choi} {et~al.}(2018){Choi}, {Somerville}, {Ostriker}, {Naab}, \& {Hirschmann}}]{Choi2018}
{Choi}, E., {Somerville}, R.~S., {Ostriker}, J.~P., {Naab}, T., \& {Hirschmann}, M. 2018, \apj, 866, 91, \dodoi{10.3847/1538-4357/aae076}

\bibitem[{{Chojnowski} \& {Keel}(2011)}]{2011AAS...21714207C}
{Chojnowski}, D., \& {Keel}, W.~C. 2011, in American Astronomical Society Meeting Abstracts, Vol. 217, American Astronomical Society Meeting Abstracts \#217, 142.07

\bibitem[{{Cid Fernandes} {et~al.}(2011){Cid Fernandes}, {Stasi{\'n}ska}, {Mateus}, \& {Vale Asari}}]{CidF2011}
{Cid Fernandes}, R., {Stasi{\'n}ska}, G., {Mateus}, A., \& {Vale Asari}, N. 2011, \mnras, 413, 1687, \dodoi{10.1111/j.1365-2966.2011.18244.x}

\bibitem[{{Colbert} {et~al.}(1996){Colbert}, {Baum}, {Gallimore}, {O'Dea}, \& {Christensen}}]{Colbert1996}
{Colbert}, E. J.~M., {Baum}, S.~A., {Gallimore}, J.~F., {O'Dea}, C.~P., \& {Christensen}, J.~A. 1996, \apj, 467, 551, \dodoi{10.1086/177633}

\bibitem[{{Condon}(1992)}]{1992ARA&A..30..575C}
{Condon}, J.~J. 1992, \araa, 30, 575, \dodoi{10.1146/annurev.aa.30.090192.003043}

\bibitem[{{Condon} \& {Ransom}(2016)}]{2016era..book.....C}
{Condon}, J.~J., \& {Ransom}, S.~M. 2016, {Essential Radio Astronomy}

\bibitem[{{Couto} {et~al.}(2013){Couto}, {Storchi-Bergmann}, {Axon}, {Robinson}, {Kharb}, \& {Riffel}}]{Couto2013}
{Couto}, G.~S., {Storchi-Bergmann}, T., {Axon}, D.~J., {et~al.} 2013, \mnras, 435, 2982, \dodoi{10.1093/mnras/stt1491}

\bibitem[{{Dannen} {et~al.}(2020){Dannen}, {Proga}, {Waters}, \& {Dyda}}]{2020ApJ...893L..34D}
{Dannen}, R.~C., {Proga}, D., {Waters}, T., \& {Dyda}, S. 2020, \apjl, 893, L34, \dodoi{10.3847/2041-8213/ab87a5}

\bibitem[{{Davidson} \& {Netzer}(1979)}]{1979RvMP...51..715D}
{Davidson}, K., \& {Netzer}, H. 1979, Reviews of Modern Physics, 51, 715, \dodoi{10.1103/RevModPhys.51.715}

\bibitem[{{de Bruyn} \& {Wilson}(1978)}]{deBruyn1978}
{de Bruyn}, A.~G., \& {Wilson}, A.~S. 1978, \aap, 64, 433

\bibitem[{{Di Matteo} {et~al.}(2005){Di Matteo}, {Springel}, \& {Hernquist}}]{2005Natur.433..604D}
{Di Matteo}, T., {Springel}, V., \& {Hernquist}, L. 2005, \nat, 433, 604, \dodoi{10.1038/nature03335}

\bibitem[{{Dimitrijevi{\'c}} {et~al.}(2007){Dimitrijevi{\'c}}, {Popovi{\'c}}, {Kova{\v{c}}evi{\'c}}, {Da{\v{c}}i{\'c}}, \& {Ili{\'c}}}]{2007MNRAS.374.1181D}
{Dimitrijevi{\'c}}, M.~S., {Popovi{\'c}}, L.~{\v{C}}., {Kova{\v{c}}evi{\'c}}, J., {Da{\v{c}}i{\'c}}, M., \& {Ili{\'c}}, D. 2007, \mnras, 374, 1181, \dodoi{10.1111/j.1365-2966.2006.11238.x}

\bibitem[{{Dopita} \& {Sutherland}(1995)}]{1995ApJ...455..468D}
{Dopita}, M.~A., \& {Sutherland}, R.~S. 1995, \apj, 455, 468, \dodoi{10.1086/176596}

\bibitem[{{Duggal} {et~al.}(2021){Duggal}, {O'Dea}, {Baum}, {Labiano}, {Morganti}, {Tadhunter}, {Worrall}, {Tremblay}, {Dicken}, \& {Capetti}}]{Chetna2021}
{Duggal}, C., {O'Dea}, C., {Baum}, S., {et~al.} 2021, Astronomische Nachrichten, 342, 1087, \dodoi{10.1002/asna.20210054}

\bibitem[{{Duggal} {et~al.}(2023){Duggal}, {O'Dea}, {Baum}, {Labiano}, {Morganti}, {Tadhunter}, {Worrall}, {Tremblay}, {Dicken}, \& {Capetti}}]{Chetna2023}
{Duggal}, C., {O'Dea}, C., {Baum}, S., {et~al.} 2023, in American Astronomical Society Meeting Abstracts, Vol. 241, American Astronomical Society Meeting Abstracts, 336.01

\bibitem[{{Emmering} {et~al.}(1992){Emmering}, {Blandford}, \& {Shlosman}}]{Emmering1992}
{Emmering}, R.~T., {Blandford}, R.~D., \& {Shlosman}, I. 1992, \apj, 385, 460, \dodoi{10.1086/170955}

\bibitem[{{Esparza-Arredondo} {et~al.}(2025){Esparza-Arredondo}, {Ramos Almeida}, {Audibert}, {Pereira-Santaella}, {Garc{\'\i}a-Bernete}, {Garc{\'\i}a-Burillo}, {Shimizu}, {Davies}, {Hermosa Mu{\~n}oz}, {Alonso-Herrero}, {Combes}, {Speranza}, {Zhang}, {Campbell}, {Bellocchi}, {Bunker}, {D{\'\i}az-Santos}, {Garc{\'\i}a-Lorenzo}, {Gonz{\'a}lez-Mart{\'\i}n}, {Hicks}, {Labiano}, {Levenson}, {Ricci}, {Rosario}, {Hoenig}, {Packham}, {Stalevski}, {Fuller}, {Izumi}, {L{\'o}pez-Rodr{\'\i}guez}, {Rigopoulou}, {Rouan}, \& {Ward}}]{Esparza-Arredondo2025}
{Esparza-Arredondo}, D., {Ramos Almeida}, C., {Audibert}, A., {et~al.} 2025, \aap, 693, A174, \dodoi{10.1051/0004-6361/202452488}

\bibitem[{{Esposito} {et~al.}(2024){Esposito}, {Alonso-Herrero}, {Garc{\'\i}a-Burillo}, {Casasola}, {Combes}, {Dallacasa}, {Davies}, {Garc{\'\i}a-Bernete}, {Garc{\'\i}a-Lorenzo}, {Hermosa Mu{\~n}oz}, {de Arriba}, {Pereira-Santaella}, {Pozzi}, {Ramos Almeida}, {Shimizu}, {Vallini}, {Bellocchi}, {Gonz{\'a}lez-Mart{\'\i}n}, {Hicks}, {H{\"o}nig}, {Labiano}, {Levenson}, {Ricci}, \& {Rosario}}]{Esposito2024}
{Esposito}, F., {Alonso-Herrero}, A., {Garc{\'\i}a-Burillo}, S., {et~al.} 2024, \aap, 686, A46, \dodoi{10.1051/0004-6361/202449245}

\bibitem[{{Fabian}(1999)}]{1999MNRAS.308L..39F}
{Fabian}, A.~C. 1999, \mnras, 308, L39, \dodoi{10.1046/j.1365-8711.1999.03017.x}

\bibitem[{{Fabian}(2012)}]{Fabian2012}
---. 2012, \araa, 50, 455, \dodoi{10.1146/annurev-astro-081811-125521}

\bibitem[{{Falc{\'o}n-Barroso} {et~al.}(2006){Falc{\'o}n-Barroso}, {Bacon}, {Bureau}, {Cappellari}, {Davies}, {de Zeeuw}, {Emsellem}, {Fathi}, {Krajnovi{\'c}}, {Kuntschner}, {McDermid}, {Peletier}, \& {Sarzi}}]{2006MNRAS.369..529F}
{Falc{\'o}n-Barroso}, J., {Bacon}, R., {Bureau}, M., {et~al.} 2006, \mnras, 369, 529, \dodoi{10.1111/j.1365-2966.2006.10261.x}

\bibitem[{{Faucher-Gigu{\`e}re} \& {Quataert}(2012)}]{Faucher-Giguere2012}
{Faucher-Gigu{\`e}re}, C.-A., \& {Quataert}, E. 2012, \mnras, 425, 605, \dodoi{10.1111/j.1365-2966.2012.21512.x}

\bibitem[{{Ferland} \& {Netzer}(1983)}]{1983ApJ...264..105F}
{Ferland}, G.~J., \& {Netzer}, H. 1983, \apj, 264, 105, \dodoi{10.1086/160577}

\bibitem[{{Ferrarese} \& {Merritt}(2000)}]{Ferrarese2000}
{Ferrarese}, L., \& {Merritt}, D. 2000, \apjl, 539, L9, \dodoi{10.1086/312838}

\bibitem[{{Finlez} {et~al.}(2022){Finlez}, {Treister}, {Bauer}, {Keel}, {Koss}, {Nagar}, {Sartori}, {Maksym}, {Venturi}, {Tub{\'\i}n}, \& {Harvey}}]{2022ApJ...936...88F}
{Finlez}, C., {Treister}, E., {Bauer}, F., {et~al.} 2022, \apj, 936, 88, \dodoi{10.3847/1538-4357/ac854e}

\bibitem[{{Fiore} {et~al.}(2017){Fiore}, {Feruglio}, {Shankar}, {Bischetti}, {Bongiorno}, {Brusa}, {Carniani}, {Cicone}, {Duras}, {Lamastra}, {Mainieri}, {Marconi}, {Menci}, {Maiolino}, {Piconcelli}, {Vietri}, \& {Zappacosta}}]{2017A&A...601A.143F}
{Fiore}, F., {Feruglio}, C., {Shankar}, F., {et~al.} 2017, \aap, 601, A143, \dodoi{10.1051/0004-6361/201629478}

\bibitem[{{Gaibler} {et~al.}(2012){Gaibler}, {Khochfar}, {Krause}, \& {Silk}}]{Gaibler2012}
{Gaibler}, V., {Khochfar}, S., {Krause}, M., \& {Silk}, J. 2012, \mnras, 425, 438, \dodoi{10.1111/j.1365-2966.2012.21479.x}

\bibitem[{{Gallimore} {et~al.}(2006){Gallimore}, {Axon}, {O'Dea}, {Baum}, \& {Pedlar}}]{2006AJ....132..546G}
{Gallimore}, J.~F., {Axon}, D.~J., {O'Dea}, C.~P., {Baum}, S.~A., \& {Pedlar}, A. 2006, \aj, 132, 546, \dodoi{10.1086/504593}

\bibitem[{{Garrington} {et~al.}(1988){Garrington}, {Leahy}, {Conway}, \& {Laing}}]{1988Natur.331..147G}
{Garrington}, S.~T., {Leahy}, J.~P., {Conway}, R.~G., \& {Laing}, R.~A. 1988, \nat, 331, 147, \dodoi{10.1038/331147a0}

\bibitem[{{Girdhar}(2022)}]{2022hypa.confE...2G}
{Girdhar}, A. 2022, in Hypatia Colloquium 2022, 2, \dodoi{10.5281/zenodo.7104492}

\bibitem[{{G{\"u}ltekin} {et~al.}(2009){G{\"u}ltekin}, {Richstone}, {Gebhardt}, {Lauer}, {Tremaine}, {Aller}, {Bender}, {Dressler}, {Faber}, {Filippenko}, {Green}, {Ho}, {Kormendy}, {Magorrian}, {Pinkney}, \& {Siopis}}]{Gutekin2009}
{G{\"u}ltekin}, K., {Richstone}, D.~O., {Gebhardt}, K., {et~al.} 2009, \apj, 698, 198, \dodoi{10.1088/0004-637X/698/1/198}

\bibitem[{{Hardcastle} \& {Croston}(2020)}]{Hardcastle2020}
{Hardcastle}, M.~J., \& {Croston}, J.~H. 2020, \nar, 88, 101539, \dodoi{10.1016/j.newar.2020.101539}

\bibitem[{{Harrison} {et~al.}(2014){Harrison}, {Alexander}, {Mullaney}, \& {Swinbank}}]{Harrison2014}
{Harrison}, C.~M., {Alexander}, D.~M., {Mullaney}, J.~R., \& {Swinbank}, A.~M. 2014, \mnras, 441, 3306, \dodoi{10.1093/mnras/stu515}

\bibitem[{{Harrison} {et~al.}(2018){Harrison}, {Costa}, {Tadhunter}, {Fl{\"u}tsch}, {Kakkad}, {Perna}, \& {Vietri}}]{2018NatAs...2..198H}
{Harrison}, C.~M., {Costa}, T., {Tadhunter}, C.~N., {et~al.} 2018, Nature Astronomy, 2, 198, \dodoi{10.1038/s41550-018-0403-6}

\bibitem[{{Harrison} {et~al.}(2015){Harrison}, {Thomson}, {Alexander}, {Bauer}, {Edge}, {Hogan}, {Mullaney}, \& {Swinbank}}]{2015ApJ...800...45H}
{Harrison}, C.~M., {Thomson}, A.~P., {Alexander}, D.~M., {et~al.} 2015, \apj, 800, 45, \dodoi{10.1088/0004-637X/800/1/45}

\bibitem[{{Harvey} {et~al.}(2022){Harvey}, {Maksym}, {Keel}, {Koss}, {Bennert}, {Chojnowski}, {Treister}, {Finlez}, {Lintott}, {Moiseev}, {Simmons}, {Sartori}, \& {Urry}}]{2022arXiv220805915H}
{Harvey}, T., {Maksym}, W.~P., {Keel}, W., {et~al.} 2022, arXiv e-prints, arXiv:2208.05915.
\newblock \doarXiv{2208.05915}

\bibitem[{{Helou} {et~al.}(1988){Helou}, {Khan}, {Malek}, \& {Boehmer}}]{Helou1988}
{Helou}, G., {Khan}, I.~R., {Malek}, L., \& {Boehmer}, L. 1988, \apjs, 68, 151, \dodoi{10.1086/191285}

\bibitem[{{Holt} {et~al.}(2008){Holt}, {Tadhunter}, \& {Morganti}}]{2008MNRAS.387..639H}
{Holt}, J., {Tadhunter}, C.~N., \& {Morganti}, R. 2008, \mnras, 387, 639, \dodoi{10.1111/j.1365-2966.2008.13089.x}

\bibitem[{{Ishibashi} \& {Courvoisier}(2011)}]{Ishibashi2011}
{Ishibashi}, W., \& {Courvoisier}, T.~J.~L. 2011, \aap, 525, A118, \dodoi{10.1051/0004-6361/201014987}

\bibitem[{{Ishibashi} {et~al.}(2013){Ishibashi}, {Fabian}, \& {Canning}}]{2013MNRAS.431.2350I}
{Ishibashi}, W., {Fabian}, A.~C., \& {Canning}, R.~E.~A. 2013, \mnras, 431, 2350, \dodoi{10.1093/mnras/stt333}

\bibitem[{{Jarvis} {et~al.}(2019){Jarvis}, {Harrison}, {Thomson}, {Circosta}, {Mainieri}, {Alexander}, {Edge}, {Lansbury}, {Molyneux}, \& {Mullaney}}]{2019MNRAS.485.2710J}
{Jarvis}, M.~E., {Harrison}, C.~M., {Thomson}, A.~P., {et~al.} 2019, \mnras, 485, 2710, \dodoi{10.1093/mnras/stz556}

\bibitem[{{Jarvis} {et~al.}(2021){Jarvis}, {Harrison}, {Mainieri}, {Alexander}, {Arrigoni Battaia}, {Calistro Rivera}, {Circosta}, {Costa}, {De Breuck}, {Edge}, {Girdhar}, {Kakkad}, {Kharb}, {Lansbury}, {Molyneux}, {Mukherjee}, {Mullaney}, {Farina}, {Silpa}, {Thomson}, \& {Ward}}]{2021MNRAS.503.1780J}
{Jarvis}, M.~E., {Harrison}, C.~M., {Mainieri}, V., {et~al.} 2021, \mnras, 503, 1780, \dodoi{10.1093/mnras/stab549}

\bibitem[{{Jiang} {et~al.}(2010){Jiang}, {Ciotti}, {Ostriker}, \& {Spitkovsky}}]{Jiang2010}
{Jiang}, Y.-F., {Ciotti}, L., {Ostriker}, J.~P., \& {Spitkovsky}, A. 2010, \apj, 711, 125, \dodoi{10.1088/0004-637X/711/1/125}

\bibitem[{{J{\'o}zsa} {et~al.}(2009){J{\'o}zsa}, {Garrett}, {Oosterloo}, {Rampadarath}, {Paragi}, {van Arkel}, {Lintott}, {Keel}, {Schawinski}, \& {Edmondson}}]{2009A&A...500L..33J}
{J{\'o}zsa}, G.~I.~G., {Garrett}, M.~A., {Oosterloo}, T.~A., {et~al.} 2009, \aap, 500, L33, \dodoi{10.1051/0004-6361/200912402}

\bibitem[{{Kaasinen} {et~al.}(2017){Kaasinen}, {Bian}, {Groves}, {Kewley}, \& {Gupta}}]{2017MNRAS.465.3220K}
{Kaasinen}, M., {Bian}, F., {Groves}, B., {Kewley}, L.~J., \& {Gupta}, A. 2017, \mnras, 465, 3220, \dodoi{10.1093/mnras/stw2827}

\bibitem[{{Kakkad} {et~al.}(2018){Kakkad}, {Groves}, {Dopita}, {Thomas}, {Davies}, {Mainieri}, {Kharb}, {Scharw{\"a}chter}, {Hampton}, \& {Ho}}]{2018A&A...618A...6K}
{Kakkad}, D., {Groves}, B., {Dopita}, M., {et~al.} 2018, \aap, 618, A6, \dodoi{10.1051/0004-6361/201832790}

\bibitem[{{Kakkad} {et~al.}(2022){Kakkad}, {Sani}, {Rojas}, {Mallmann}, {Veilleux}, {Bauer}, {Ricci}, {Mushotzky}, {Koss}, {Ricci}, {Treister}, {Privon}, {Nguyen}, {B{\"a}r}, {Harrison}, {Oh}, {Powell}, {Riffel}, {Stern}, {Trakhtenbrot}, \& {Urry}}]{2022MNRAS.511.2105K}
{Kakkad}, D., {Sani}, E., {Rojas}, A.~F., {et~al.} 2022, \mnras, 511, 2105, \dodoi{10.1093/mnras/stac103}

\bibitem[{{Kauffmann} {et~al.}(2003){Kauffmann}, {Heckman}, {Tremonti}, {Brinchmann}, {Charlot}, {White}, {Ridgway}, {Brinkmann}, {Fukugita}, {Hall}, {Ivezi{\'c}}, {Richards}, \& {Schneider}}]{2003MNRAS.346.1055K}
{Kauffmann}, G., {Heckman}, T.~M., {Tremonti}, C., {et~al.} 2003, \mnras, 346, 1055, \dodoi{10.1111/j.1365-2966.2003.07154.x}

\bibitem[{{Keel} {et~al.}(2011){Keel}, {Lintott}, {Schawinski}, {Bennert}, {Thomas}, {Manning}, {Chojnowski}, {van Arkel}, {Lynn}, \& {Galaxy Zoo Team}}]{2011AAS...21714208K}
{Keel}, W.~C., {Lintott}, C., {Schawinski}, K., {et~al.} 2011, in American Astronomical Society Meeting Abstracts, Vol. 217, American Astronomical Society Meeting Abstracts \#217, 142.08

\bibitem[{{Keel} {et~al.}(2012{\natexlab{a}}){Keel}, {Lintott}, {Schawinski}, {Bennert}, {Thomas}, {Manning}, {Chojnowski}, {van Arkel}, \& {Lynn}}]{2012AJ....144...66K}
{Keel}, W.~C., {Lintott}, C.~J., {Schawinski}, K., {et~al.} 2012{\natexlab{a}}, \aj, 144, 66, \dodoi{10.1088/0004-6256/144/2/66}

\bibitem[{{Keel} {et~al.}(2012{\natexlab{b}}){Keel}, {Chojnowski}, {Bennert}, {Schawinski}, {Lintott}, {Lynn}, {Pancoast}, {Harris}, {Nierenberg}, {Sonnenfeld}, \& {Proctor}}]{2012MNRAS.420..878K}
{Keel}, W.~C., {Chojnowski}, S.~D., {Bennert}, V.~N., {et~al.} 2012{\natexlab{b}}, \mnras, 420, 878, \dodoi{10.1111/j.1365-2966.2011.20101.x}

\bibitem[{{Keel} {et~al.}(2015){Keel}, {Maksym}, {Bennert}, {Lintott}, {Chojnowski}, {Moiseev}, {Smirnova}, {Schawinski}, {Urry}, {Evans}, {Pancoast}, {Scott}, {Showley}, \& {Flatland}}]{2015AJ....149..155K}
{Keel}, W.~C., {Maksym}, W.~P., {Bennert}, V.~N., {et~al.} 2015, \aj, 149, 155, \dodoi{10.1088/0004-6256/149/5/155}

\bibitem[{{Keel} {et~al.}(2017){Keel}, {Lintott}, {Maksym}, {Bennert}, {Chojnowski}, {Moiseev}, {Smirnova}, {Schawinski}, {Sartori}, {Urry}, {Pancoast}, {Schirmer}, {Scott}, {Showley}, \& {Flatland}}]{2017ApJ...835..256K}
{Keel}, W.~C., {Lintott}, C.~J., {Maksym}, W.~P., {et~al.} 2017, \apj, 835, 256, \dodoi{10.3847/1538-4357/835/2/256}

\bibitem[{{Kellermann} {et~al.}(1989){Kellermann}, {Sramek}, {Schmidt}, {Shaffer}, \& {Green}}]{1989AJ.....98.1195K}
{Kellermann}, K.~I., {Sramek}, R., {Schmidt}, M., {Shaffer}, D.~B., \& {Green}, R. 1989, \aj, 98, 1195, \dodoi{10.1086/115207}

\bibitem[{{Kennicutt}(1998)}]{1998ARA&A..36..189K}
{Kennicutt}, Robert~C., J. 1998, \araa, 36, 189, \dodoi{10.1146/annurev.astro.36.1.189}

\bibitem[{{Kewley} {et~al.}(2001){Kewley}, {Dopita}, {Sutherland}, {Heisler}, \& {Trevena}}]{2001ApJ...556..121K}
{Kewley}, L.~J., {Dopita}, M.~A., {Sutherland}, R.~S., {Heisler}, C.~A., \& {Trevena}, J. 2001, \apj, 556, 121, \dodoi{10.1086/321545}

\bibitem[{{Kharb} {et~al.}(2006){Kharb}, {O'Dea}, {Baum}, {Colbert}, \& {Xu}}]{Kharb2006}
{Kharb}, P., {O'Dea}, C.~P., {Baum}, S.~A., {Colbert}, E.~J.~M., \& {Xu}, C. 2006, \apj, 652, 177, \dodoi{10.1086/507945}

\bibitem[{{Kharb} {et~al.}(2016){Kharb}, {Srivastava}, {Singh}, {Gallimore}, {Ishwara-Chandra}, \& {Ananda}}]{2016MNRAS.459.1310K}
{Kharb}, P., {Srivastava}, S., {Singh}, V., {et~al.} 2016, \mnras, 459, 1310, \dodoi{10.1093/mnras/stw699}

\bibitem[{{King} \& {Pounds}(2015)}]{King2015}
{King}, A., \& {Pounds}, K. 2015, \araa, 53, 115, \dodoi{10.1146/annurev-astro-082214-122316}

\bibitem[{{Kozlova} {et~al.}(2020){Kozlova}, {Moiseev}, \& {Smirnova}}]{2020CoSka..50..309K}
{Kozlova}, D.~V., {Moiseev}, A.~V., \& {Smirnova}, A.~A. 2020, Contributions of the Astronomical Observatory Skalnate Pleso, 50, 309, \dodoi{10.31577/caosp.2020.50.1.309}

\bibitem[{{Krause}(2023)}]{Krause2023}
{Krause}, M. G.~H. 2023, Galaxies, 11, 29, \dodoi{10.3390/galaxies11010029}

\bibitem[{{Krolik} \& {Vrtilek}(1984)}]{Krolik1984}
{Krolik}, J.~H., \& {Vrtilek}, J.~M. 1984, \apj, 279, 521, \dodoi{10.1086/161916}

\bibitem[{{Laing}(1988)}]{1988Natur.331..149L}
{Laing}, R.~A. 1988, \nat, 331, 149, \dodoi{10.1038/331149a0}

\bibitem[{{Lanz} {et~al.}(2015){Lanz}, {Ogle}, {Evans}, {Appleton}, {Guillard}, \& {Emonts}}]{Lanz2015}
{Lanz}, L., {Ogle}, P.~M., {Evans}, D., {et~al.} 2015, \apj, 801, 17, \dodoi{10.1088/0004-637X/801/1/17}

\bibitem[{{Lena} {et~al.}(2015){Lena}, {Robinson}, {Storchi-Bergman}, {Schnorr-M{\"u}ller}, {Seelig}, {Riffel}, {Nagar}, {Couto}, \& {Shadler}}]{Lena2015}
{Lena}, D., {Robinson}, A., {Storchi-Bergman}, T., {et~al.} 2015, \apj, 806, 84, \dodoi{10.1088/0004-637X/806/1/84}

\bibitem[{{Lintott} {et~al.}(2009){Lintott}, {Schawinski}, {Keel}, {van Arkel}, {Bennert}, {Edmondson}, {Thomas}, {Smith}, {Herbert}, {Jarvis}, {Virani}, {Andreescu}, {Bamford}, {Land}, {Murray}, {Nichol}, {Raddick}, {Slosar}, {Szalay}, \& {Vandenberg}}]{2009MNRAS.399..129L}
{Lintott}, C.~J., {Schawinski}, K., {Keel}, W., {et~al.} 2009, \mnras, 399, 129, \dodoi{10.1111/j.1365-2966.2009.15299.x}

\bibitem[{{Mahony} {et~al.}(2016){Mahony}, {Oonk}, {Morganti}, {Tadhunter}, {Bessiere}, {Short}, {Emonts}, \& {Oosterloo}}]{2016MNRAS.455.2453M}
{Mahony}, E.~K., {Oonk}, J.~B.~R., {Morganti}, R., {et~al.} 2016, \mnras, 455, 2453, \dodoi{10.1093/mnras/stv2456}

\bibitem[{{Marecki} {et~al.}(2016){Marecki}, {Jamrozy}, \& {Machalski}}]{Marecki2016}
{Marecki}, A., {Jamrozy}, M., \& {Machalski}, J. 2016, \mnras, 463, 338, \dodoi{10.1093/mnras/stw2006}

\bibitem[{{Marecki} {et~al.}(2021){Marecki}, {Jamrozy}, {Machalski}, \& {Pajdosz-{\'S}mierciak}}]{Marecki2021}
{Marecki}, A., {Jamrozy}, M., {Machalski}, J., \& {Pajdosz-{\'S}mierciak}, U. 2021, \mnras, 501, 853, \dodoi{10.1093/mnras/staa3632}

\bibitem[{{Marecki} {et~al.}(2023){Marecki}, {Sebastian}, \& {Ishwara-Chandra}}]{Marecki2023}
{Marecki}, A., {Sebastian}, B., \& {Ishwara-Chandra}, C.~H. 2023, \mnras, 518, L83, \dodoi{10.1093/mnrasl/slac105}

\bibitem[{{McCarthy} {et~al.}(2010){McCarthy}, {Schaye}, {Ponman}, {Bower}, {Booth}, {Dalla Vecchia}, {Crain}, {Springel}, {Theuns}, \& {Wiersma}}]{McCarthy2010}
{McCarthy}, I.~G., {Schaye}, J., {Ponman}, T.~J., {et~al.} 2010, \mnras, 406, 822, \dodoi{10.1111/j.1365-2966.2010.16750.x}

\bibitem[{{McCarthy} {et~al.}(1995){McCarthy}, {Spinrad}, \& {van Breugel}}]{McCarthy1995}
{McCarthy}, P.~J., {Spinrad}, H., \& {van Breugel}, W. 1995, \apjs, 99, 27, \dodoi{10.1086/192178}

\bibitem[{{Merloni} \& {Heinz}(2007)}]{2007MNRAS.381..589M}
{Merloni}, A., \& {Heinz}, S. 2007, \mnras, 381, 589, \dodoi{10.1111/j.1365-2966.2007.12253.x}

\bibitem[{{Merritt} \& {Ekers}(2002)}]{2002Sci...297.1310M}
{Merritt}, D., \& {Ekers}, R.~D. 2002, Science, 297, 1310, \dodoi{10.1126/science.1074688}

\bibitem[{{Mingozzi} {et~al.}(2019){Mingozzi}, {Cresci}, {Venturi}, {Marconi}, {Mannucci}, {Perna}, {Belfiore}, {Carniani}, {Balmaverde}, {Brusa}, {Cicone}, {Feruglio}, {Gallazzi}, {Mainieri}, {Maiolino}, {Nagao}, {Nardini}, {Sani}, {Tozzi}, \& {Zibetti}}]{2019A&A...622A.146M}
{Mingozzi}, M., {Cresci}, G., {Venturi}, G., {et~al.} 2019, \aap, 622, A146, \dodoi{10.1051/0004-6361/201834372}

\bibitem[{{Molina} {et~al.}(2018){Molina}, {Eracleous}, {Barth}, {Maoz}, {Runnoe}, {Ho}, {Shields}, \& {Walsh}}]{2018ApJ...864...90M}
{Molina}, M., {Eracleous}, M., {Barth}, A.~J., {et~al.} 2018, \apj, 864, 90, \dodoi{10.3847/1538-4357/aad5ed}

\bibitem[{{Morganti} {et~al.}(2015){Morganti}, {Oosterloo}, {Oonk}, {Frieswijk}, \& {Tadhunter}}]{Morganti2015}
{Morganti}, R., {Oosterloo}, T., {Oonk}, J.~B.~R., {Frieswijk}, W., \& {Tadhunter}, C. 2015, \aap, 580, A1, \dodoi{10.1051/0004-6361/201525860}

\bibitem[{{Morganti} {et~al.}(2016){Morganti}, {Veilleux}, {Oosterloo}, {Teng}, \& {Rupke}}]{Morganti2016}
{Morganti}, R., {Veilleux}, S., {Oosterloo}, T., {Teng}, S.~H., \& {Rupke}, D. 2016, \aap, 593, A30, \dodoi{10.1051/0004-6361/201628978}

\bibitem[{{Mukherjee} {et~al.}(2016){Mukherjee}, {Bicknell}, {Sutherland}, \& {Wagner}}]{2016MNRAS.461..967M}
{Mukherjee}, D., {Bicknell}, G.~V., {Sutherland}, R., \& {Wagner}, A. 2016, \mnras, 461, 967, \dodoi{10.1093/mnras/stw1368}

\bibitem[{{Mukherjee} {et~al.}(2018){Mukherjee}, {Wagner}, {Bicknell}, {Morganti}, {Oosterloo}, {Nesvadba}, \& {Sutherland}}]{2018MNRAS.476...80M}
{Mukherjee}, D., {Wagner}, A.~Y., {Bicknell}, G.~V., {et~al.} 2018, \mnras, 476, 80, \dodoi{10.1093/mnras/sty067}

\bibitem[{{Mullaney} {et~al.}(2013){Mullaney}, {Alexander}, {Fine}, {Goulding}, {Harrison}, \& {Hickox}}]{Mullaney2013}
{Mullaney}, J.~R., {Alexander}, D.~M., {Fine}, S., {et~al.} 2013, \mnras, 433, 622, \dodoi{10.1093/mnras/stt751}

\bibitem[{{Murray} {et~al.}(2005){Murray}, {Quataert}, \& {Thompson}}]{2005ApJ...618..569M}
{Murray}, N., {Quataert}, E., \& {Thompson}, T.~A. 2005, \apj, 618, 569, \dodoi{10.1086/426067}

\bibitem[{{Nandi} {et~al.}(2019){Nandi}, {Saikia}, {Roy}, {Dabhade}, {Wadadekar}, {Larsson}, {Baes}, {Chandola}, \& {Singh}}]{Nandi2019}
{Nandi}, S., {Saikia}, D.~J., {Roy}, R., {et~al.} 2019, \mnras, 486, 5158, \dodoi{10.1093/mnras/stz1184}

\bibitem[{{Napier} {et~al.}(1983){Napier}, {Thompson}, \& {Ekers}}]{Napier1983}
{Napier}, P.~J., {Thompson}, A.~R., \& {Ekers}, R.~D. 1983, IEEE Proceedings, 71, 1295

\bibitem[{{Nelson} \& {Whittle}(1996)}]{1996ApJ...465...96N}
{Nelson}, C.~H., \& {Whittle}, M. 1996, \apj, 465, 96, \dodoi{10.1086/177405}

\bibitem[{{Nesvadba} {et~al.}(2017{\natexlab{a}}){Nesvadba}, {De Breuck}, {Lehnert}, {Best}, \& {Collet}}]{Nesvadba2017}
{Nesvadba}, N.~P.~H., {De Breuck}, C., {Lehnert}, M.~D., {Best}, P.~N., \& {Collet}, C. 2017{\natexlab{a}}, \aap, 599, A123, \dodoi{10.1051/0004-6361/201528040}

\bibitem[{{Nesvadba} {et~al.}(2017{\natexlab{b}}){Nesvadba}, {Drouart}, {De Breuck}, {Best}, {Seymour}, \& {Vernet}}]{2017A&A...600A.121N}
{Nesvadba}, N.~P.~H., {Drouart}, G., {De Breuck}, C., {et~al.} 2017{\natexlab{b}}, \aap, 600, A121, \dodoi{10.1051/0004-6361/201629357}

\bibitem[{{Nims} {et~al.}(2015){Nims}, {Quataert}, \& {Faucher-Gigu{\`e}re}}]{Nims2015}
{Nims}, J., {Quataert}, E., \& {Faucher-Gigu{\`e}re}, C.-A. 2015, \mnras, 447, 3612, \dodoi{10.1093/mnras/stu2648}

\bibitem[{{O'Dea} {et~al.}(2009){O'Dea}, {Daly}, {Kharb}, {Freeman}, \& {Baum}}]{Odea2009}
{O'Dea}, C.~P., {Daly}, R.~A., {Kharb}, P., {Freeman}, K.~A., \& {Baum}, S.~A. 2009, \aap, 494, 471, \dodoi{10.1051/0004-6361:200809416}

\bibitem[{{O'Dea} \& {Owen}(1987)}]{Odea1987}
{O'Dea}, C.~P., \& {Owen}, F.~N. 1987, \apj, 316, 95, \dodoi{10.1086/165182}

\bibitem[{{Ogle} {et~al.}(2025){Ogle}, {Sebastian}, {Aravindan}, {McDonald}, {Canalizo}, {Ashby}, {Azadi}, {Antonucci}, {Barthel}, {Baum}, {Birkinshaw}, {Carilli}, {Chiaberge}, {Duggal}, {Gebhardt}, {Hyman}, {Kuraszkiewicz}, {Lopez-Rodriguez}, {Medling}, {Miley}, {Omoruyi}, {O'Dea}, {Perley}, {Perley}, {Perlman}, {Reynaldi}, {Singha}, {Sparks}, {Tremblay}, {Wilkes}, {Willner}, \& {Worrall}}]{Ogle2025}
{Ogle}, P.~M., {Sebastian}, B., {Aravindan}, A., {et~al.} 2025, arXiv e-prints, arXiv:2502.06603, \dodoi{10.48550/arXiv.2502.06603}

\bibitem[{{Osterbrock} \& {Ferland}(2006)}]{2006agna.book.....O}
{Osterbrock}, D.~E., \& {Ferland}, G.~J. 2006, {Astrophysics of gaseous nebulae and active galactic nuclei}

\bibitem[{{P{\'e}rez-Torres} \& {Alberdi}(2007)}]{Perez2007}
{P{\'e}rez-Torres}, M.~A., \& {Alberdi}, A. 2007, \mnras, 379, 275, \dodoi{10.1111/j.1365-2966.2007.11944.x}

\bibitem[{{Perley} {et~al.}(2011){Perley}, {Chandler}, {Butler}, \& {Wrobel}}]{Perley2011}
{Perley}, R.~A., {Chandler}, C.~J., {Butler}, B.~J., \& {Wrobel}, J.~M. 2011, \apjl, 739, L1, \dodoi{10.1088/2041-8205/739/1/L1}

\bibitem[{{Perna} {et~al.}(2020){Perna}, {Arribas}, {Catal{\'a}n-Torrecilla}, {Colina}, {Bellocchi}, {Fluetsch}, {Maiolino}, {Cazzoli}, {Hern{\'a}n Caballero}, {Pereira Santaella}, {Piqueras L{\'o}pez}, \& {Rodr{\'\i}guez del Pino}}]{2020A&A...643A.139P}
{Perna}, M., {Arribas}, S., {Catal{\'a}n-Torrecilla}, C., {et~al.} 2020, \aap, 643, A139, \dodoi{10.1051/0004-6361/202038328}

\bibitem[{{Peterson} {et~al.}(2003){Peterson}, {Kahn}, {Paerels}, {Kaastra}, {Tamura}, {Bleeker}, {Ferrigno}, \& {Jernigan}}]{2003ApJ...590..207P}
{Peterson}, J.~R., {Kahn}, S.~M., {Paerels}, F.~B.~S., {et~al.} 2003, \apj, 590, 207, \dodoi{10.1086/374830}

\bibitem[{{Rao} {et~al.}(2023){Rao}, {Kharb}, {Rubinur}, {Silpa}, {Roy}, {Sebastian}, {Singh}, {Baghel}, {Manna}, \& {Ishwara-Chandra}}]{Rao2023}
{Rao}, V.~V., {Kharb}, P., {Rubinur}, K., {et~al.} 2023, \mnras, 524, 1615, \dodoi{10.1093/mnras/stad1901}

\bibitem[{{Rau} \& {Cornwell}(2011)}]{RauCornwell11}
{Rau}, U., \& {Cornwell}, T.~J. 2011, \aap, 532, A71, \dodoi{10.1051/0004-6361/201117104}

\bibitem[{{Riffel} {et~al.}(2014){Riffel}, {Storchi-Bergmann}, \& {Riffel}}]{Riffel2014}
{Riffel}, R.~A., {Storchi-Bergmann}, T., \& {Riffel}, R. 2014, \apjl, 780, L24, \dodoi{10.1088/2041-8205/780/2/L24}

\bibitem[{{Rosario} {et~al.}(2013){Rosario}, {Burtscher}, {Davies}, {Genzel}, {Lutz}, \& {Tacconi}}]{Rosario2013}
{Rosario}, D.~J., {Burtscher}, L., {Davies}, R., {et~al.} 2013, \apj, 778, 94, \dodoi{10.1088/0004-637X/778/2/94}

\bibitem[{{Rose} {et~al.}(2018){Rose}, {Tadhunter}, {Ramos Almeida}, {Rodr{\'\i}guez Zaur{\'\i}n}, {Santoro}, \& {Spence}}]{2018MNRAS.474..128R}
{Rose}, M., {Tadhunter}, C., {Ramos Almeida}, C., {et~al.} 2018, \mnras, 474, 128, \dodoi{10.1093/mnras/stx2590}

\bibitem[{{Rubinur} {et~al.}(2017){Rubinur}, {Das}, {Kharb}, \& {Honey}}]{Rubinur2017}
{Rubinur}, K., {Das}, M., {Kharb}, P., \& {Honey}, M. 2017, \mnras, 465, 4772, \dodoi{10.1093/mnras/stw2981}

\bibitem[{{Sambruna} {et~al.}(2001){Sambruna}, {Netzer}, {Kaspi}, {Brandt}, {Chartas}, {Garmire}, {Nousek}, \& {Weaver}}]{Sambruna2001}
{Sambruna}, R.~M., {Netzer}, H., {Kaspi}, S., {et~al.} 2001, \apjl, 546, L13, \dodoi{10.1086/318068}

\bibitem[{{Sanders} {et~al.}(2016){Sanders}, {Shapley}, {Kriek}, {Reddy}, {Freeman}, {Coil}, {Siana}, {Mobasher}, {Shivaei}, {Price}, \& {de Groot}}]{2016ApJ...816...23S}
{Sanders}, R.~L., {Shapley}, A.~E., {Kriek}, M., {et~al.} 2016, \apj, 816, 23, \dodoi{10.3847/0004-637X/816/1/23}

\bibitem[{{Sartori} {et~al.}(2016){Sartori}, {Schawinski}, {Koss}, {Treister}, {Maksym}, {Keel}, {Urry}, {Lintott}, \& {Wong}}]{2016MNRAS.457.3629S}
{Sartori}, L.~F., {Schawinski}, K., {Koss}, M., {et~al.} 2016, \mnras, 457, 3629, \dodoi{10.1093/mnras/stw230}

\bibitem[{{Sarzi} {et~al.}(2006){Sarzi}, {Falc{\'o}n-Barroso}, {Davies}, {Bacon}, {Bureau}, {Cappellari}, {de Zeeuw}, {Emsellem}, {Fathi}, {Krajnovi{\'c}}, {Kuntschner}, {McDermid}, \& {Peletier}}]{2006MNRAS.366.1151S}
{Sarzi}, M., {Falc{\'o}n-Barroso}, J., {Davies}, R.~L., {et~al.} 2006, \mnras, 366, 1151, \dodoi{10.1111/j.1365-2966.2005.09839.x}

\bibitem[{{Schawinski} {et~al.}(2007){Schawinski}, {Thomas}, {Sarzi}, {Maraston}, {Kaviraj}, {Joo}, {Yi}, \& {Silk}}]{2007MNRAS.382.1415S}
{Schawinski}, K., {Thomas}, D., {Sarzi}, M., {et~al.} 2007, \mnras, 382, 1415, \dodoi{10.1111/j.1365-2966.2007.12487.x}

\bibitem[{{Schawinski} {et~al.}(2010){Schawinski}, {Evans}, {Virani}, {Urry}, {Keel}, {Natarajan}, {Lintott}, {Manning}, {Coppi}, {Kaviraj}, {Bamford}, {J{\'o}zsa}, {Garrett}, {van Arkel}, {Gay}, \& {Fortson}}]{2010ApJ...724L..30S}
{Schawinski}, K., {Evans}, D.~A., {Virani}, S., {et~al.} 2010, \apjl, 724, L30, \dodoi{10.1088/2041-8205/724/1/L30}

\bibitem[{{Schaye} \& {Dalla Vecchia}(2008)}]{2008MNRAS.383.1210S}
{Schaye}, J., \& {Dalla Vecchia}, C. 2008, \mnras, 383, 1210, \dodoi{10.1111/j.1365-2966.2007.12639.x}

\bibitem[{{Schaye} {et~al.}(2015){Schaye}, {Crain}, {Bower}, {Furlong}, {Schaller}, {Theuns}, {Dalla Vecchia}, {Frenk}, {McCarthy}, {Helly}, {Jenkins}, {Rosas-Guevara}, {White}, {Baes}, {Booth}, {Camps}, {Navarro}, {Qu}, {Rahmati}, {Sawala}, {Thomas}, \& {Trayford}}]{Schaye2015}
{Schaye}, J., {Crain}, R.~A., {Bower}, R.~G., {et~al.} 2015, \mnras, 446, 521, \dodoi{10.1093/mnras/stu2058}

\bibitem[{{Schiano}(1986)}]{1986ApJ...302...81S}
{Schiano}, A.~V.~R. 1986, \apj, 302, 81, \dodoi{10.1086/163975}

\bibitem[{{Schmitt} {et~al.}(2003){Schmitt}, {Donley}, {Antonucci}, {Hutchings}, {Kinney}, \& {Pringle}}]{Schmitt2003}
{Schmitt}, H.~R., {Donley}, J.~L., {Antonucci}, R.~R.~J., {et~al.} 2003, \apj, 597, 768, \dodoi{10.1086/381224}

\bibitem[{{Sebastian} {et~al.}(2018){Sebastian}, {Ishwara-Chandra}, {Joshi}, \& {Wadadekar}}]{Sebastian2018}
{Sebastian}, B., {Ishwara-Chandra}, C.~H., {Joshi}, R., \& {Wadadekar}, Y. 2018, \mnras, 473, 4926, \dodoi{10.1093/mnras/stx2631}

\bibitem[{{Sebastian} {et~al.}(2019){Sebastian}, {Kharb}, {O'Dea}, {Gallimore}, \& {Baum}}]{Sebastian2019}
{Sebastian}, B., {Kharb}, P., {O'Dea}, C.~P., {Gallimore}, J.~F., \& {Baum}, S.~A. 2019, \mnras, 490, L26, \dodoi{10.1093/mnrasl/slz136}

\bibitem[{{Sebastian} {et~al.}(2020){Sebastian}, {Kharb}, {O'Dea}, {Gallimore}, \& {Baum}}]{2020MNRAS.499..334S}
---. 2020, \mnras, 499, 334, \dodoi{10.1093/mnras/staa2473}

\bibitem[{{Sebastian} {et~al.}(2024){Sebastian}, {Caproni}, {Kharb}, {Nayana}, {Ali}, {Rubinur}, {O'Dea}, {Baum}, \& {Nandi}}]{Sebastian2024}
{Sebastian}, B., {Caproni}, A., {Kharb}, P., {et~al.} 2024, \mnras, 530, 4902, \dodoi{10.1093/mnras/stae546}

\bibitem[{{Shih} \& {Stockton}(2014)}]{2014ApJ...786....3S}
{Shih}, H.-Y., \& {Stockton}, A. 2014, \apj, 786, 3, \dodoi{10.1088/0004-637X/786/1/3}

\bibitem[{{Silk} \& {Mamon}(2012)}]{Silk2012}
{Silk}, J., \& {Mamon}, G.~A. 2012, Research in Astronomy and Astrophysics, 12, 917, \dodoi{10.1088/1674-4527/12/8/004}

\bibitem[{{Silpa} {et~al.}(2022){Silpa}, {Kharb}, {Harrison}, {Girdhar}, {Mukherjee}, {Mainieri}, \& {Jarvis}}]{2022MNRAS.513.4208S}
{Silpa}, S., {Kharb}, P., {Harrison}, C.~M., {et~al.} 2022, \mnras, 513, 4208, \dodoi{10.1093/mnras/stac1044}

\bibitem[{{Singh} {et~al.}(2013){Singh}, {van de Ven}, {Jahnke}, {Lyubenova}, {Falc{\'o}n-Barroso}, {Alves}, {Cid Fernandes}, {Galbany}, {Garc{\'\i}a-Benito}, {Husemann}, {Kennicutt}, {Marino}, {M{\'a}rquez}, {Masegosa}, {Mast}, {Pasquali}, {S{\'a}nchez}, {Walcher}, {Wild}, {Wisotzki}, \& {Ziegler}}]{Singh2013}
{Singh}, R., {van de Ven}, G., {Jahnke}, K., {et~al.} 2013, \aap, 558, A43, \dodoi{10.1051/0004-6361/201322062}

\bibitem[{{Singha} {et~al.}(2023){Singha}, {O'Dea}, \& {Baum}}]{Singha2023}
{Singha}, M., {O'Dea}, C.~P., \& {Baum}, S.~A. 2023, Galaxies, 11, 85, \dodoi{10.3390/galaxies11040085}

\bibitem[{{Singha} {et~al.}(2022){Singha}, {Husemann}, {Urrutia}, {O'Dea}, {Scharw{\"a}chter}, {Gaspari}, {Combes}, {Nevin}, {Terrazas}, {P{\'e}rez-Torres}, {Rose}, {Davis}, {Tremblay}, {Neumann}, {Smirnova-Pinchukova}, \& {Baum}}]{Singha2022CARS}
{Singha}, M., {Husemann}, B., {Urrutia}, T., {et~al.} 2022, \aap, 659, A123, \dodoi{10.1051/0004-6361/202040122}

\bibitem[{{Smith} {et~al.}(2022){Smith}, {Krause}, {Hardcastle}, \& {Drake}}]{2022MNRAS.514.3879S}
{Smith}, D.~J.~B., {Krause}, M.~G., {Hardcastle}, M.~J., \& {Drake}, A.~B. 2022, \mnras, 514, 3879, \dodoi{10.1093/mnras/stac1568}

\bibitem[{{Spoon} \& {Holt}(2009)}]{Spoon2009}
{Spoon}, H.~W.~W., \& {Holt}, J. 2009, \apjl, 702, L42, \dodoi{10.1088/0004-637X/702/1/L42}

\bibitem[{{Steiman-Cameron} {et~al.}(1992){Steiman-Cameron}, {Kormendy}, \& {Durisen}}]{1992AJ....104.1339S}
{Steiman-Cameron}, T.~Y., {Kormendy}, J., \& {Durisen}, R.~H. 1992, \aj, 104, 1339, \dodoi{10.1086/116323}

\bibitem[{{Stocke} {et~al.}(1992){Stocke}, {Morris}, {Weymann}, \& {Foltz}}]{Stocke1992}
{Stocke}, J.~T., {Morris}, S.~L., {Weymann}, R.~J., \& {Foltz}, C.~B. 1992, \apj, 396, 487, \dodoi{10.1086/171735}

\bibitem[{{Storey} \& {Zeippen}(2000)}]{2000MNRAS.312..813S}
{Storey}, P.~J., \& {Zeippen}, C.~J. 2000, \mnras, 312, 813, \dodoi{10.1046/j.1365-8711.2000.03184.x}

\bibitem[{{Sutherland} {et~al.}(2018){Sutherland}, {Dopita}, {Binette}, \& {Groves}}]{Sutherl&Dopita2018}
{Sutherland}, R., {Dopita}, M., {Binette}, L., \& {Groves}, B. 2018, {MAPPINGS V: Astrophysical plasma modeling code}, Astrophysics Source Code Library, record ascl:1807.005

\bibitem[{{Sutherland} \& {Bicknell}(2007)}]{Sutherland&Bicknell2007}
{Sutherland}, R.~S., \& {Bicknell}, G.~V. 2007, \apjs, 173, 37, \dodoi{10.1086/520640}

\bibitem[{{Tadhunter} {et~al.}(1989){Tadhunter}, {Fosbury}, \& {Quinn}}]{1989MNRAS.240..225T}
{Tadhunter}, C.~N., {Fosbury}, R.~A.~E., \& {Quinn}, P.~J. 1989, \mnras, 240, 225, \dodoi{10.1093/mnras/240.2.225}

\bibitem[{{Thompson} {et~al.}(2015){Thompson}, {Fabian}, {Quataert}, \& {Murray}}]{2015MNRAS.449..147T}
{Thompson}, T.~A., {Fabian}, A.~C., {Quataert}, E., \& {Murray}, N. 2015, \mnras, 449, 147, \dodoi{10.1093/mnras/stv246}

\bibitem[{{Vazdekis} {et~al.}(2015){Vazdekis}, {Coelho}, {Cassisi}, {Ricciardelli}, {Falc{\'o}n-Barroso}, {S{\'a}nchez-Bl{\'a}zquez}, {La Barbera}, {Beasley}, \& {Pietrinferni}}]{2015MNRAS.449.1177V}
{Vazdekis}, A., {Coelho}, P., {Cassisi}, S., {et~al.} 2015, \mnras, 449, 1177, \dodoi{10.1093/mnras/stv151}

\bibitem[{{Veilleux}(1991)}]{Veilleux1991}
{Veilleux}, S. 1991, \apj, 369, 331, \dodoi{10.1086/169765}

\bibitem[{{Veilleux} \& {Osterbrock}(1987)}]{1987ApJS...63..295V}
{Veilleux}, S., \& {Osterbrock}, D.~E. 1987, \apjs, 63, 295, \dodoi{10.1086/191166}

\bibitem[{{Venturi} {et~al.}(2021){Venturi}, {Cresci}, {Marconi}, {Mingozzi}, {Nardini}, {Carniani}, {Mannucci}, {Marasco}, {Maiolino}, {Perna}, {Treister}, {Bland-Hawthorn}, \& {Gallimore}}]{2021A&A...648A..17V}
{Venturi}, G., {Cresci}, G., {Marconi}, A., {et~al.} 2021, \aap, 648, A17, \dodoi{10.1051/0004-6361/202039869}

\bibitem[{{Venturi} {et~al.}(2023){Venturi}, {Treister}, {Finlez}, {D'Ago}, {Bauer}, {Harrison}, {Ramos Almeida}, {Revalski}, {Ricci}, {Sartori}, {Girdhar}, {Keel}, \& {Tub{\'\i}n}}]{Venturi2023}
{Venturi}, G., {Treister}, E., {Finlez}, C., {et~al.} 2023, \aap, 678, A127, \dodoi{10.1051/0004-6361/202347375}

\bibitem[{{Veron} \& {Veron-Cetty}(1995)}]{1995A&A...296..315V}
{Veron}, P., \& {Veron-Cetty}, M.~P. 1995, \aap, 296, 315

\bibitem[{{Villar-Mart{\'\i}n} {et~al.}(2017){Villar-Mart{\'\i}n}, {Emonts}, {Cabrera Lavers}, {Tadhunter}, {Mukherjee}, {Humphrey}, {Rodr{\'\i}guez Zaur{\'\i}n}, {Ramos Almeida}, {P{\'e}rez Torres}, \& {Bessiere}}]{2017MNRAS.472.4659V}
{Villar-Mart{\'\i}n}, M., {Emonts}, B., {Cabrera Lavers}, A., {et~al.} 2017, \mnras, 472, 4659, \dodoi{10.1093/mnras/stx2209}

\bibitem[{{Virtanen} {et~al.}(2020){Virtanen}, {Gommers}, {Oliphant}, {Haberland}, {Reddy}, {Cournapeau}, {Burovski}, {Peterson}, {Weckesser}, {Bright}, {van der Walt}, {Brett}, {Wilson}, {Millman}, {Mayorov}, {Nelson}, {Jones}, {Kern}, {Larson}, {Carey}, {Polat}, {Feng}, {Moore}, {VanderPlas}, {Laxalde}, {Perktold}, {Cimrman}, {Henriksen}, {Quintero}, {Harris}, {Archibald}, {Ribeiro}, {Pedregosa}, {van Mulbregt}, \& {SciPy 1. 0 Contributors}}]{2020NatMe..17..261V}
{Virtanen}, P., {Gommers}, R., {Oliphant}, T.~E., {et~al.} 2020, Nature Methods, 17, 261, \dodoi{10.1038/s41592-019-0686-2}

\bibitem[{{Wang}(2008)}]{Wang2008}
{Wang}, J.-M. 2008, \apjl, 682, L81, \dodoi{10.1086/590928}

\bibitem[{{Whittle} \& {Wilson}(2004)}]{2004AJ....127..606W}
{Whittle}, M., \& {Wilson}, A.~S. 2004, \aj, 127, 606, \dodoi{10.1086/380940}

\bibitem[{{Zakamska} \& {Greene}(2014)}]{Zakamska2014}
{Zakamska}, N.~L., \& {Greene}, J.~E. 2014, \mnras, 442, 784, \dodoi{10.1093/mnras/stu842}

\bibitem[{{Zhang} {et~al.}(2024){Zhang}, {Packham}, {Hicks}, {Davies}, {Shimizu}, {Alonso-Herrero}, {Hermosa Mu{\~n}oz}, {Garc{\'\i}a-Bernete}, {Pereira-Santaella}, {Audibert}, {L{\'o}pez-Rodr{\'\i}guez}, {Bellocchi}, {Bunker}, {Combes}, {D{\'\i}az-Santos}, {Gandhi}, {Garc{\'\i}a-Burillo}, {Garc{\'\i}a-Lorenzo}, {Gonz{\'a}lez-Mart{\'\i}n}, {Imanishi}, {Labiano}, {Leist}, {Levenson}, {Ramos Almeida}, {Ricci}, {Rigopoulou}, {Rosario}, {Stalevski}, {Ward}, {Esparza-Arredondo}, {Delaney}, {Fuller}, {Haidar}, {H{\"o}nig}, {Izumi}, \& {Rouan}}]{Zhang2024}
{Zhang}, L., {Packham}, C., {Hicks}, E. K.~S., {et~al.} 2024, \apj, 974, 195, \dodoi{10.3847/1538-4357/ad6a4b}

\bibitem[{{Zubovas} \& {King}(2012)}]{Zubovas2012}
{Zubovas}, K., \& {King}, A. 2012, \apjl, 745, L34, \dodoi{10.1088/2041-8205/745/2/L34}

\end{thebibliography}
\bibliographystyle{aasjournal}







\appendix

\renewcommand{\thefigure}{A\arabic{figure}} 
\begin{figure}
\centering
\includegraphics[height=7cm,trim = 0 0 0 0 ]{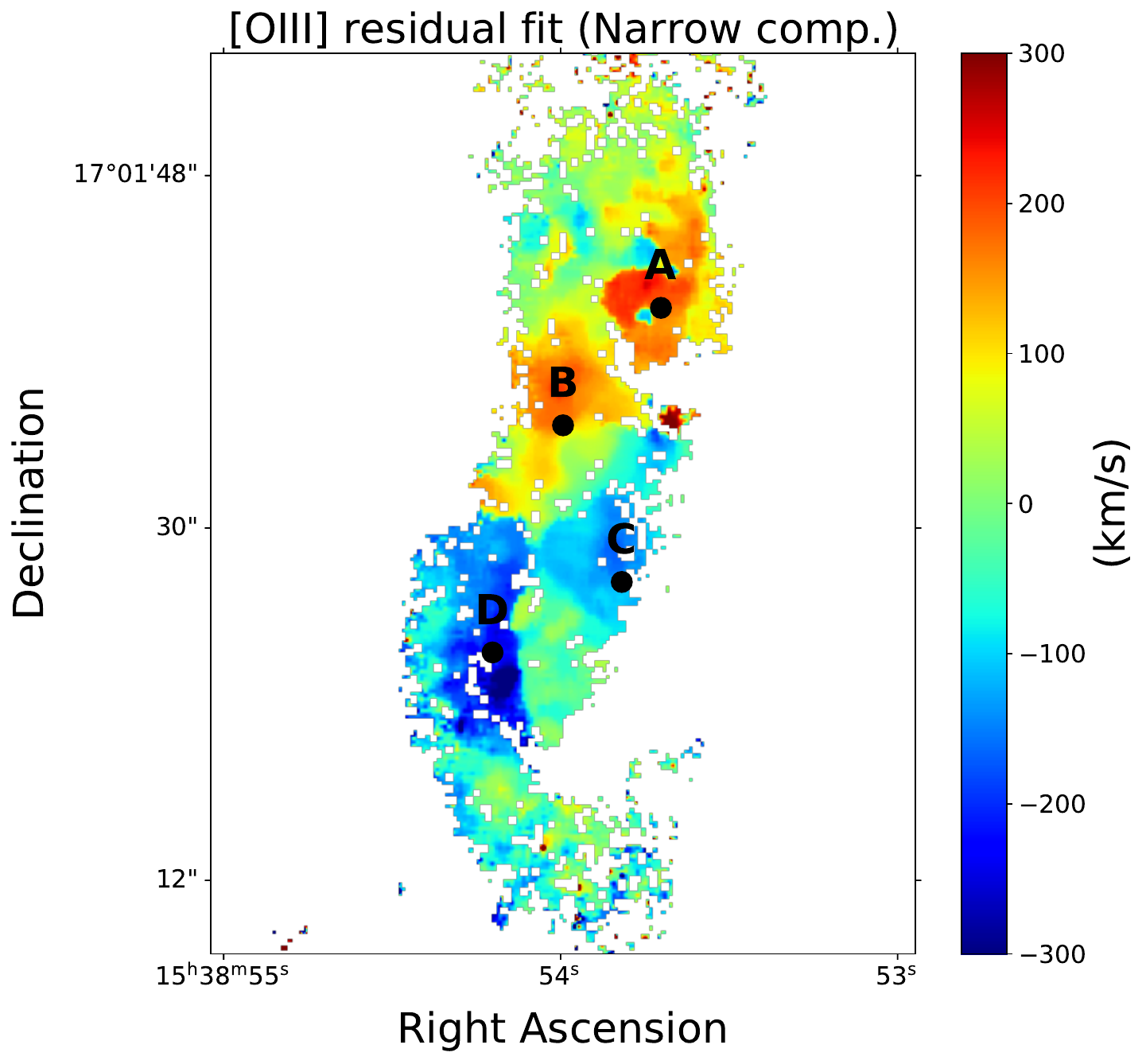}

\caption
{Pixel locations used for representing the double Gaussian fitting. The map used is the [O\,{\small{III}}] residual fit for Gaussian 1.
}
\label{spaxel}
\end{figure}

\renewcommand{\thefigure}{A\arabic{figure}} 
\begin{figure*}
\centering

\includegraphics[height=6cm,trim = 0 0 0 0 ]{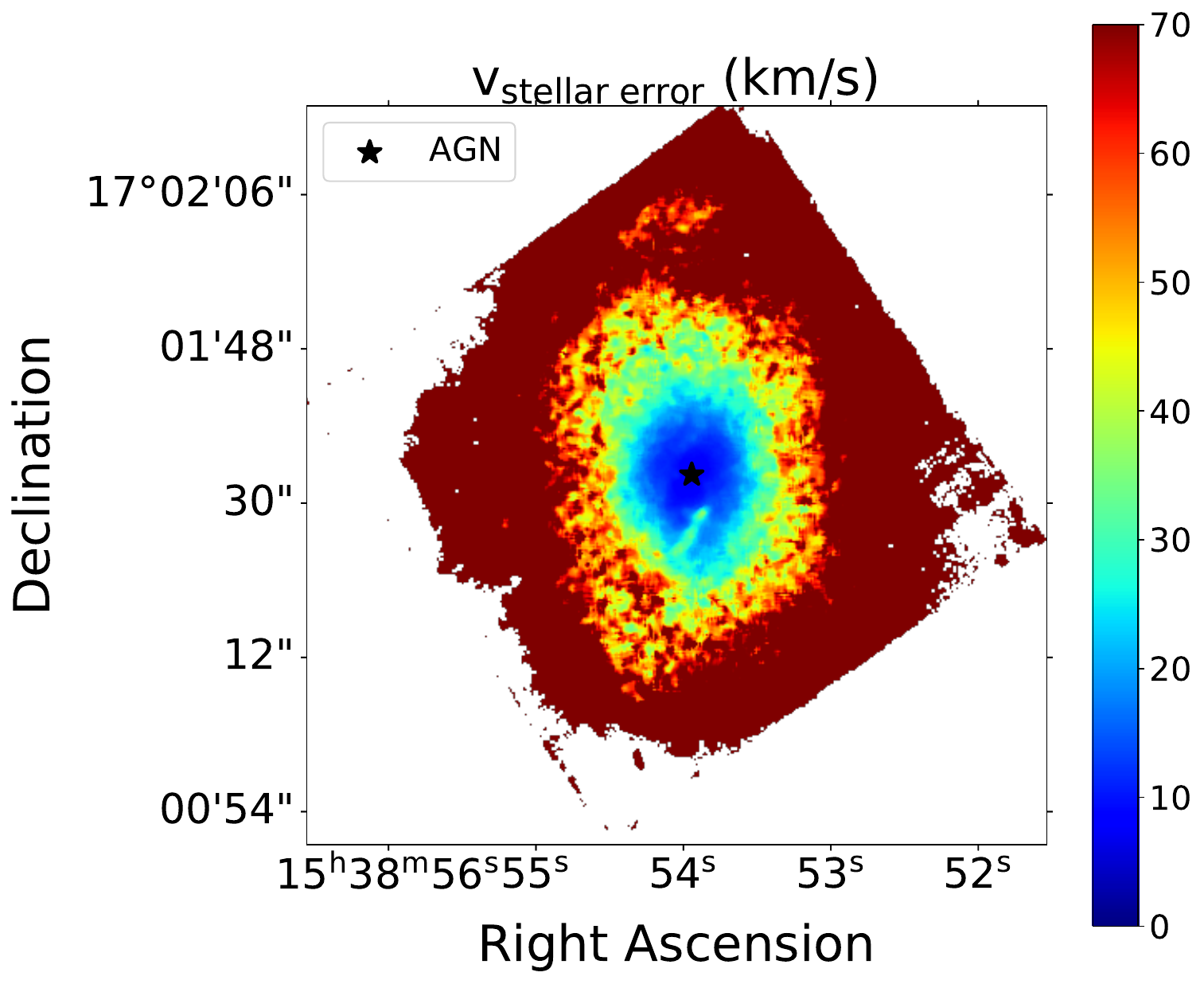}
\includegraphics[height=6cm,trim = 0 0 0 0 ]{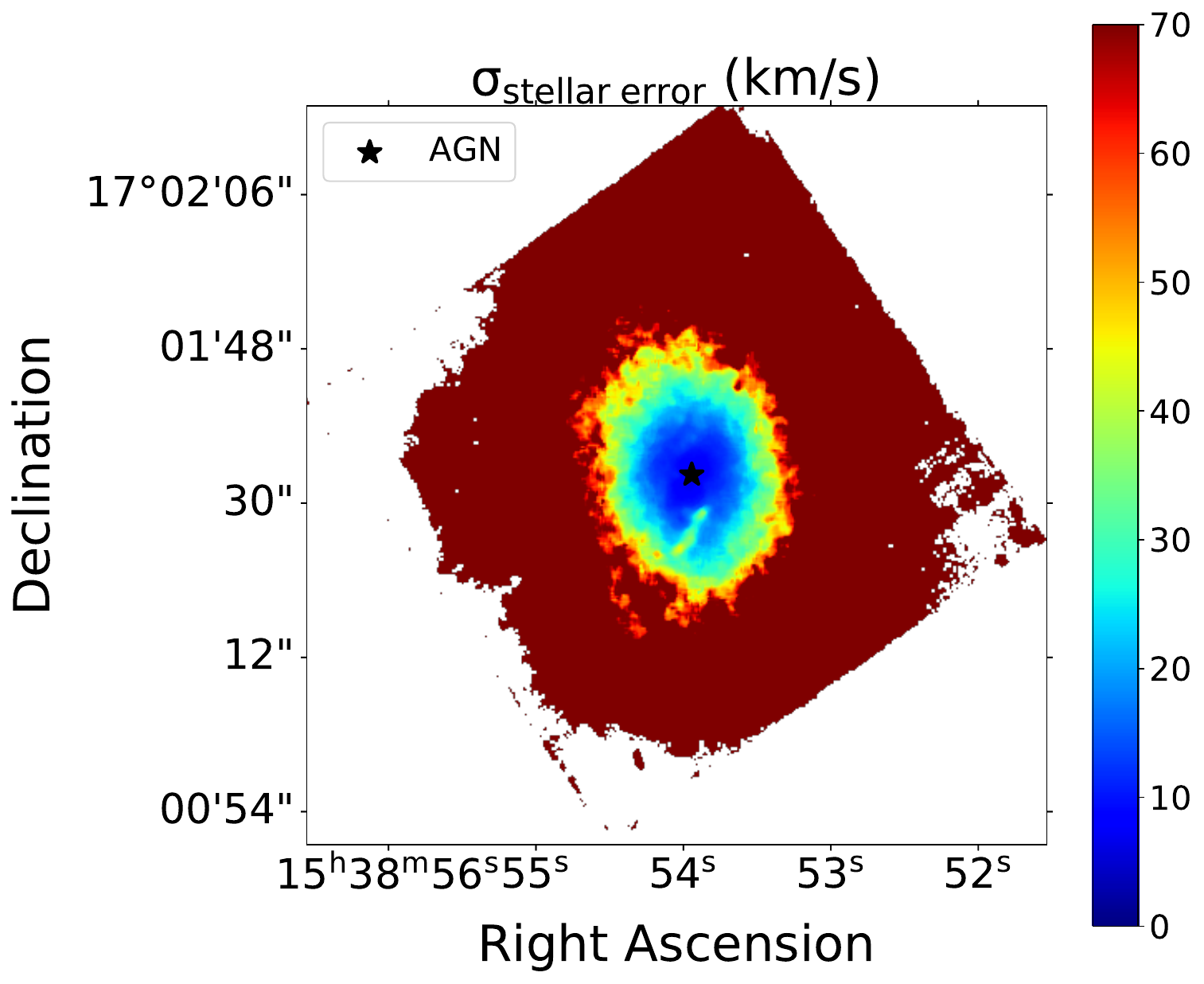}

\caption
{Error maps for line-of-sight stellar velocity and stellar velocity dispersion. 
}
\label{stellar_err_map}
\end{figure*}

\renewcommand{\thefigure}{A\arabic{figure}} 

\end{document}